\documentclass[a4paper,11pt]{article}
\usepackage{jheppub}
\usepackage{silence}
\usepackage{bm}
\usepackage[missing=]{gitinfo2}
\usepackage{xcolor}
\definecolor{darkred}{rgb}{0.5,0.15,0.15}
\hypersetup{colorlinks=true,urlcolor=darkred,linkcolor=darkred,citecolor=darkred}
\usepackage{soul}
\usepackage{graphicx}
\usepackage{epsfig}
\usepackage{amsmath}
\usepackage{amssymb}
\usepackage{amsthm}
\usepackage{indentfirst}
\usepackage{xspace}
\usepackage{multirow}
\usepackage{hyperref}
\usepackage{verbatim}
\usepackage{subcaption}
\usepackage{tikz-cd}
\usepackage{capt-of}
\usepackage{color}

\usepackage{tikz}
\usetikzlibrary{calc}   
\usepackage{varioref}

\usetikzlibrary{topaths}
\usetikzlibrary{decorations}
\usetikzlibrary{decorations.pathmorphing}

\numberwithin{equation}{section}

\newcommand{\CN}{{\cal N}}

\newcommand{\be}{\begin{equation}}
\newcommand{\ee}{\end{equation}}
\newcommand{\ba}{\begin{aligned}}
\newcommand{\ea}{\end{aligned}}

\newcommand{\cM}{\ensuremath{\mathcal M}}

\newcommand{\cX}{\ensuremath{\mathcal X}}

\newcommand{\cW}{\ensuremath{\mathcal W}}

\def\IF{{\mathbb F}}

\newcommand{\CB}{{\cal B}}
\def\IR{{\mathbb R}}
\def\IZ{{\mathbb Z}}

\newcommand{\R}{\ensuremath{\mathbb R}}
\newcommand{\C}{\ensuremath{\mathbb C}}

\newcommand{\Z}{\ensuremath{\mathbb Z}}

\newcommand{\half}{\ensuremath{\frac{1}{2}}}

\newcommand{\N}{{\mathcal N}}

\newcommand{\ri}{{\rm i}}
\newcommand{\rd}{{\rm d}}

\newcommand{\I}{{\mathrm i}}
\newcommand{\E}{{\mathrm e}}
\newcommand{\de}{\mathrm{d}}

\newcommand{\adi}{{\tt a}_{\rm D}}
\newcommand{\ai}{{\tt a}}

\newcommand{\xadi}{\re^{{\tt a}_{\rm D}}}

\newcommand{\abs}[1]{\lvert#1\rvert}

\newcommand{\IP}[1]{\langle#1\rangle}

\newcommand{\eps}{\epsilon}

\newcommand{\ti}[1]{\textit{#1}}

\DeclareMathOperator{\im}{Im}
\DeclareMathOperator{\re}{{\rm e}}
\DeclareMathOperator{\Tr}{Tr}

\usepackage{color}

\usepackage{scalerel,stackengine}
\stackMath
\newcommand\reallywidehat[1]{%
\savestack{\tmpbox}{\stretchto{%
  \scaleto{%
    \scalerel*[\widthof{\ensuremath{#1}}]{\kern-.6pt\bigwedge\kern-.6pt}%
    {\rule[-\textheight/2]{1ex}{\textheight}}
  }{\textheight}%
}{0.5ex}}%
\stackon[1pt]{#1}{\tmpbox}%
}
\begin{document}

\abstract{ We study in detail
the Schr\"{o}dinger equation corresponding to the four dimensional $SU(2)$ $\CN=2$ SQCD theory with one flavour. 
We calculate the {Voros symbols}, or {quantum periods}, in four different ways: Borel summation of
the WKB series, direct computation of Wronskians of exponentially decaying 
solutions, the TBA equations of Gaiotto-Moore-Neitzke/Gaiotto,
and instanton counting. 
We make computations by all of these methods, finding good agreement. We also study the exact
quantization condition for the spectrum, and we compute 
the Fredholm determinant of the inverse of the Schr\"odinger operator 
using the TS/ST correspondence and Zamolodchikov's TBA, again finding good agreement.
In addition, we  explore two aspects of the relationship between singularities of the Borel transformed
WKB series and BPS states: BPS states of the 4d theory are related to singularities in the Borel transformed WKB series
for the quantum periods, and BPS states of a coupled 2d+4d system are related to singularities in the Borel transformed WKB
series for local solutions of the Schr\"odinger equation.}

\preprint{{\flushright
UTTG 03-2021\\
CERN-TH-2021-071\\
}}

\title{Exact WKB methods in $SU(2)$  $N_f=1$}
\date{{{\tiny \color{gray} \tt \gitAuthorIsoDate}}
{{\tiny \color{gray} \tt \gitAbbrevHash}}}
\author[1,2]{Alba Grassi}
\author[3]{Qianyu Hao}
\author[4]{Andrew Neitzke}
\affiliation[1]{Section de Math\'{e}matiques, Universit\'{e} de Gen\`{e}ve, 1211 Gen\`{e}ve 4, Switzerland}
\affiliation[2]{Theoretical Physics Department, CERN, 1211 Geneva 23, Switzerland}
\affiliation[3]{Department of Physics, University of Texas at Austin}
\affiliation[4]{Department of Mathematics, Yale University}

\maketitle

\setcounter{page}{1}

\section{Introduction}
\label{sec:intro}

This paper concerns a detailed exploration of the relation between Schr\"odinger 
equations and $\N=2$ supersymmetric gauge theories, from several different points of view.
We focus on one particular example, the $SU(2)$ theory with one fundamental hypermultiplet.
In this introduction we briefly recall the main players in this story, 
describe the example we study, and indicate some of the highlights.

\subsection{Quantum periods} \label{sec:quantum-periods}

Consider a Schr\"odinger equation, of the form 
\begin{equation} \label{eq:schrodinger-eq}
	\left( - \hbar^2 \partial_x^2 + V(x) - E \right) \psi(x,\hbar) = 0,
\end{equation}
with a confining potential $V(x)$. 
Then a well-known problem is to identify the bound state energies, i.e. the values
$E(\hbar)$ for which there exists a solution $\psi(x,\hbar)$ on the real line $x \in \R$ such
that $\psi(x,\hbar)$ decays 
exponentially as $x \to \infty$ and as $x \to -\infty$.

More generally, we may consider an equation \eqref{eq:schrodinger-eq} where $x$ and $\hbar$ 
are allowed to be complex, and $V(x)$ is a holomorphic or meromorphic function.
Then the potential may not be confining on the real line, but we can pick a more general contour where it is
confining,
and then ask for the existence of a solution which is exponentially decaying at both ends of the contour. This
problem again determines some discrete energies $E(\hbar)$, which we may think of as generalized bound states,
or resonances.

One of the important insights of the exact WKB method \cite{bpv,voros-quartic,voros, ddpham,reshyper}
is that the (generalized) bound state energies can be usefully
expressed as solutions of an \ti{exact quantization condition}.
The exact quantization condition is an algebraic equation
in terms of quantities $\cX_\gamma = \exp \Pi_\gamma$
labeled by cycles $\gamma$ on the spectral curve  \cite{bpv}
\begin{equation} \label{eq:spectral-curve-intro}
	\Sigma = \{ y^2 = V(x) - E \}.
\end{equation}
The $\Pi_\gamma$ are the \ti{quantum periods}.\footnote{In the exact WKB terminology these quantities are also called Voros symbols. More recently they have been called spectral coordinates,
and (in some cases) identified with 
Fock-Goncharov coordinates, Fenchel-Nielsen coordinates or higher analogues thereof on moduli spaces
of flat connections; see \cite{in-exactwkb} for a clear account of the relation between
quantum periods and Fock-Goncharov coordinates.} Up to a factor of $\hbar$,
they can be thought of as an $\hbar$-deformation of the 
classical periods $Z_\gamma = \oint_\gamma \sqrt{V(x)-E} \, \de x$;
indeed,
they are defined\footnote{Strictly speaking we take only the even part of this series, but this only affects the quantum
periods by a simple shift: see \eqref{mon}.} by
Borel summation of the series $\hbar^{-1} \oint_\gamma Y(x,\hbar) \de x$, 
where $Y(x,\hbar) = \sqrt{V(x) - E} + \sum_{n \ge 1} \hbar^n Y_n $ is 
the exponent in the WKB construction of solutions to 
the Schr\"odinger equation \eqref{eq:schrodinger-eq}.

There is a connection between some Schr\"odinger equations
\eqref{eq:schrodinger-eq} and $\N=2$ supersymmetric gauge theories 
placed in the Nekrasov-Shatashvili (NS) limit of the $\Omega$-background.
In this connection the spectral curve 
\eqref{eq:spectral-curve-intro} is identified as the Seiberg-Witten (SW)
curve of the field theory, $\sqrt{V(x)-E} \, \de x$ is the
Seiberg-Witten differential,
and $\hbar$ is identified
with the $\Omega$-background parameter $\eps_1$.
In particular, the classical periods $Z_\gamma$ 
control the central charges and masses of BPS states, while the 
quantum periods $\Pi_\gamma$ can be computed
in terms of the Nekrasov-Shatashvili instanton partition function.\footnote{The $\cX_\gamma$ can also be 
understood as vacuum expectation values of supersymmetric IR line defects, in a certain scaling limit called ``conformal limit'' \cite{Gaiotto:2014bza,Hollands:2019wbr}.
In the present paper, though, we define the $\cX_\gamma$ directly as functions of $\hbar$, rather than defining them as conformal limits of line defect vevs.}
This connection has been derived from various different points of view, including
direct gauge theory computations and class $S$ constructions using the AGT correspondence; see
 e.g.
\cite{ns,nrs,Nekrasov:2010ka,Drukker:2009id,Alday:2009fs,Maruyoshi:2010iu,Jeong:2018qpc,mirmor}.
Some recent papers which discuss quantum periods and their relation to four dimensional $\N=2$ theories are
\cite{kpt,Emery:2019znd,Dumas:2020zoz,Ito:2018eon,Ito:2019twh,Hollands:2019wbr,Fioravanti:2019awr,Yan:2020kkb,gm3,Hatsuda_2018,Imaizumi:2021cxf,Emery:2020qqu,Aminov:2020yma,Ito:2021boh}.

\subsection{The \texorpdfstring{$SU(2)$ $N_f = 1$}{SU(2) Nf=1} equation}\label{sec:our-example}

In this paper we consider a specific example of \eqref{eq:schrodinger-eq}:
\begin{equation}
\label{eq:our-eq-intro}
\left(-\hbar^2\partial_x^2+\left(\half \Lambda^2 \re^{-x}+2 m \Lambda \re^{x} - \Lambda^2 \re^{2x}\right)-E\right)\psi(x)=0. 
\end{equation}
This equation corresponds to 
$\N=2$ supersymmetric Yang-Mills theory with gauge group 
$SU(2)$ and $N_f = 1$;\footnote{There are many different choices of convention for writing the equation
corresponding to this theory; our convention matches that of \cite{Maruyoshi:2010iu}.}
$\Lambda \in \C^\times$ is the dynamical scale, $m \in \C$ is
the hypermultiplet mass, and $E \in \C$ parameterizes the
Coulomb branch.

\subsection{Computing the quantum periods}

Much of the utility of the quantum periods comes from the fact that they can be understood
and computed from many different points of view.
In Sections \ref{sec:borel}-\ref{sec:QPI} below we compute the quantum periods of 
the equation \eqref{eq:our-eq-intro}, at various points
of the parameter space, in four different ways:
\begin{enumerate}
	\item \ti{Pad\'{e}-Borel summation} (\autoref{sec:borel}): 
	As we have explained above, the $\Pi_\gamma$ are defined 
	by Borel summation of integrals $\hbar^{-1} \oint_\gamma Y(x,\hbar) \, \de x$.
	We compute this series numerically to high order in $\hbar$,
	and then use the method of Pad\'{e}-Borel summation to produce 
	numerical approximations $\Pi^{\mathrm{PB}}_\gamma$ 
	of $\Pi_\gamma$.
	\item \ti{Wronskians} (\autoref{sec:ssection}): The equation \eqref{eq:our-eq-intro} admits
	three distinguished solutions $\psi_1$, $\psi_2$, $\psi_3$ 
	which have exponential decay along three distinguished directions
	in the $x$-plane; e.g. when $\hbar, \Lambda > 0$ these directions are $x \to -\infty$, $x \to \infty - \half \I \pi$, 
	$x \to \infty + \half \I \pi$ respectively.
	The $\cX_\gamma$ can be expressed directly
	in terms of combinations
	of Wronskians of the solutions $\psi_i$ and their images under the 
	monodromy operation of continuation $x \to x+2 \pi \I$; see e.g. \eqref{s0}-\eqref{s2} below for
	examples of such formulas.\footnote{One can think of these combinations of Wronskians as slight generalizations of the 
	Fock-Goncharov or Fenchel-Nielsen coordinates on moduli spaces of flat connections.}
	By direct numerical
	integration of \eqref{eq:our-eq-intro} we can thus compute
	numerical approximations $\cX^{\mathrm{SS}}_\gamma$ 
	of $\cX_\gamma$.
	\item \ti{TBA equations} (\autoref{sec:GMNTBA}): The $\cX_\gamma$ are (conjecturally) solutions of 
	certain integral equations given in \cite{gmn,Gaiotto:2014bza}, closely
	related to the ODE/IM correspondence \cite{ddt}. These equations are
	formulated in terms of the classical periods (central charges) 
	and the BPS state spectrum of the theory. 
	We solve these integral
	equations numerically, for parameters lying in the strong coupling 
	region (in this region the BPS spectrum is finite, which makes the
	equations much simpler to deal with), and thus obtain numerical
	approximations $\Pi^{\mathrm{TBA}}_\gamma$ of $\Pi_\gamma$. 
	\item \ti{Instanton counting} (\autoref{sec:QPI}): As we have recalled above, 
	the $\Pi_\gamma$ can be computed
	from the Nekrasov-Shatashvili instanton partition function in the 
	$SU(2)$ $N_f=1$ theory.
	We carry out this computation numerically, summing the instanton
	series to sufficiently high order in $\Lambda$, 
	to obtain approximations $\Pi^{\mathrm{inst}}_\gamma$
	of $\Pi_\gamma$.
\end{enumerate}

Various relations between these methods have been proven or conjectured in the literature. Although the details are complex
 one can roughly summarize as follows:
 \begin{itemize}
 \item The equivalence between methods 1 and 2 is a theorem, following from results in the exact WKB literature, as explained 
 most precisely in \cite{in-exactwkb}; see also \cite{Hollands:2019wbr} for a different route to deriving this equivalence, closer
 to our point of view in this paper. The
 recent \cite{nikolaev2020exact} gives a sharp treatment of the necessary facts from exact WKB.
 \item The equivalence between methods 2 and 3 is proposed in \cite{Gaiotto:2009hg,Gaiotto:2014bza} and closely related
 to the ODE-IM correspondence \cite{ddt}; it is not yet
 a mathematical theorem as far as we know.
 \item The equivalence between methods 2 and 4 would follow from a combination of the proposal of \cite{nrs} (proven in some cases in \cite{Jeong:2018qpc}) and the results in \cite{Hollands:2013qza,Hollands:2019wbr}.
 \end{itemize}
 Combining these theorems and conjectures one arrives at the conclusion that all four methods should give the same results, but it is
 not yet a theorem in full generality.
 Moreover, the translation between the various methods is somewhat
intricate, particularly as their equivalence is formulated in different chambers of the moduli space.. Thus in this work we set ourselves the task of working out the translation 
carefully in the case of the $SU(2)$ $N_f=1$ theory, computing at various points in the parameter space
with all four methods, and comparing the results.
We find good agreement: within the precision we are able to obtain,
the results match.
See \autoref{tab:introresultsnew} for some sample numerical results.

Various numerical comparisons of this sort have been made before in the literature;
for example see \cite{Grassi:2019coc} for comparisons between methods 1, 3, 4 in the pure $SU(2)$ theory, and 
\cite{ddt,Dorey:1998pt,Dumas:2020zoz} for methods 2, 3 in Argyres-Douglas theories (note method 4 is not available in Argyres-Douglas
theories at the moment, since it relies on the Lagrangian description of the theory). See also \cite{Yan:2020kkb} for
computations in the pure $SU(3)$ theory using method 2 and comparisons to the WKB series.

In making these comparisons, we need to be careful about the analytic structure of the quantum periods.
As we review in \autoref{analyticstructures}, there is a wall-and-chamber structure
in the parameter space; the quantum periods are analytic in each chamber but jump when one
crosses a wall. Thus, in comparing a result computed in one chamber with the analytic continuation
of a result computed in a different chamber, one has to take account of extra contributions
from the walls separating the chambers. These extra contributions take the form of Kontsevich-Soibelman/cluster
transformations associated to BPS states of the bulk $\N=2$ theory; their concrete form is in \eqref{eq:pi-transform}-\eqref{eq:pi-transform3}
below. We use them explicitly in \autoref{sec:comparisons} when comparing instanton counting expressions with other methods.

\subsection{Fredholm determinant}

Another approach to studying the Schr\"odinger equation \eqref{eq:our-eq-intro} runs through 
the
\ti{Fredholm determinant}
\begin{equation}
 D(E,\hbar)	= \det \left(1 - \frac{E}{{\mathrm O}}\right)
\end{equation}
where ${\mathrm O}$ is the Schr\"odinger operator, with trace class inverse, acting on $L^2$ functions along the contour where we have
a confining potential. This determinant has zeros exactly 
at the (generalized) bound state energies, so computing it in particular determines these energies.
Moreover, it was argued in \cite{Grassi:2019coc} that 
in some examples $D(E,\hbar)$ can be computed explicitly by using a particular limit  of the TS/ST correspondence \cite{cgm, ghm}, which allows to express  Fredholm determinants using topological string data. This construction has been tested in several works (see for instance \cite{mmrev} and reference therein). A proof in one particular example and in a special limit was given in \cite{bgt} using Painlev\'e equations. Recently  more  general corollaries of this construction have been proven in \cite{Doran:2021hcy}.
 Nevertheless a  mathematical proof in full generality is  still missing.

In \autoref{sec:fredholm}
we apply this approach in the example of \eqref{eq:our-eq-intro}.
In particular, we give a formula \eqref{sd1} for the Fredholm determinant in terms
of the instanton-counting quantities introduced in \autoref{sec:QPI}.

We also consider a TBA integral equation for $D(E,\hbar)$,
introduced by Zamolodchikov in \cite{post-zamo}. 
In general this TBA equation seems to be unrelated to the system of TBA equations
obeyed by the quantum periods. However, there is one special situation where there is a relation.
Namely, suppose we set $E = 0$ and $m = 0$. In that case 
the TBA equations obeyed by the quantum periods simplify, reducing to a single equation, which matches with the Zamolodchikov TBA. 
We describe this in \autoref{Zamolodchikov}. A very similar
phenomenon was noticed in the pure $SU(2)$ theory in \cite{Grassi:2019coc}. We do not have a conceptual
explanation for it, in either case; it would be very interesting to find one.

\subsection{Quantization condition}

The equation \eqref{eq:our-eq-intro} does not have a confining potential on the real line if we take $\Lambda$ and $\hbar$
to be real.
However, if we take $\hbar$ real, 
$\arg \Lambda = -\frac{\pi}{6}$, and look along the line $\im x = - \frac{\pi}{3}$, then we do 
have a confining potential (which may be real or complex, depending on the phase of $m$),
so we can formulate the bound state problem.
It turns out that for $m \in \I \R_{>0}$, the case of a real convex confining potential, this condition 
can be formulated very simply: it is
\begin{equation} \label{eq:qc-intro}
\cX_{\gamma_{[1,0,0]}} = -1
\end{equation}
where $\gamma_{[1,0,0]}$ is the cycle connecting the two classical turning 
points, shown in \autoref{quantizationcd} below.

In this paper we give two different routes to deriving
this exact quantization condition. One, in \autoref{sec:ssection}, goes through the connection between
quantum periods and Wronskians of solutions.
The other, in \autoref{sec:fredholm}, uses the connection between quantum periods and Fredholm determinants,
combined with the relation between Fredholm determinants and instanton counting.
Happily, both methods independently give the quantization condition \eqref{eq:qc-intro};
this provides a nice cross-check of our computations.

\subsection{Comments}

We comment on a few interesting points and future directions here:

\begin{itemize}
	\item The result of instanton counting agrees directly with
	the quantum period only at a particular locus of the parameter space,
	which we call the instanton locus. This locus lies on one of the walls in the
	wall-and-chamber decomposition of the parameter space.
	 The results of
	\cite{nrs,Hollands:2013qza}, 
	translated to our language, amount to a determination of the instanton locus 
	in the $SU(2)$ $N_f=4$ theory, and similarly \cite{Grassi:2019coc} determines
	the chamber in the pure $SU(2)$ theory. 
	For the $SU(3)$ $N_f=6$ theory 
	the instanton locus was found in \cite{Hollands:2017ahy}.
	In all of these examples, the instanton
	locus lies at weak coupling, with $\hbar$ chosen so that $\arg \hbar$ is the 
	phase of the central charge of the vectormultiplet.
	In this paper we identify the instanton
	locus in the $SU(2)$ $N_f = 1$ theory.
	We find that this locus is determined by requiring that the parameters
	 $(\Lambda, m)$ define a positive, convex and confining potential along some line in the $x$-plane. 
	 As for the $\hbar$ parameter, we find that it again follows the same pattern as above.

For generic complex values of the parameters, instanton counting will agree with Borel summation only after we implement an appropriate Kontsevich-Soibelman (KS) transformation.\footnote{In the pure $SU(2)$ case, the examples considered  in \cite{Grassi:2019coc} take place at the instanton locus, where such a transformation is not needed.} This is described in \autoref{sec:comparisons}. 

	\item We work out explicitly the transformation relating
	the quantum periods at weak coupling to the analytic continuation of the quantum 
	periods from strong coupling; see \eqref{xxs-1}-\eqref{xxs-3}. This transformation can be viewed
	as a relation between specific sorts of
	Fock-Goncharov-type and Fenchel-Nielsen-type coordinates, generalizing a similar (simpler) one which
	appears in the pure $SU(2)$ theory and was used recently in \cite{Grassi:2019coc,Coman:2020qgf}.

	\item As we remarked above, the singularities of the Borel transform of $\Pi_\gamma$
	are expected to appear at the central charges of BPS particles in the $\N=2$ field theory.
	This phenomenon was explored numerically in \cite{Grassi:2019coc} for the pure $SU(2)$
	theory, and we do the same in the $SU(2)$ $N_f=1$ theory here.
	We also discuss a related statement for the singularities of the Borel transform of the
	WKB solutions of the Schr\"odinger equation \eqref{eq:our-eq-intro}: namely, 
	these appear at the
	central charges of BPS solitons in a certain 2d-4d coupled system. (In practical terms this means
	considering integrals of the Seiberg-Witten differential along open paths instead of closed ones.) 
	This is discussed in \autoref{padeborelsol} and Appendix \ref{localsolpure}.
	
	\item The coupled 2d-4d system should also lead to other methods of computing local solutions of
	the Schr\"odinger equation directly, by appropriate 2d-4d extensions of the methods we use
	here to compute quantum periods. Indeed, an extension of the TBA integral equations
	which should compute the solutions
	has been described in \cite{Gaiotto:2014bza}; it arises as the conformal limit of the 
	2d-4d equations of \cite{Gaiotto:2011tf}. One should also be able to compute the solutions by a version of the Nekrasov-Shatashvili instanton counting in the presence of the surface defect, see for instance \cite{Maruyoshi:2010iu} or \cite{Jeong:2018qpc} for more recent work.
	It would be interesting to work this out carefully 
	in some examples.

\item In this paper we study the four-dimensional field-theoretic framework which, on the operator theory side, corresponds to studying a differential equation. 
It would be interesting to generalise this approach to the five-dimensional field theory setup and the corresponding difference equations, considered e.g. in \cite{acdkv}. One would expect this to make contact with exponential networks \cite{eager,Longhi:2021qvz,longhi,Banerjee:2020moh} and BPS states of local CY threefolds \cite{dfr}, see also \cite{Closset:2019juk,Bonelli:2020dcp}. It would also be interesting to understand the structure of the exact solution for the spectral theory of such difference equations proposed in \cite{ghm,cgm,wzh,mz-wv}.

\end{itemize}

\subsection*{Acknowledgements}

We thank Daniele Gregori, Jie Gu, Lotte Hollands, Ahsan Khan, Marcos Mari\~no, and Fei Yan for helpful discussions.
The work of AN is supported by National Science Foundation grant 2005312 (DMS).
The work of AG is partially supported by the Fonds National Suisse, Grant No.~185723  
and by the NCCR ``The Mathematics of Physics'' (SwissMAP).

\begin{table}[!ht]
\begin{tabular}{ |p{1cm}||p{4.5cm}|p{4.5cm}|p{4.5cm}| }
 \hline 
 \hline
 \multicolumn{4}{|c|}{\large{Strong coupling region}} \\
 \hline
 \hline
  \multicolumn{4}{|c|}{$m=-1/10$, $\Lambda=1$, $u=0$,  $\hbar=1+\ri/10$} \\
 \hline
$\gamma$ & $[-1,-1,1]$ & $[1,0,0]$ & $[0,1,1]$\\
\hline
$\Pi_{\gamma}^{\rm PB}$ & $2.31044583$ & $1.7968493$  & $-2.86310004138$ \\
                   & $+ 3.06848675\ri$ & $- 3.5249441\ri$ & $+ 0.33203784\ri$ \\
 \hline
$\Pi^{\rm TBA}_{\gamma}$ & $2.3105$ & $1.797$  & $-2.863$ \\
                   & $+ 3.068\ri$ & $-3.5249\ri$ & $+ 0.3320\ri$ \\
 \hline
$\Pi^{\rm SS}_{\gamma}$  & $2.31044583$ & $1.7968493$  & $-2.8631000414 $ \\
                 & $+ 3.06848675\ri$ & $- 3.5249441\ri$ & $+ 0.33203784 \ri$ \\ \hline
 \multicolumn{4}{|c|}{$m=-1/10$, $\Lambda=1$, $u=0$,  $\hbar=\frac{8}{5}(1+\ri/10)$} \\
 \hline
$\gamma$ & $[-1,-1,1]$ & $[1,0,0]$ & $[0,1,1]$\\
\hline
$\Pi_{\gamma}^{\rm PB}$ & $1.251837$ & $1.10078$  & $-1.5749919$ \\
                   & $+2.10014\ri$ & $-2.4030\ri$ & $+0.225114\ri$ \\
\hline
$\Pi_{\gamma}^{\rm inst}$ & $1.2518373$ & $ 1.1007766$  & $-1.5749919$ \\
                   & $2.1001386\ri$ & $-2.4030151\ri$ & $+0.22511425\ri$ \\                  
 \hline
$\Pi^{\rm TBA}_{\gamma}$ & $1.252$ & $1.1008$  & $-1.575$ \\
                   & $+2.100\ri$ & $-2.403\ri$ & $+0.2251\ri$ \\
 \hline
$\Pi^{\rm SS}_{\gamma}$  & $1.251837$ & $1.10078$  & $-1.5749919$ \\
                 & $+2.10014\ri$ & $-2.4030\ri$ & $+0.225114\ri$ \\ \hline

\hline
 \multicolumn{4}{|c|}{$m= -1/10$, $\Lambda=1/5$, $u=0$, $\hbar=1+\ri/10$} \\
 \hline
$ \gamma$&  $[-1,-1,1]$ & $[1,0,0]$ & $[0,1,1]$ \\
\hline
$\Pi^{\rm inst}_\gamma$&$-0.07209621704459846369$ &$0.79394041891943969067$ &$0.52235090845775015459$  \\
&$+1.05629021332321667555\ri$ &$-1.25702427315714059939\ri$ &$+0.07631454880066478568\ri$ \\
\hline
$\Pi^{\rm TBA}_\gamma$ & $-0.072$ & $0.794$&$0.522$\\
 & $+1.056\ri$ & $-1.257\ri$ &$+0.0763\ri$\\
 \hline
$\Pi^{\rm SS}_\gamma$&$-0.07209622$  & $0.793940419$&$0.52235091$\\
& $+1.0562902\ri$ & $-1.25702427\ri$ &$+0.07631455\ri$\\
 \hline
  \multicolumn{4}{|c|}{$m=0$, $\Lambda=1/5$, $u=0$,  $\hbar=1+\ri/10$} \\
 \hline
$\gamma$ & $[-1,-1,1]$ & $[1,0,0]$ & $[0,1,1]$ \\
 \hline
$\Pi^{\rm inst}_{\gamma}$ & $-0.40014633959247147382$    & $0.25765417201941334603$ & $0.14249216757305812778$\\
  & $+1.05116204874070340420 \ri$ & $-1.16973711207642800270 \ri$ & $+0.11857506333572459850\ri$\\
 \hline
$\Pi^{\rm TBA}_{\gamma}$&  $-0.400 $     & $0.258$    & $0.142$\\
        &     $+1.051\ri $    & $-1.170\ri$& $+0.119\ri$\\
 \hline
$\Pi^{\rm SS}_{\gamma}$& $-0.4001463396$         & $0.25765417202$   & $0.14249216757$\\
  & $+1.051162049\ri $  & $-1.169737112\ri$ & $+0.118575063\ri$\\
 \hline
  \multicolumn{4}{|c|}{$m=0$, $\Lambda=1/5$, $u=0$,  $\hbar=\frac{1}{3}(1+\ri/10)$} \\
 \hline
$\gamma$ & $[-1,-1,1]$ & $[1,0,0]$ & $[0,1,1]$ \\
 \hline
 $\Pi^{\rm PB}_{\gamma}$ & $0.940082$    &  $0.78280$ & $-1.7228798$\\
  & $+2.03841 \ri$ & $-2.2823 \ri$ & $+0.243848\ri$\\
  \hline
$\Pi^{\rm inst}_{\gamma}$ & $0.9400824$    & $0.782797$ & $-1.7228798$\\
  & $ +2.03840655 \ri$ & $-2.28225498 \ri$ & $+0.2438484\ri$\\
 \hline
$\Pi^{\rm TBA}_{\gamma}$&  $0.940$     & $0.7828$    & $-1.723$\\
        &     $+2.038\ri $    & $-2.282\ri$& $+0.2438\ri$\\
 \hline
$\Pi^{\rm SS}_{\gamma}$& $0.940082$    & $0.78280 $ & $-1.7228798$\\
  & $+2.03841 \ri$ & $-2.2823 \ri$ & $+0.243848\ri$\\
 \hline
 \multicolumn{4}{|r|}{{Continued on next page}} \\
 \hline
\end{tabular}

\end{table}
\clearpage

\begin{table}[!ht]
\begin{tabular}{ |p{1cm}||p{3.6cm}|p{4cm}|p{5.5cm}| }
 \hline
 \hline
 \multicolumn{4}{|c|}{\large{Weak coupling region}} \\
 \hline
 \hline
  \multicolumn{4}{|c|}{$m=0$, $\Lambda=1/5$, $u=2$,  $\hbar=1+\ri/10$}\\
  \hline
 $\gamma$ & $[-1,-1,1]$ & $[1,0,0]$ & $[0,2,0]$\\
 \hline
$\Pi_{\gamma}^{\rm PB}$ &$3.562524034409252$& $8.8794275059299982$ & $-24.8839030806784999581424$\\
                   &$-26.953094797469923\ri$~& $+25.708899620419279\ri$& $+2.488390354101287262159325\ri$\\
  \hline
$\Pi^{\rm inst}_{\gamma}$   &$3.562524034409252 $         & $8.8794275059299982 $      & $-24.8839030806784999581424$\\
                 & $-26.953094797469923\ri $     &$+ 25.708899620419279 \ri$ &$+2.488390354101287262159325\ri$\\

  \hline
$\Pi^{\rm SS}_{\gamma}$                        & $3.562524034$   &  $8.8794275059$     & $-24.8839030807$\\
                    &$-26.9530947975\ri$& $+25.7088996204 \ri$  &  $+2.488390354101\ri$\\
  \hline

   \multicolumn{4}{|c|}{$m=0$, $\Lambda=1/5$, $u=2$,  $\hbar=1$}  \\
\hline
 $\gamma$ & $[-1,-1,1]$ & $[1,0,0]$ & $[0,2,0]$ \\
 \hline
$\Pi_{\gamma}^{\rm PB}$ &$6.283185528939026$& $6.283185528939026$ & $-25.132742115756103$\\
                   &$-26.595529611194577\ri$& $+26.595529611194577\ri$& $ 627145534$\\
\hline
$\Pi^{\rm inst}_{\gamma}$  &$6.283185528939026$& $6.283185528939026$ & $-25.132742115756103$\\
 &$-26.5955296111946\ri$& $+26.595529611194577\ri$& 627145534057 \\
\hline
$\Pi^{\rm SS}_{\gamma}$ &$6.2831855289$& $6.2831855289$ & $-25.13274211576$\\
 &$-26.5955296112\ri$& $+26.5955296112\ri$& \\
 \hline 

  \multicolumn{4}{|c|}{$m=\ri/10,~\Lambda=\re^{-\frac{\pi\ri}{6}},~ u=2,~ \hbar=1$}\\
  \hline
 $\gamma$ & $[-1,-1,1]$ & $[1,0,0]$ & $[0,2,0]$\\
 \hline
$\Pi_{\gamma}^{\rm PB}$ & $12.5407604031$ & $6.948499 \ri  $ & $-25.08152080613$\\
&$ -7.57682 \ri$&&\\
  \hline
$\Pi^{\rm inst}_{\gamma}$   & $12.5407604031$ & $6.9484991958 \ri$  & $-25.08152080613$\\
&$-7.576818\ri$&&\\
  \hline
$\Pi^{\rm SS}_{\gamma}$  & $12.540760403$ &  $ 6.9484991958 \ri$     & $-25.0815208061$\\
&$-7.576818\ri$&&\\
 \hline 
 \hline
\end{tabular}
\caption{Summary of results of numerical computation of $\Pi_\gamma$ for various moduli, charges and values of $\hbar$. $\Pi_\gamma^{\rm PB}$, $\Pi_\gamma^{\rm SS}$, $\Pi^{\rm TBA}_{\gamma}$, $\Pi^{\rm inst}_{\gamma}$ are the quantum periods we obtain by Pad\'{e}-Borel summation, small section, TBA and instanton counting methods respectively; all of them are approximations to the true $\Pi_\gamma$.
For each point of parameter space which we study, we list results from all methods which we were able to apply at that point. We emphasize that $\Pi_\gamma$ denotes the canonical piecewise-analytic function 
discussed in \autoref{analyticstructures} below;
e.g. this is the quantity one would obtain directly by Borel summation without any
further transformations.}
\label{tab:introresultsnew}

\end{table}

\section{Analytic structures}
\label{analyticstructures}
In what follows we are going to compute the quantum periods by various
different methods and compare them.
To make this comparison correctly one needs to take account of
a certain wall-and-chamber structure in the parameter space,
which we review here.

\subsection{Walls and chambers}

We consider a complex four-dimensional 
parameter space, consisting of the complex couplings
$(m, \Lambda)$, the Coulomb branch modulus $u$, 
and a complex mass parameter $\hbar \neq 0$,
which could be interpreted
as an $\Omega$-background parameter.

In this space one has various real-codimension-1 
walls\footnote{The walls have many different names in the literature. They have
been called ``BPS walls'' \cite{Gaiotto:2010be} and ``walls of the first kind'' \cite{ks}.
Their projections to fixed $\vartheta$ are also identified with the walls
of the scattering diagram in the sense of \cite{MR2181810,MR2846484}.
The projections of the walls 
to fixed $(u,m,\Lambda)$
are the ``BPS rays'' \cite{Gaiotto:2010be} or ``active rays'' \cite{Bridgeland_2018}. 
These walls are \ti{not} the same as the walls of marginal stability for
4d bulk BPS states, aka ``walls of the second kind'' in \cite{ks}.
Indeed, each of the BPS walls is labeled by a single BPS state
or a collection of BPS states with mutually local charges, while the
walls of marginal stability are places where mutually non-local BPS states interact.}
defined as follows. A point $(m, \Lambda, u, \hbar)$ is on a wall if and only
if the 4d field theory with couplings $(m, \Lambda)$, at the point $u$ of
its Coulomb branch, has a BPS one-particle state for which the central charge 
$Z$ has $\arg(-Z) = \arg(\hbar)$, i.e. $Z / \hbar \in \R_-$.
In this case we say that the wall \ti{supports} the BPS state.
Generically, each wall will support only BPS states which are mutually
local with one another, i.e. their charges obey $\IP{\gamma,\gamma'} = 0$.
The walls divide the parameter-space up into chambers.

\begin{figure}
     \centering
     \begin{subfigure}[b]{0.47\textwidth}
         \centering
         \includegraphics[width=0.7\textwidth]{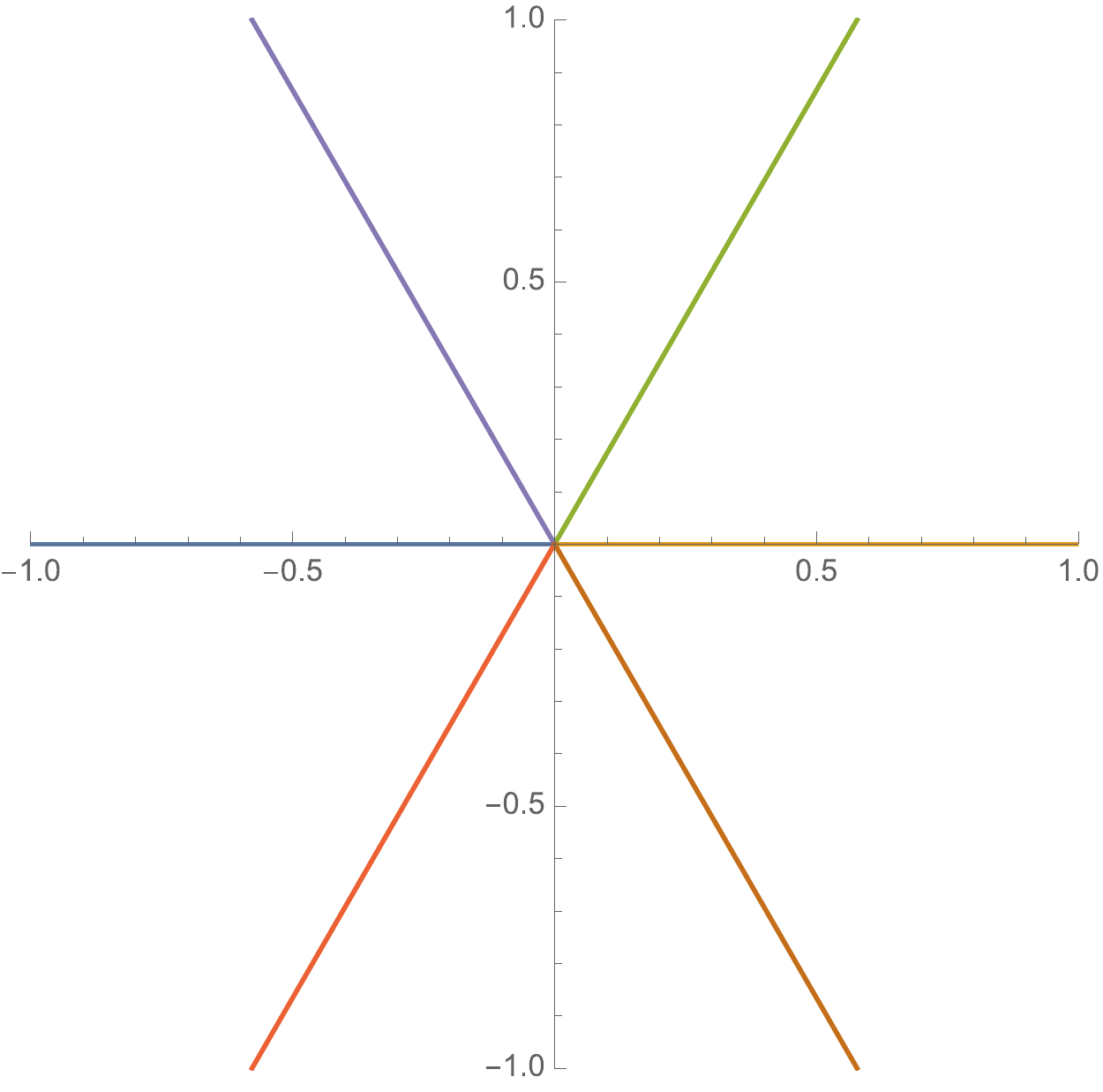}
         \caption{The walls in the $\hbar$-plane at $m=0$, $\Lambda=1/5$, $u=0$ (strong coupling).}
         \label{BPSrayst}
     \end{subfigure}
     \hfill
          \begin{subfigure}[b]{0.47\textwidth}
         \centering
         \includegraphics[width=0.7\textwidth]{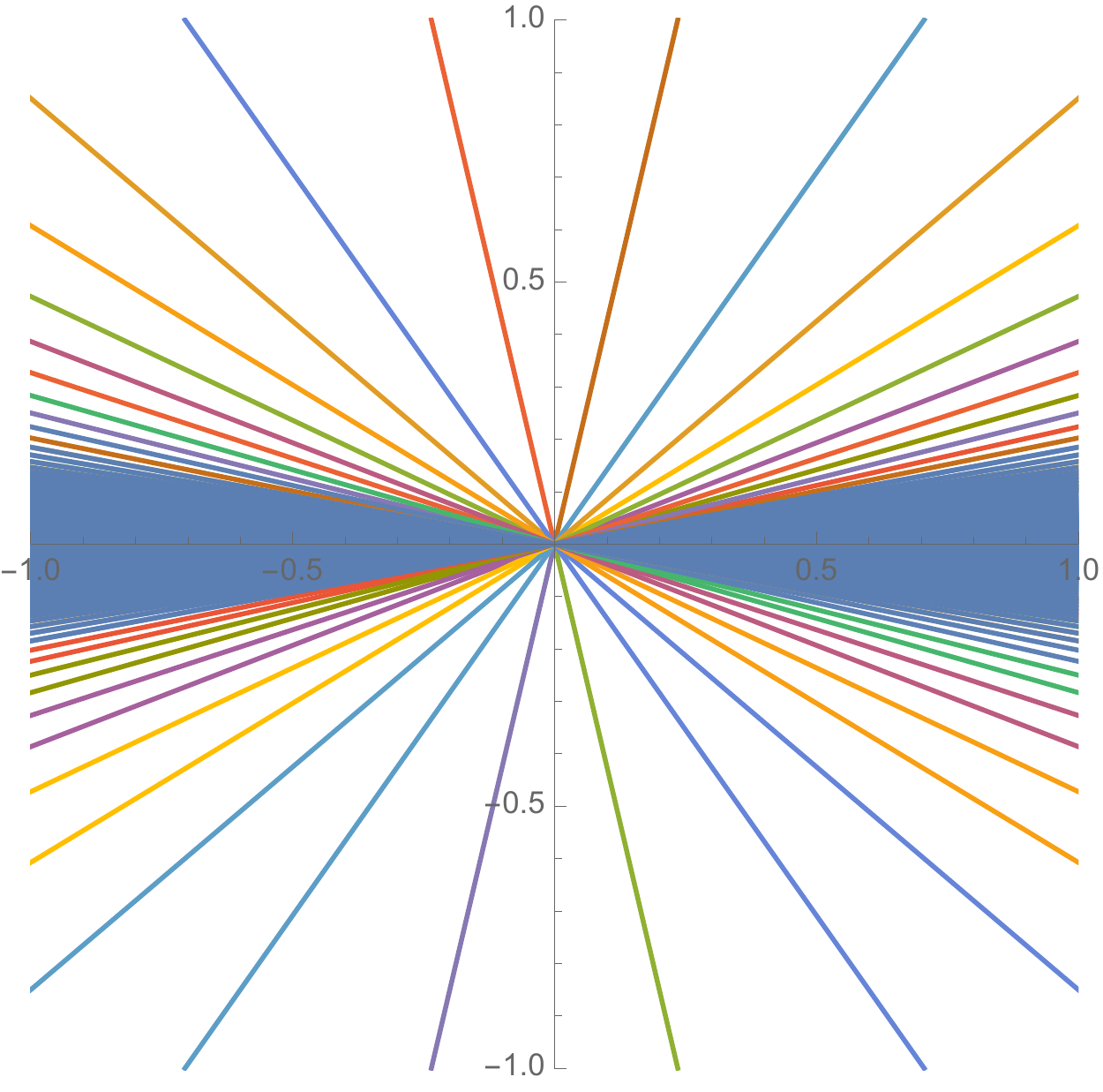}
         \caption{The walls in the $\hbar$-plane at $m=0$, $\Lambda=1/5$, $u=2$ (weak coupling).}
         \label{BPSraywk}
     \end{subfigure}
            \caption{The strong coupling BPS spectrum consists of 3 hypermultiplets and their antiparticles, while the weak coupling BPS spectrum consists of 1 vectormultiplet and infinitely many hypermultiplets. There is a wall in the direction 
            $\vartheta$ whenever there exists a 
            BPS particle of charge $\gamma$ with $\vartheta=\arg(Z_\gamma)$. (Generically
            each wall corresponds to a single particle, but in \autoref{BPSraywk}
            each of the horizontal rays actually represents several BPS particles with the same phase:
            a vectormultiplet and two hypermultiplets. This accidental degeneracy would be broken
            for $m \neq 0$.) 
            In \autoref{BPSraywk}, the rays accumulate around $\vartheta=0$, which is the phase
            corresponding to the vectormultiplet. This kind of accumulation 
            happens generically in theories with 
            higher-spin BPS particles.}
        \label{BPSray}
\end{figure}

To get some feeling for this structure let us consider slicing it along
various directions. If we hold the parameters $(m, \Lambda, u)$ fixed and let only 
$\hbar$ vary, then the walls reduce to rays in the $\hbar$-plane.
Each BPS state which exists in the theory at $(m, \Lambda, u)$ 
is supported by a ray in the $\hbar$-plane whose angle is $\arg(-Z)$.
Since $Z$ is determined by the electromagnetic and flavor charge
$\gamma$ of the state, we may also denote it as $Z_\gamma$.
See \autoref{BPSray} for the walls in 
the $\hbar$-plane at two different points in $(m, \Lambda, u)$ space.

\begin{figure} 
     \centering
         \includegraphics[width=0.9\textwidth]{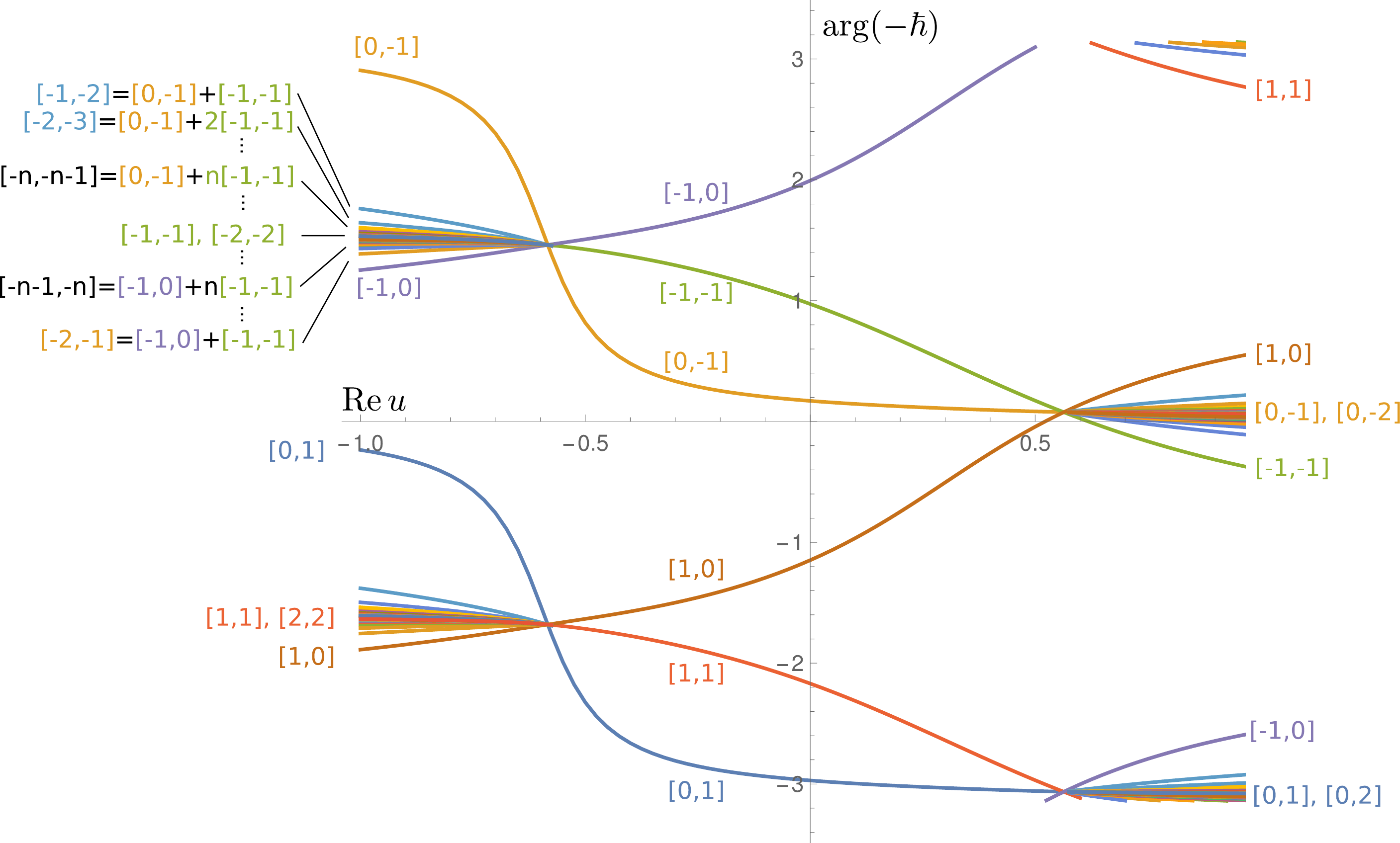} 
 
  \caption{The BPS walls when we fix $\Lambda=1,\ m=0,\ {\rm Im}(u)=\frac{\ri}{10}$. The horizontal axis is ${\rm Re}(u)$; the vertical axis is $\arg(-\hbar) \in [-\pi,\pi]$. 
  There are 2 special points at around $u_1\approx-0.6+\frac{\ri}{10}$ and $u_2\approx0.675+\frac{\ri}{10}$ where the wall of marginal stability is encountered. For ${\rm Re}(u) \in (u_1,u_2)$ (strong coupling) we have 3 hypermultiplets and their antiparticles, so there are 6 walls in total shown in this region. For $u<u_1$ or $u>u_2$ (weak coupling), we have infinitely many hypermultiplets. The hypermultiplet walls accumulate around the vectormultiplet wall. 
  We list the electromagnetic charges of the particles next to their corresponding walls.}
        \label{argz_reuwhole}
\end{figure}

As $(m, \Lambda, u)$ are varied, the quantities $Z_\gamma$ vary,
and thus the rays move in the $\hbar$-plane. To capture this behavior
it is useful to draw a different slicing, where we show one real direction in
the $(m, \Lambda, u)$ parameter space, and also show the phase $\arg \hbar$.
See \autoref{argz_reuwhole} for an example of this in the $SU(2)$ $N_f=1$
theory, and \autoref{su2argz_reu} for an example in the 
simpler pure $SU(2)$ theory.
Each of these figures can be thought of as a graph of the evolution of the phases
of the central charges of BPS particles as we move along a curve in the Coulomb
branch. Note again that when the phases collide some BPS bound states can form
or decay, and thus we have different numbers of BPS states in different
regions of the Coulomb branch: finitely many at strong coupling,
infinitely many at weak coupling.

Finally we could consider holding $(m, \Lambda, \hbar)$ fixed and letting
$u$ vary. In that case we get a collection of walls on the Coulomb branch.
See \autoref{su2counter} for an example in the pure $SU(2)$ theory.

\subsection{Chamber structure for quantum periods}

The importance of this chamber structure for us is that 
the quantum periods 
$\Pi_\gamma(m, \Lambda, u, \hbar)$
are \ti{piecewise} analytic functions on the parameter space: precisely, 
they are analytic in each chamber, but jump at the walls.
This structure shows up in a different way in each method of computing
the quantum periods:
\begin{enumerate}
\item In the Borel summation method, the jumps occur because of the need to move a contour of integration across a singularity in the Borel plane.
\item In the Wronskian or ``small section'' method, the jumps arise as follows. Each 
$(m, \Lambda, u, \hbar)$ determines a spectral network drawn on the 
$z$-plane. The spectral network can be used to give a formula for
$\cX_\gamma = \exp \Pi_\gamma$ in terms of Wronskians of solutions of the ODE 
\eqref{eq:our-eq-intro}. The precise expression for $\cX_\gamma$  in terms of Wronskians
depends on the topology of the spectral network, 
which in turn depends on which chamber $(m, \Lambda, u, \hbar)$ is in.
\item In the TBA method, the jumps arise again from the need to move a contour of integration across a singularity, this time a singularity
in the TBA integration kernel.
\item In the instanton counting method, one does not see the jumps directly.
Rather, instanton counting produces multivalued analytic functions, which 
can be understood as analytic continuations of
the quantum periods from a specific locus in the parameter-space.
\end{enumerate}

To keep track of the analytic structure, and having in mind the possibility of analytic
continuation from one chamber to another, 
it is convenient to introduce new functions $\Pi_\gamma^{(m_0,\Lambda_0,u_0,\hbar_0)} (m,\Lambda,u,\hbar)$
which are defined by analytic continuation of $\Pi_\gamma$ from
a base-point $(m_0,\Lambda_0,u_0,\hbar_0)$.
Thus our convention is: when we write $\Pi_\gamma(m,\Lambda,u,\hbar)$ without a basepoint
superscript, we mean the canonical
piecewise-analytic function; when we write it with a basepoint superscript, we mean the
function defined by analytic continuation from the basepoint.\footnote{In so doing we should be careful to specify the path we follow for the analytic continuation: indeed the continued functions are not necessarily
single-valued, whereas the original function $\Pi_\gamma$ is completely
single-valued in all parameters 
(but, as we have emphasized, only piecewise analytic).} Since the dependence on $\hbar_0$ is only through
$\arg \hbar_0$ we also sometimes use the notation $\vartheta = \arg \hbar_0$ and write e.g.
$\Pi_\gamma^\vartheta$ instead of $\Pi_\gamma^{\hbar_0}$.

The question now naturally arises: what is the relation between the analytic
continuations from different chambers?
To understand this, it is sufficient to understand what happens for two chambers separated
by a single wall. To be definite, let us hold $(m,\Lambda,u)$ fixed for a moment.
Then the wall is just a ray in the $\hbar$-plane, at some $\arg \hbar = \vartheta_*$. 
Now we consider the relation between $\Pi_\gamma^\vartheta$ and $\Pi_\gamma^{\vartheta'}$
for two different phases $\vartheta$, $\vartheta'$ with
$\vartheta < \vartheta_* < \vartheta'$. In this case
the relation takes the form \cite{Gaiotto:2009hg,Gaiotto:2012rg}
\begin{equation} \label{eq:pi-transform}
	\Pi_\mu^{\vartheta'} - \Pi_\mu^{\vartheta} = \sum_\gamma \IP{\mu,\gamma} \Omega(\gamma) \log(1 - \sigma_{\mathrm{can}}(\gamma) \cX_\gamma^{\vartheta}).
\end{equation}
Here:
\begin{itemize} \item The sum runs over the charges $\gamma$ 
of BPS states supported at the wall.
\item $\Omega(\gamma)$ is the BPS index (second helicity supertrace)
counting BPS states of charge $\gamma$.
\item $\sigma_{\mathrm{can}}$ is a factor defined in \cite{Gaiotto:2009hg}, with the property
that $\sigma_{\mathrm{can}}(\gamma) = -1$ whenever $\gamma$ is the charge of a BPS hypermultiplet,
and $\sigma_{\mathrm{can}}(\gamma) = 1$ whenever $\gamma$ is the charge of a BPS vectormultiplet.
\end{itemize}

At first \eqref{eq:pi-transform} seems to have 
a puzzling asymmetry: why did we write $\cX_\gamma^\vartheta$ on the right
instead of $\cX_\gamma^{\vartheta'}$? But since the BPS states supported at a generic point of the wall are 
all mutually local,\footnote{The BPS states supported at the wall all have the same 
phase for their central charge; at a generic point of the wall, this condition implies that their electromagnetic
charges all lie on a common rank-1 sublattice, and thus they are mutually local.}
\eqref{eq:pi-transform} says that 
$\Pi_\mu^{\vartheta} = \Pi_\mu^{\vartheta'}$ when $\mu$ is any charge supported at the wall. 
Thus we have $\cX_\gamma^\vartheta = \cX_\gamma^{\vartheta'}$, so it did not matter which 
we wrote on the RHS.

The simplest and most generic case is a wall supporting
a single BPS hypermultiplet of charge $\gamma$; in that case we have $\Omega(\gamma) = 1$, 
$\sigma_{\mathrm{can}}(\gamma) = -1$, and thus
the transformation becomes \footnote{This is closely related to the  Delabaere-Pham discontinuity formula \cite{dpham}.}
\begin{equation}\label{eq:pi-transform2}
	\Pi_\mu^{\vartheta'} - \Pi_\mu^{\vartheta} = \IP{\mu,\gamma} \log(1 + \cX_\gamma^\vartheta).
\end{equation}

So far we have been discussing $\Pi_\gamma$ in the open chambers, but it will be important below
sometimes to consider $\Pi_\gamma$ on the walls as well. On a wall, we define $\Pi_\gamma$ to be
the average of the limits of its values from the two sides of the wall. 
Thus, the analogue of \eqref{eq:pi-transform} for
comparing the value on the wall to the value on one side of the wall just involves an extra factor $\half$:
\begin{equation}\label{eq:pi-transform3}
	\Pi_\mu^{\vartheta_*} - \Pi_\mu^{\vartheta} = \half \sum_\gamma \IP{\mu,\gamma} \Omega(\gamma) \log(1 - \sigma_{\mathrm{can}}(\gamma) \cX_\gamma^\vartheta).
\end{equation}

More generally we may consider varying all the parameters
$(m, \Lambda, u, \hbar)$. For example, looking at 
\autoref{argz_reuwhole}, we see that while holding $(m,\Lambda)$ fixed, we could cross any given
wall either by varying $\arg \hbar$ while holding $u$ fixed or by varying $u$ while holding
$\arg \hbar$ fixed. Irrespective of which parameters we vary, the transformation associated
to the wall is the same, given by \eqref{eq:pi-transform}.

\section{Borel summation}
\label{sec:borel}

In this section, we briefly review the exact WKB method and Pad\'{e}-Borel summation, applied to the 
Schr\"{o}dinger equation 
\begin{equation}
\label{dex2}
\left(-\hbar^2\partial_x^2+P(x)\right)\psi(x)=\left(-\hbar^2\partial_x^2+(\frac{\Lambda^2 \re^{-x}}{2}+2 m \Lambda \re^{x} - \Lambda^2 \re^{2x})-E\right)\psi(x)=0 ,
\end{equation}
where
\begin{equation} \label{P-cylinder}
P(x)=\left(\frac{\Lambda^2 \re^{-x}}{2}+2 m \Lambda \re^{x} - \Lambda^2 \re^{2x}\right)-E.
\end{equation}
For the moment, we do not impose any reality conditions on the parameters $(m,\Lambda,E)$ 
or the variable $x$ in \eqref{dex2}.
Sometimes it is also useful to make the coordinate transformation \be z=\re^x \ \ee and redefine $\psi(z)=( z)^{-\frac{1}{2}}\psi(z)$, after which \eqref{dex2}
becomes
 \begin{equation}
 \label{dez}
 \left(-z^2\hbar^2\partial_z^2 +(\frac{\Lambda^2 }{2z}+2 m \Lambda z - \Lambda^2 z^2)-(E+\frac{\hbar^2}{4})\right)\psi(z)=0.
 \end{equation}
Here $z$ is a coordinate on the punctured Riemann surface $C=\mathbb{CP}^1\backslash\{0,\infty\}$. 
In the rest of the manuscript we go from $x$ to $z$ interchangeably. Which variable we are using should be clear from the context.

\subsection{All-orders WKB}
\label{allorder}
The starting point is the all-orders WKB analysis.
Let us make the following ansatz for a solution of \eqref{dex2}:
\begin{equation}
\label{ansatzx}
\psi(x)=\exp\left(\frac{1}{\hbar}\int_{x_0}^xY(x,\hbar) \de x\right).
\end{equation}
Then \eqref{dex2} implies that $Y(x,\hbar)$ should satisfy the Ricatti equation
\begin{equation}\label{ricattix}
-\left(Y(x,\hbar)\right)^2-\hbar {\frac{\rd}{\rd x}} Y(x,\hbar)+P(x)=0\, .
\end{equation}
We can solve \eqref{ricattix} formally as a power series in $\hbar$ using the ansatz
\begin{equation}\label{wkbY}
Y(x,\hbar) = \sum_{n=0}^\infty Y^{n}(x)\hbar^{n}\, ,
\end{equation}
where $Y^{0}=\pm\sqrt{P(x)}$.
We denote the two choices of sign by $Y^{0,(1)}$ and $Y^{0,(2)}$.
All the higher order terms depend on which $Y^{0,(i)}$ we choose, so we denote the solutions \be\label{psidef} \psi^{(i)}(x)=\exp\left(\frac{1}{\hbar}\int_{x_0}^xY^{(i)}(x,\hbar) \de x\right).\ee
It is convenient to split the formal series expansion \eqref{wkbY} into even and odd components as
\be \label{eodef}\ba Y_{\rm even}=& \sum_{n=0}^\infty Y^{2n}(x)\hbar^{2n}\, ,\\
Y_{\rm odd}=&\sum_{n=0}^\infty Y^{2n+1}(x)\hbar^{2n+1}.\ea\ee
Then one finds that
\be \label{oddeven}Y_{\rm odd} = -{\frac{\hbar}{2}}{\frac{\rd}{\rd x}}\log Y_{\rm even}.\ee

\subsection{Borel summation of the local solutions}
An important point is that the all-orders WKB method does not provide actual analytic solutions, but formal power series. 
Indeed the coefficients in \eqref{wkbY} grow factorially as
\be Y^{n} \sim n!  \quad \text{for}\quad n\gg 1.\ee
Therefore, \eqref{wkbY} is purely a formal expression with zero radius of convergence. 
Borel summation gives a way to convert this type of asymptotic series 
into an analytic function (for $\hbar$ lying in some half-plane). This works as follows. We 
consider the Borel transform of the formal series $Y(z,\hbar)$,
 \begin{equation}
 \label{boreltr}
 \hat{Y}(z,\zeta)=\sum_{n=0}^\infty \frac{Y^n(z)}{(n)!}\zeta^{n},
 \end{equation} 
 which is a convergent series for sufficiently small $\abs{\zeta}$.
 The Borel summation $s(Y)$ is then defined by the Laplace transform 
 \begin{equation}\label{laplace}
  s(Y)(z,\hbar)=\frac{1}{|\hbar|}\int_0^{\infty} \hat{Y}(z,{\rm e}^{\ri \arg(\hbar)}\zeta)\re^{-\zeta/|\hbar|}d\zeta. 
 \end{equation}
Note that in the integral \eqref{laplace}, we need to analytically continue $\hat{Y}(z,\zeta)$ along the integration ray,
beyond the region where the sum 
\eqref{boreltr} converges. It is known that such 
an analytic continuation exists, and the integral \eqref{laplace} converges, 
for generic choices of $\arg(\hbar)$ \cite{MR3706198,nikolaev2020exact}. 
In numerical computations, we approximate the desired 
analytic continuation
by taking finitely many terms in the series \eqref{boreltr}, and then taking 
a Pad\'{e} approximant of the resulting polynomial; this
gives a rational function, meromorphic in the whole $\zeta$-plane. Substituting this rational
function for $\hat{Y}(z,\zeta)$ in \eqref{laplace}, 
we obtain an approximation to the desired $s(Y)(z,\hbar)$.
This method of approximation is known as Pad\'{e}-Borel summation.

For some special values of $\arg(\hbar)$, it may happen that
the integral in \eqref{laplace} is not well defined, because
$\hat{Y}(z,\zeta)$ has singularities along the integration contour. 
In this case, we say that $Y(z,\hbar)$ is not Borel summable, and 
consider instead the lateral Borel summation
 \begin{equation}\label{sdeflat}
  s_{\pm}(Y)(z,\hbar)=\frac{1}{|\hbar|}\int_0^{{\rm e}^{\ri 0^{\pm}}\infty} \hat{Y}(z,{\rm e}^{\ri \arg(\hbar)}\zeta)\re^{-\zeta/|\hbar|}d\zeta. 
 \end{equation}
In this paper we will never use lateral Borel summation directly; rather we always use the 
median summation, which is defined as
 \be\label{smedian} \overline{s}(Y)(z,\hbar)= {\frac12}\left(s_{+}(Y)(z,\hbar)+s_{-}(Y)(z,\hbar)\right).\ee

 \subsection{Seiberg-Witten description of \texorpdfstring{$SU(2)$}{SU(2)} \texorpdfstring{$N_f=1$}{Nf=1}}
 \label{sec:SW}

One powerful way of understanding the singularities in the  Borel plane, and the corresponding behavior of Borel summation, is by exploiting  the connection between the operators \eqref{dex2}, \eqref{dez} and Seiberg-Witten theory, which we now review.

It is well known that the
classical limit of \eqref{dez}
  corresponds to the Seiberg-Witten curve for four dimensional $SU(2)$ theory with $N_f=1$ by identifying $E=2 u$; see for instance \cite{Maruyoshi:2010iu} which has
  the same conventions we do. We recall that this Seiberg-Witten curve can be represented as
  \begin{equation}
  \label{SWcurve}
\Sigma=\left\{\lambda^2-\left( \frac{\Lambda^2 }{2z^3}+\frac{2 m \Lambda}{z} - \Lambda^2 -\frac{2u}{z^2}\right) \de z^2=0 \right\}\subset T^* C.
 \end{equation}
From this perspective, the label $i$ for the square-root branches in \eqref{psidef}
is identified with the label of the sheets of the $2$-fold covering $\Sigma\xrightarrow{2:1}C$.
For generic $u$, the quantity $P(z)$ 
in the parentheses has 3 zeros; we call them $z_1$, $z_2$, $z_3$;
then $\{0,z_1,z_2,z_3\}$ are the branch points of the covering.

For any value of parameters, the electromagnetic and flavor charge lattice $\Gamma$ can be represented
as a rank $3$ sublattice of $H_1(\overline{\Sigma},\Z),$ where $\overline{\Sigma}$ is obtained by filling in the punctures of $\Sigma$.
We will choose a basis of $\Gamma$ for each value of parameters that we study, and thus
represent the charges concretely as $\gamma_{[m,n,\mu]}$.
In each case we fix a pure magnetic charge
\begin{equation}
\label{c1}
\gamma_{a_D}\equiv\gamma_{[1,0,0]},
\end{equation}
a pure electric charge
\begin{equation}
\label{c2}
\gamma_a\equiv\gamma_{[0,2,0]}=2\gamma_{[0,1,0]},
\end{equation}
and a pure flavor charge
\begin{equation}
\label{c3}
\gamma_{f}\equiv\gamma_{[0,0,1]}.
\end{equation}
In each case we take $\gamma_f$ to be the homology class of a loop surrounding only the irregular singularity $z=\infty$ and no other turning points or singularities.

Some examples of charges in the bases we use 
are shown in \autoref{chargecycles} and \autoref{fig:mlwkcharge}.  
Note that when the parameters $(u,m,\Lambda)$ change, the positions of branch points also change; thus we have to analytically continue the branch points and the corresponding cycles.
The specific choice of basis shown in \autoref{fig:mlwkcharge} is well adapted 
for comparison with the instanton counting method: the vectormultiplet has charge $\gamma_a$, while the derivative of the prepotential corresponds to the charge $\gamma_{a_D}$. We discuss this in more detail  in \autoref{sec:QPI}. 

We also recall some standard quantities associated to the charge lattice. 
The antisymmetric non-degenerate intersection pairing of electromagnetic charges is given by\footnote{The flavor charge doesn't contribute to the intersection number. Hence we neglect flavor charge and represent it by $*$.}
\begin{equation}
\langle\gamma,\gamma'\rangle=\langle[n_1,n_2,*],[n_1',n_2',*']\rangle=n_1n_2'-n_2n_1'\,. 
\end{equation}
The central charge corresponding to $\gamma_{[m,n,\mu]}$ is 
\begin{equation}\label{charges}
Z_{\gamma_{[m,n,\mu]}}=\oint_{\gamma_{[m,n,\mu]}} \lambda=m Z_{\gamma_{[1,0,0]}}+n Z_{\gamma_{[0,1,0]}} +\mu Z_{\gamma_{[0,0,1]}}
\end{equation}
where $\lambda$ is the Seiberg-Witten differential.

\subsubsection{A strong coupling point}
\begin{figure}[htb]
\centering
  \includegraphics[width=0.35\linewidth]{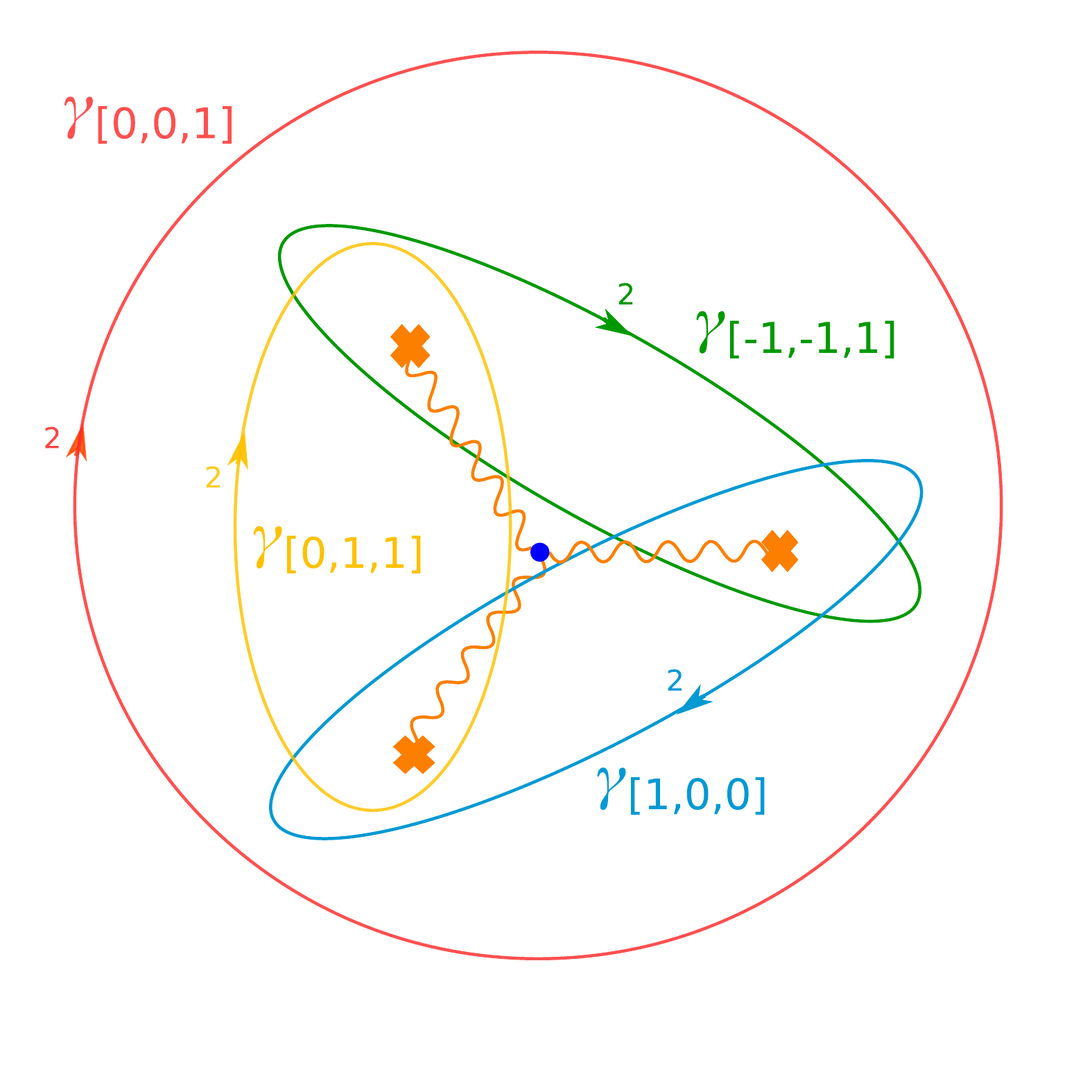}
  \caption{Charges of BPS hypermultiplets at $m=-1/10$, $\Lambda=1$ and $u=0$ as loops on the $z$ plane. We use orange crosses to denote the turning points of the SW differential. They are branch points of the covering $\Sigma \to C$. The orange wavy lines represent the branch cuts. The blue dot denotes the singularity at $z=0$. There is another singularity at $z=\infty$. In the strong coupling region, we have 3 BPS states corresponding to the blue, green and yellow cycles. The pure flavor charge is plotted as a large red loop.}
  \label{chargecycles}
\end{figure}

As an example we can take the parameters to be  $m=-1/10$, $\Lambda=1$, $u=0$. In this case we are in the strong coupling region and there are only 3 BPS hypermultiplets, with charges $\gamma_{\pm [1,0,0]}$, $\gamma_{\pm [0,1,1]}$ and $\gamma_{\pm [-1,-1,1]}$. These charges are shown in Figure \ref{chargecycles}. The corresponding central charges are
\begin{itemize}
\item
$
Z_{\gamma_{[-1,-1,1]}}=2.1922857627+ 3.0893009403\ri,
$
\item
$
Z_{\gamma_{[1,0,0]}}=2.1922857627- 3.0893009403\ri,
$
\item
$
Z_{\gamma_{[0,1,1]}}=-3.1279344639.
$
\end{itemize}
Let us discuss separately the central charge corresponding to the flavor mass $Z_{\gamma_{[0,0,1]}}$. In the limit $z\rightarrow\infty$, we have $\lambda\approx\pm\sqrt{-(\Lambda-\frac{m}{z})^2} \de z$. Hence there is a singularity at $\infty$ whose residue gives $\mp2\pi m$. Integrating around the loop $\gamma_{[0,0,1]}$ 
shown in \autoref{chargecycles} then we have
\begin{itemize}
\item $Z_{\gamma_{[0,0,1]}}= -2\pi m=\frac{1}{5}\pi.$
\end{itemize}

\subsubsection{A weak coupling point}
\label{wkBPSspectrum}
\begin{figure}
\begin{subfigure}{.3\textwidth}
  \centering
  \includegraphics[width=1.5\linewidth]{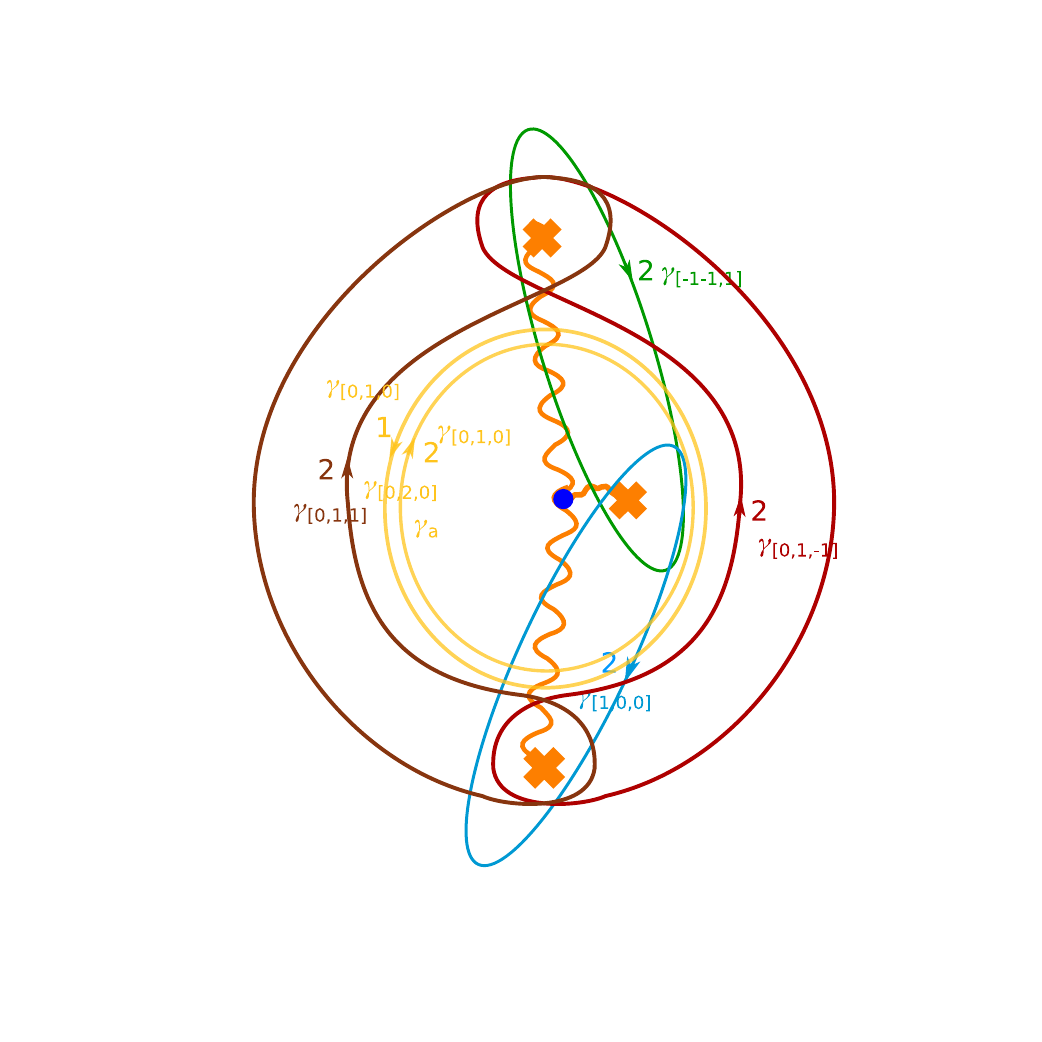}
 \caption{}
  \label{fig:mlwkcharge}
\end{subfigure}%
\begin{subfigure}{.3\textwidth}
  \centering
  \includegraphics[width=1.5\linewidth]{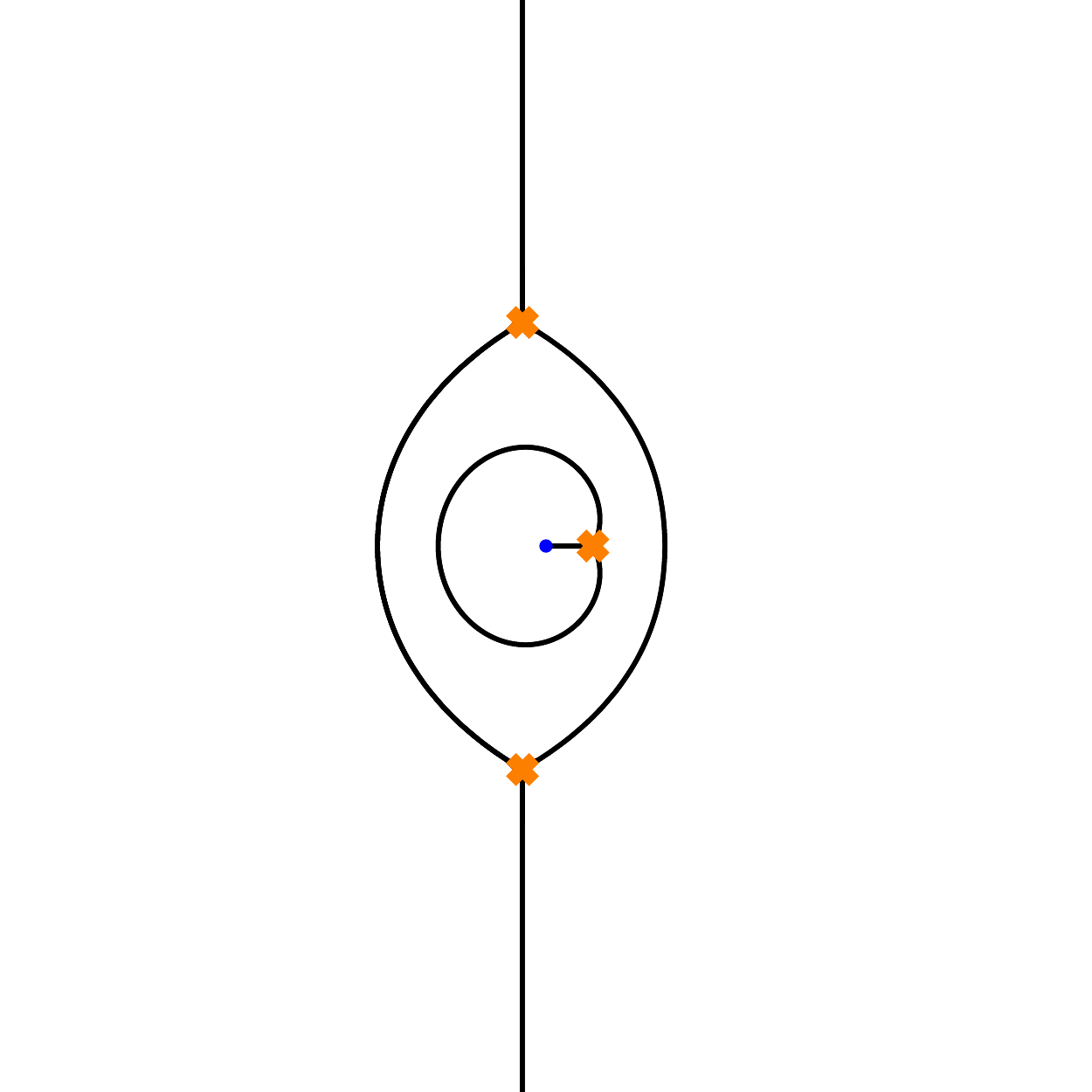}
  \caption{}
  \label{fig:SNmlwk}
\end{subfigure}%
\begin{subfigure}{.3\textwidth}
  \centering
  \includegraphics[width=1.5\linewidth]{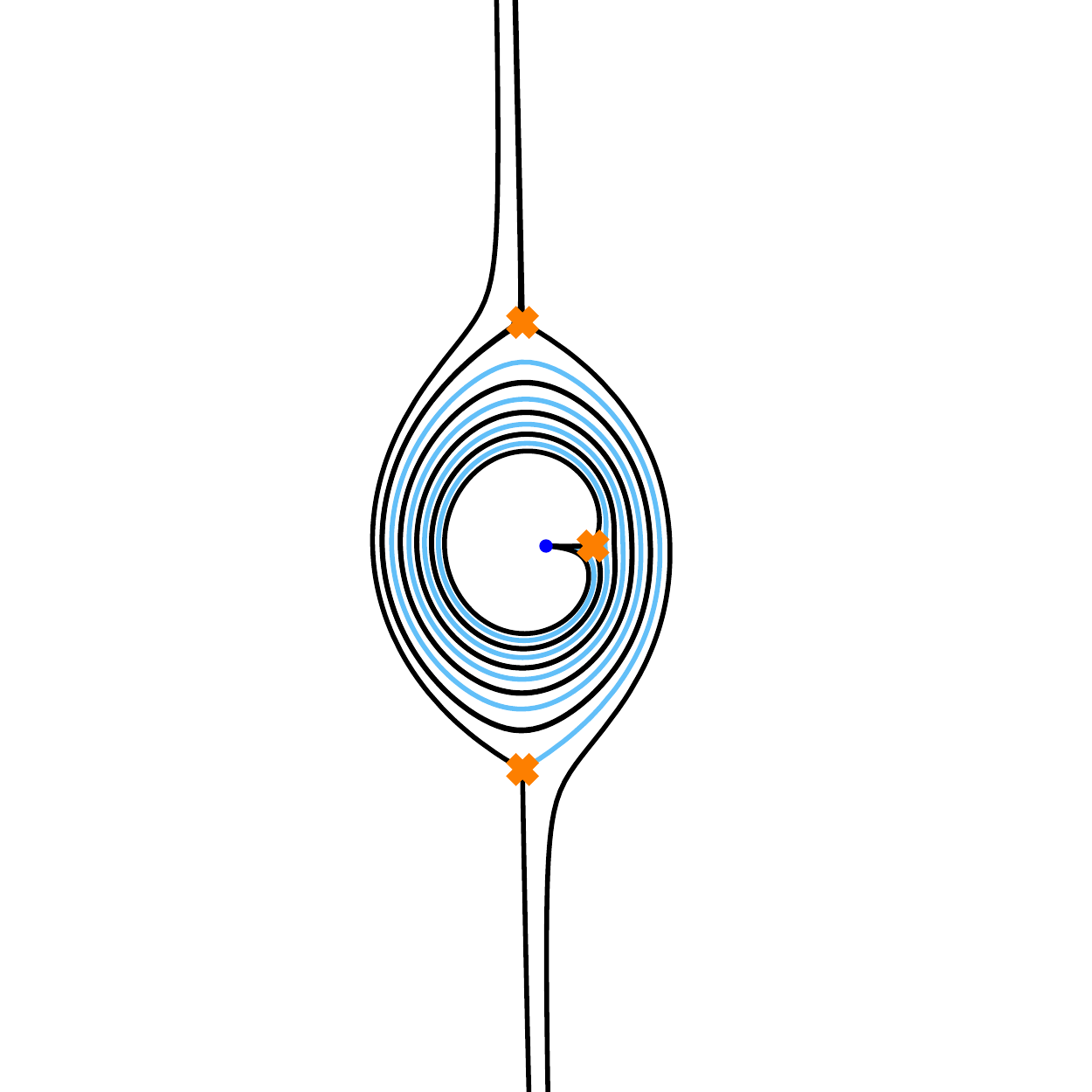}
  \caption{}
  \label{fig:SNmlwk02}
\end{subfigure}
  \caption{ In \autoref{fig:mlwkcharge} we plot the lattice charges as loops in the $z$ plane.
The yellow double loops is $\gamma_{[0,2,0]}$,  while the brown and red loops correspond to $\gamma_{[0,1,1]}$ and $\gamma_{[0,1,-1]}$. The blue loop is  $\gamma_{[1,0,0]}$ while the green loop is $\gamma_{[-1,-1,1]}$. Notice that the blue and the green loops are the analytic continuation from the charges in the strong coupling region, see \autoref{chargecycles}. So they maintain their charge labeling.
  In \autoref{fig:SNmlwk} we show  $\mathcal{W}^{\vartheta=0, m=0}$ in  the $z$ plane. This   corresponds to  the degenerate spectral network for  the BPS vectormultiplet $\gamma_a$ and the 2 BPS hypermultiplets $\gamma_{[0,1,1]}$ and $\gamma_{[0,1,-1]}$.  There are infinite many BPS hypermultiplets besides $\gamma_{[0,1,1]}$ and $\gamma_{[0,1,-1]}$. We can see this from \autoref{fig:SNmlwk02} which shows how the spectral network looks when we slightly vary $\vartheta$. This spiraling walls will be more and more less spiraling when $\vartheta$ is varying away from $\vartheta=0$. In this process, there will be infinite saddle connections corresponding to hypermultiplets. These infinite saddle connections correspond to the hypermultiplets $\gamma_{a_D}+n \gamma_a$ and $\gamma_{[-1,-1,1]}+n \gamma_a$, where $n=0,1,...$ is the winding number of the paths inside the ring domain in Figure \ref{fig:SNmlwk}. }
  \label{fig:mlwkcharges}
\end{figure}

Let us now consider an example inside the weak coupling region where we have infinitely
many BPS states. 
We can take  $u=2,\Lambda=\frac{1}{5}$ and $ m=0$. There, the BPS spectrum of the theory consists of:

\begin{itemize}
\item A vectormultiplet with charge $\gamma_a\equiv\gamma_{[0,2,0]}$, whose central charge is $Z_{\gamma_a}=-25.13274240682$.

\item Two hypermultiplets with charges $\gamma_{[0,1,1]}$ and $\gamma_{[0,1,-1]}$. Each of them has central charge $\frac{1}{2}Z_{\gamma_a}$\footnote{In the massive case, their central charges would be $Z_{\gamma_{[0,1,\pm1]}}={\frac12}Z_{\gamma_a}\mp 2 \pi m$.}. Their two
charges sum to the charge of the vectormultiplet.

\item Hypermultiplets with charges $\gamma_{[1,2n,0]}$, $n\in\mathbb{Z}$. We name the lightest hypermultiplet in this infinite tower by $\gamma_{a_D}\equiv\gamma_{[1,0,0]}$. Its central charge is $Z_{\gamma_{a_D}}=6.28318560170+ 26.72137744495\ri$.

\item Hypermultiplets with charges $\gamma_{[-1,-1+2n,1]}$, $n\in\mathbb{Z}$. The lightest hypermultiplet has $Z_{\gamma_{[-1,-1,1]}}=6.28318560170 -  26.72137744495\ri$.
\end{itemize}
We show the charges of the BPS states $\gamma_a$, $\gamma_{a_D}$, $\gamma_{[0,1,1]}$, $\gamma_{[0,1,-1]}$,  and $\gamma_{[-1,-1,1]}$ in \autoref{fig:mlwkcharges}.

\subsection{Stokes graphs and Borel poles for the local solutions}
\label{padeborelsol}

To discuss Borel summation of the all-orders WKB series, it is useful to first introduce $\vartheta$-Stokes graphs $\mathcal{W}^{\vartheta}$. (If $\vartheta$ is not specified, we assume $\vartheta = \arg(\hbar)$.)

The $\vartheta$-Stokes graph is made of $\vartheta$-Stokes curves on the punctured sphere $C$. Each $\vartheta$-Stokes curve of type $ij$ (where $(i,j) = (1,2)$ or $(2,1)$) 
is an oriented trajectory starting at a turning point, along which the 1-form
$\re^{-\ri\vartheta}(Y^{0,(i)}-Y^{0,(j)}) \de z$ is real and positive. 
Stokes graphs are also known as spectral networks, and we will use both names in this paper. 
Some examples of Stokes graphs appear in \autoref{QPFGf}, \autoref{resolst}, \autoref{QPFN}, \autoref{quantizationcd} below.

The local WKB solutions in each domain of $C\backslash\mathcal{W}^\vartheta$ are defined by \begin{equation} \label{eq:psi-borel}
\psi^{(i)}(z)=\re^{\frac{1}{\hbar} \int_{z_0}^z s(Y^{(i)})(z,\hbar) \de z}.
\end{equation}
The space of solutions of \eqref{dez} is a 2-dimensional vector space; $\psi^{(1)}$ and $\psi^{(2)}$ form a basis of this vector space 
in each domain of $C \setminus \mathcal{W}^\vartheta$. 
The solution $\psi^{(i)}$ jumps at a Stokes curve of type $ij$ (while $\psi^{(j)}$
does not jump there.)
This jumping is a manifestation of the Stokes phenomenon.

To understand this jumping phenomenon better, note that
the Borel summation in \eqref{eq:psi-borel} is only well defined if the Borel transform 
$\hat Y^{(i)}(z,\zeta)$ has no singularities in the $\zeta$-plane along
the ray of integration $(0,\re^{\I \vartheta} \infty)$. The 
positions of the singularities of $\hat Y^{(i)}(z,\zeta)$ in the $\zeta$-plane
depend on $z$ (as well as $E$, $\Lambda$, $m$).
It is known \cite{MR3706198,nikolaev2020exact} that
$\hat Y^{(i)}(z,\zeta)$ can only have singularities
along the rays
$\zeta \in (0,\re^{\I \vartheta} \infty)$ 
for specific $\vartheta$, namely those $\vartheta$ such that $z$ 
lies on a $\vartheta-$Stokes curve of type $ij$.
This gives an explanation of the fact 
that $\psi^{(i)}(z)$ is well defined only for $z$ away from 
$\vartheta$-Stokes curves of type $ij$.

So far we have described the arguments of singularities of $\hat Y^{(i)}$ 
in the Borel plane, but not their
magnitudes. We can make a more precise statement: 
if $z$ lies on a $\vartheta$-Stokes 
curve of type $ij$, then
$\hat Y^{(i)}(z,\zeta)$ has a singularity at \be \label{polesSol}\zeta = 2 \int_{z_0}^{z} \lambda\, ,\ee
where $z_0$ is the branch point where the Stokes curve begins.
This statement appears in the exact WKB literature: see in particular \cite{kawai2005algebraic},
where it is explained in terms of the rules for propagation of singularities
derived from microlocal analysis.

We made some numerical verifications of this statement in the $SU(2)$ $N_f=1$ theory, by choosing a
point $z$ and computing Pad\'e approximants to partial sums of the series
$\hat{Y}^{(i)}(z,\zeta)$ up to $N$ terms. In the $N \to \infty$ limit,
the poles of the Pad\'e approximant lie along
curves in the $\zeta$-plane, and the endpoint of each curve is one of the singularities 
of the true $\hat{Y}^{(i)}(z,\zeta)$. At finite $N$ one sees some approximation
to this structure, and can read off an approximation of the position of the
desired singularities. We find that they indeed lie at the expected positions
\eqref{polesSol}.
See
\autoref{polestructure} and \autoref{polestructurewk} for 
two examples.
In Appendix \ref{localsolpure} we present some similar checks in the pure $SU(2)$ theory.

In terms of quantum field theory, the quantity $2 \int_{z_0}^{z} \lambda$ 
has a simple meaning, as follows. We consider the 4d $SU(2)$ $N_f=1$ theory
coupled to a certain $\N=(2,2)$ supersymmetric surface defect $S(z)$, where the 
parameter $z \in C$ is identified as the coupling of the defect $S(z)$,
as discussed in \cite{Gaiotto:2009fs,Gaiotto:2011tf}. 
This surface defect has $2$ vacua corresponding
to the sheets $i$, $j$. Then $z$ lies on the $\vartheta$-Stokes
curve of type $ij$ exactly if the defect $S(z)$ admits a BPS soliton
with phase $\vartheta$ which interpolates
from vacuum $i$ to vacuum $j$.
In this case $2 \int_{z_0}^z \lambda$ is the central charge of the BPS soliton.
In short: singularities of the 
Borel transform $\hat Y^{(i)}(z,\zeta)$
appear at the central charges of BPS solitons in the surface defect theory
$S(z)$.
The number of singularities depends on the value of $z$; this is the wall-crossing
phenomenon for BPS solitons \cite{CecottiVafa,Gaiotto:2011tf}.

This proposal is parallel to the proposal of \cite{Grassi:2019coc} that
singularities in the Borel transform of the quantum periods appear
at the central charges of BPS particles in the 4d theory. (We will explore that
phenomenon in the $SU(2)$ $N_f=1$ theory below.) 
Indeed, taking the conformal limit of the analysis of \cite{CecottiVafa,Gaiotto:2011tf,gmn},
one concludes that Stokes jumps of the local solutions are induced by BPS solitons,
in complete parallel to the way that Stokes jumps of the quantum periods
are induced by 4d BPS particles.

\begin{figure}
\begin{centering}
 \includegraphics[width=\linewidth]{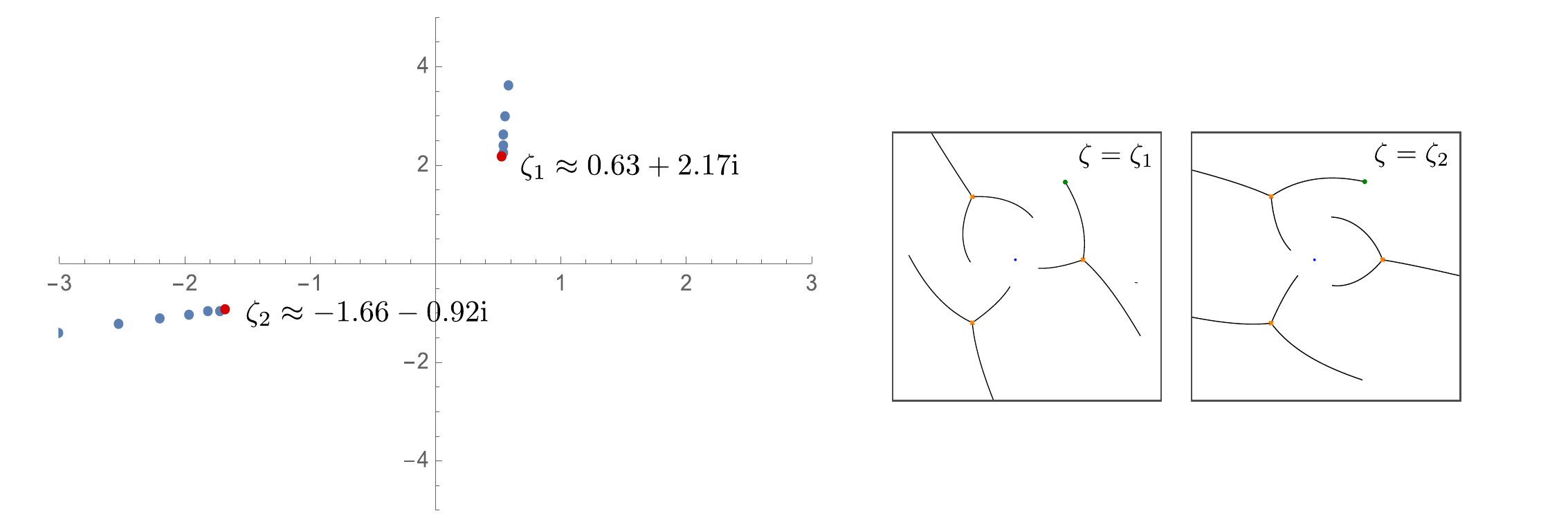}
  \caption[Caption for LOF]{
  Left: singularities of the Pad\'e-Borel transform of ${Y}^{(2)}$ in \eqref{wkbY}  for $u=0, m=-1/10$, $\Lambda=1$ and   $z=\re^{\ri}$, with $N = 40$ terms in the series.
    The leading singularities $ \zeta_i$, marked in red, match well with \eqref{polesSol}; thus as expected 
    they are the central charges of BPS solitons on the
    surface defect with parameter $z$. (For $Y^{(1)}$ the singularities would be at the opposite points $\zeta = -\zeta_i$.)
     Right: for fixed $\zeta = \zeta_i$ we plot a cutoff version 
     of the Stokes graphs with $\vartheta = \arg(\zeta)$, 
     plotting the Stokes curves only up
     to $|2\int_{z_0}^{z} \lambda| = \abs{\zeta}$.
    As explained around \eqref{polesSol}, the cutoff Stokes curves $\cW^{\vartheta=\arg(\zeta_i)}$ run exactly up to the point $z=\re^\ri$, which is plotted as a green dot. 
}
\label{polestructure}
\end{centering}
\end{figure}

\begin{figure}
\begin{centering}
 \includegraphics[width=\linewidth]{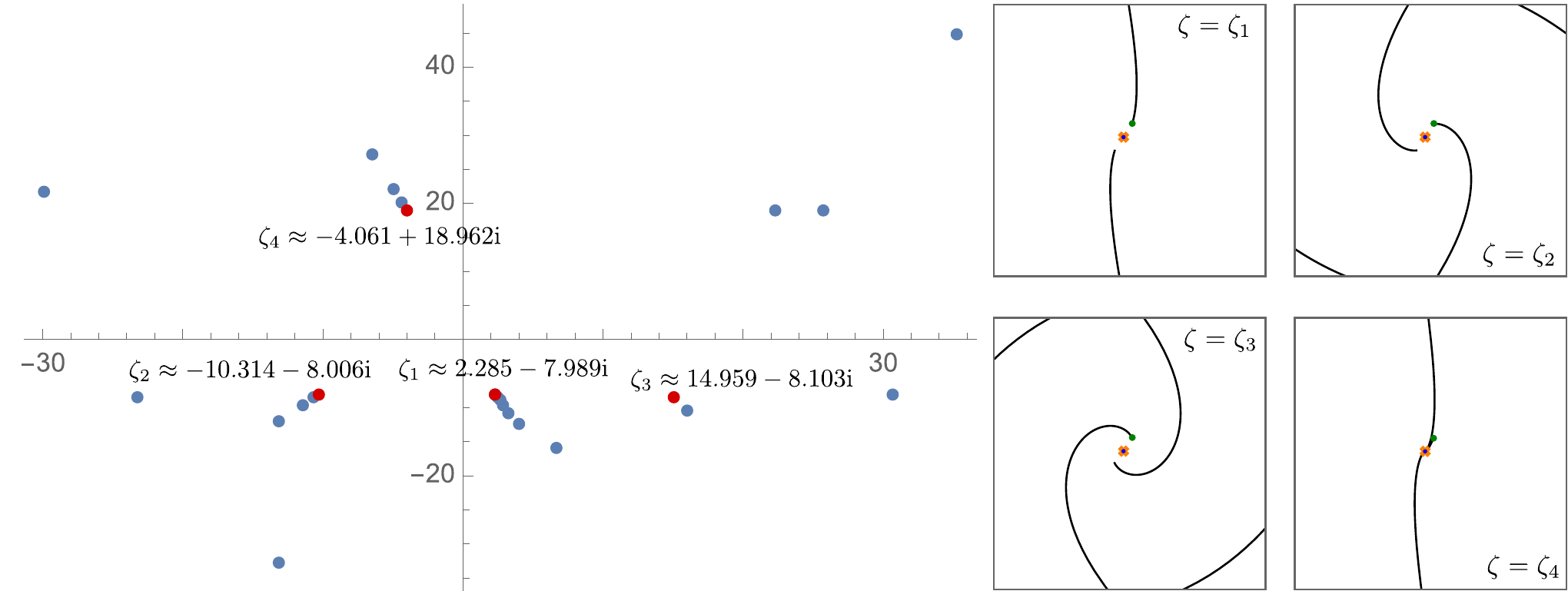}
  \caption[Caption for LOF]{
  Left: singularities of Pad\'e-Borel transform of $Y^{(2)}$ in \eqref{wkbY} for $u=2$, $m=0$, $\Lambda=1/5$ and $z=\re^{\ri}$, with $N = 50$ terms in the series.
  There are infinitely many poles of $\hat{Y}^{(2)}$, corresponding to the central charges of solitons supported at $z$. Only finitely many of them are visible in our
  approximation. (Again, for $Y^{(1)}$ the singularities would be at the opposite points $\zeta = -\zeta_i$.)
  Right: cutoff Stokes graphs at $\zeta = \zeta_1, \zeta_2, \zeta_3,\zeta_4$,
  as in \autoref{polestructure} above.
}
\label{polestructurewk}
\end{centering}
\end{figure}

 \subsection{Borel summation for quantum periods}
 \label{subsec:QPborel}
 Given $\gamma\in \Gamma$,
 we define the WKB quantum period $\Pi_\gamma^{\rm WKB}(\hbar)$
 as the integral of the even part of the WKB series \eqref{eodef} along $\gamma$:
  \be \label{wkbs}\Pi_\gamma^{\rm WKB}(\hbar)=\frac{1}{\hbar}\sum_{n=0}^\infty \Pi_\gamma^{n}\hbar^{2n}=\frac{1}{\hbar}\sum_{n=0}^\infty\left(\oint_\gamma Y^{2n}(z)\de z\right)\hbar^{2n}.\ee
   In particular, 
  \be \Pi_{\gamma_{}}^{0} =Z_{\gamma_{}}.\ee   

We are now ready to define the quantum periods, by Borel summation:
 \be \label{wkbqm}\Pi_\gamma(\hbar)= \overline{s}\left(\Pi_\gamma^{\rm WKB}\right)(\hbar).\ee

\subsection{The one-loop sign} \label{sec:one-loop-sign}

One might wonder why we do not define the WKB period as the integral of the full WKB series \eqref{wkbY}. The odd part is a total derivative, according to \eqref{oddeven}. Hence the only contribution of the odd part comes from the monodromy of 
$\log Y_{\mathrm{even}}$ around the contour of integration on $\Sigma$. 
By expanding the log one sees that this 
monodromy comes just from the leading
term in $Y_{\mathrm{even}}$, i.e. it is the same as the monodromy of
$\log \sqrt{P}$. Since $\sqrt{P}$ is single-valued on $\Sigma$, this
monodromy is a shift by $2 \pi \I \omega$ for some $\omega \in \Z$.
Thus we have
\be \label{mon}\oint Y_{\rm odd}= -\frac{\hbar}{2} \oint \de \log\sqrt{P} = -\hbar \pi \ri \omega. \ee
Including this contribution would lead to an extra sign
$\exp\left( \hbar^{-1} \oint Y_{\rm odd} \right) = (-1)^\omega$ in the exponentiated WKB series.
This sign is subtle, for two distinct reasons:
\begin{itemize}
  \item It picks up a factor $-1$ when the contour of integration on $\Sigma$ is moved across
  a branch point, so it is not (quite) a function of a homology class $\gamma \in H_1(\Sigma,\Z)$.
  \item It is not coordinate invariant: for a loop which goes around the cylinder, 
  computing in the $z$-plane and the $x$-plane lead to different signs.
\end{itemize}
For our immediate purpose,
it is convenient simply to avoid these issues, by taking only the even part as 
we did in \eqref{wkbs}; then the resulting $\Pi_\gamma$
is canonical, well defined as a function of the homology class (charge) $\gamma$,
and additive.
However, we will meet this sign again in \autoref{rulesw} below,
and we discuss it from a more invariant point of view in  \autoref{ssabeli}.

\subsection{Pad\'e-Borel computation of quantum periods}

 The coefficients $\Pi^n_\gamma$ in \eqref{wkbqm} can be efficiently computed by using the differential operator technique, parallel to \cite{Grassi:2019coc,huangNS,Huang:2014nwa}.
We first note that
\footnote{We have been neglecting moduli parameters in the equation and solutions in former discussion. Recall we have Coulomb branch vev $u$ 
and flavor mass $m$. 
So $Y^{2n}(z)$ is actually $Y^{2n}(z,u,m)$.}
the $Y^{2n}(z,u,m)\de z$ term can be expressed as 
\begin{equation}
Y^{2n}(z,u,m)\de z=b_0^nY^0(z,u,m)\de z+b_1^n\frac{\partial Y^0(z,u,m)}{\partial m}\de z+b_2^n\frac{\partial Y^0(z,u,m)}{\partial u}\de z+\de (\cdots)
\end{equation}
 where  $\de(\cdots)$ denotes a total derivative term in $z$.
Therefore 
$\Pi^n_\gamma$ satisfies
\begin{equation}
\label{diffopr}
\Pi^n_\gamma(z,u,m)=b_0^n\Pi^0_\gamma(z,u,m)+b_1^n\frac{\partial \Pi^0_\gamma(z,u,m)}{\partial m}+b_2^n\frac{\partial \Pi^0_\gamma(z,u,m)}{\partial u}.
\end{equation}
We also emphasize that the quantum period for the flavor charge $\Pi_{\gamma_{[0,0,1]}}$ is not subject to quantum corrections, as discussed for instance in \cite{Huang:2014nwa} (see also \autoref{quantummass} for a direct computation).
In other words,
\be\label{pnm}\Pi_{\gamma_{[0,0,1]}}(\hbar)=\frac{Z_{\gamma_{[0,0,1]}}}{\hbar}=-\frac{2\pi m}{\hbar}\,.\ee
In particular $ \Pi^n_{\gamma_f}=0$,   $\forall n>0 $, which implies the constraints\footnote{We tested explicitly that this relation holds also for the other periods.}
 \be \label{b1b0}b_1^n=-m b_0^n \,.\ee

\begin{table}[h!]
\centering
\renewcommand{\arraystretch}{2}
\begin{tabular}{c c c c}
 &$i=0$&$i=1$&$i=2$\\
 \hline
$b_i^1$&$\frac{25}{6734}$&$\frac{5}{13468}$&$\frac{-3371}{80808}$\\
$ b_i^2$&$\frac{121523975}{196306106724}$&$\frac{24304795}{392612213448}$&$-\frac{1738734625}{1046965902528}$\\
 $b_i^3$&$\frac{14649990947544027025}{1744761644092944719424}$&$\frac{2929998189508805405}{3489523288185889438848}$&$-\frac{2245318130362254755}{2326348858790592959232}$\\
$ b_i^4$&$\frac{1156307121570685086685480765}{79119280552841605513208588544}$&$\frac{231261424314137017337096153}{158238561105683211026417177088}$&$-\frac{1685881772274283567722957525}{140656498760607298690148601856}$\\
\hline
\end{tabular}
\caption{The first few $b_i^n$, for $u=0$, $m=-1/10$, $\Lambda=1$.}
\label{diffop}
\end{table}
Given the form of the relation \eqref{diffopr} it is relatively straightforward
to compute the coefficients $b_i^n$ up to $n = 80$, at any particular 
point $(\Lambda,m,u)$;
we show samples of the first few $b_i^n$ in \autoref{diffop}.
Then we compute the three
quantities $\Pi_\gamma^0$, $\partial_u \Pi_\gamma^0$ and $\partial_m \Pi_\gamma^0$ by direct numerical integration. Thus we obtain $80$ terms of 
the series $\Pi_\gamma^{\mathrm{WKB}}$. Then following the
same Pad\'{e}-Borel summation technique discussed in \autoref{padeborelsol}, we get the
approximate quantum period $\Pi^{\rm PB}_{\gamma}$. 
In this way one can get many digits of precision; see \autoref{tab:introresultsnew} for 
some sample results.

\begin{figure}
\begin{subfigure}{.3\textwidth}
  \centering
\includegraphics[width=\textwidth]{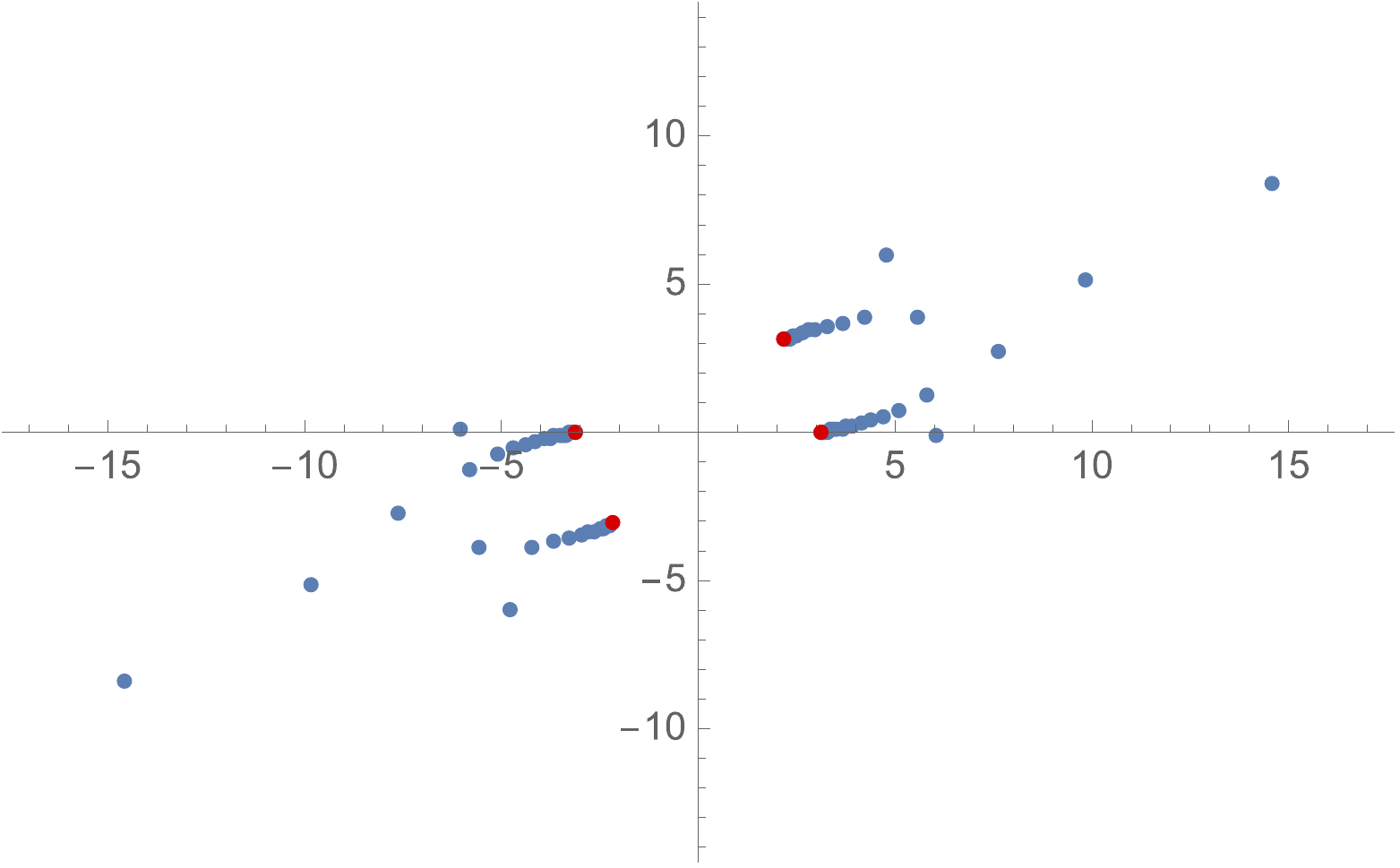}
 \caption{}
  \label{fig:qppole11}
\end{subfigure}
\begin{subfigure}{.3\textwidth}
  \centering
\includegraphics[width=\textwidth]{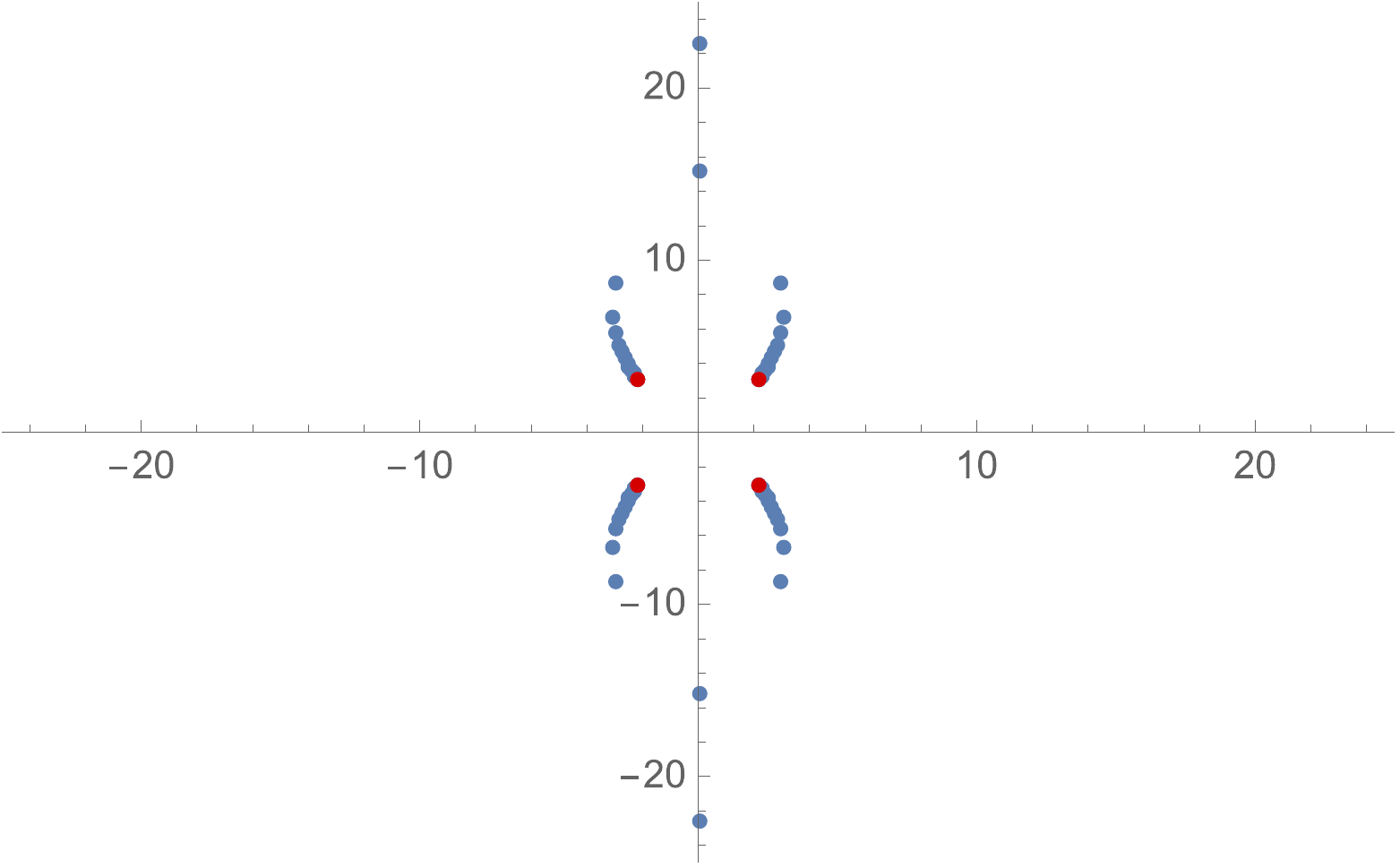}
 \caption{}
  \label{fig:qppole12}
\end{subfigure}
\begin{subfigure}{.3\textwidth}
  \centering
\includegraphics[width=\textwidth]{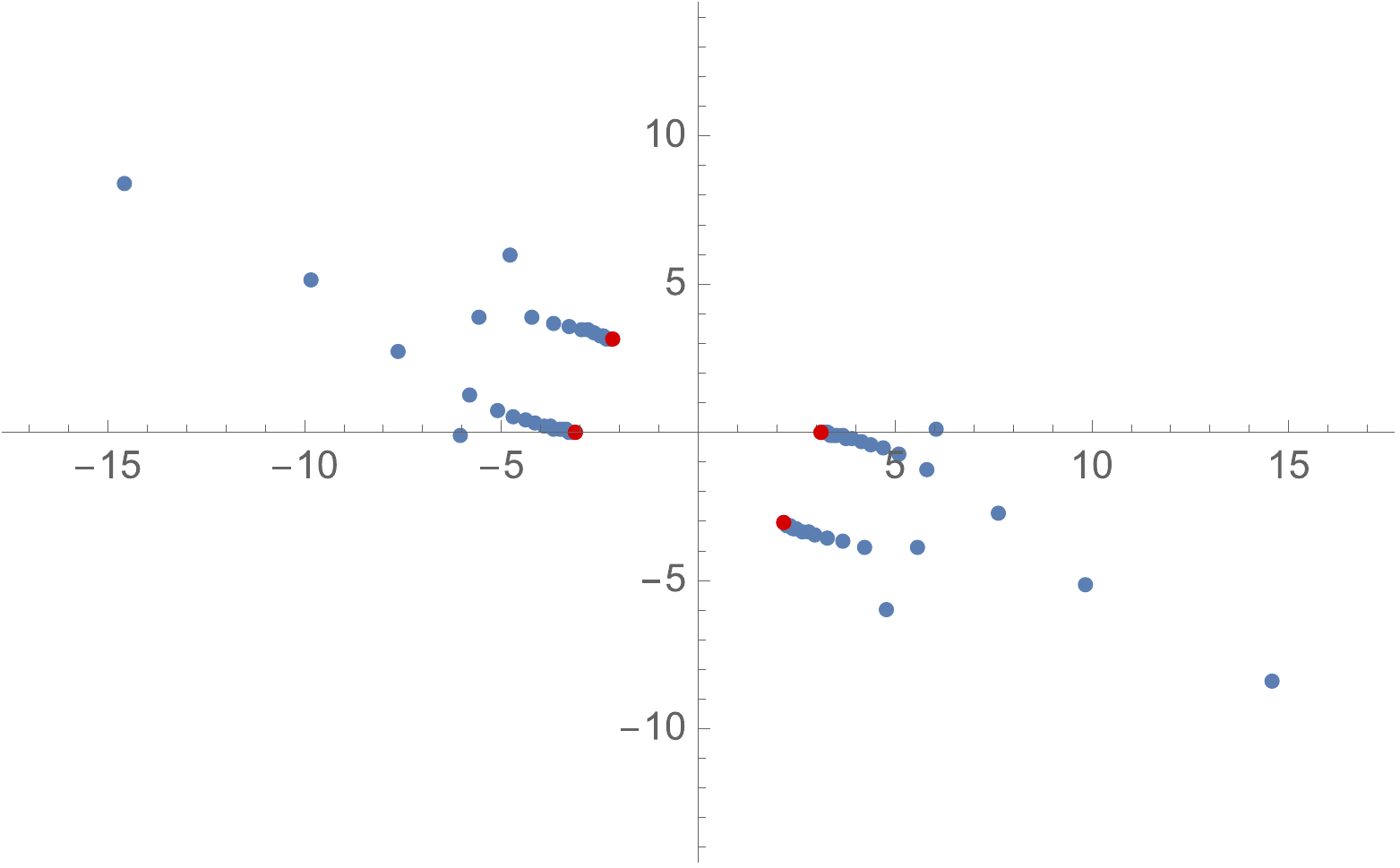}
 \caption{}
  \label{fig:qppole13}
\end{subfigure}
\\
\begin{subfigure}{.3\textwidth}
  \centering
    \includegraphics[width=\textwidth]{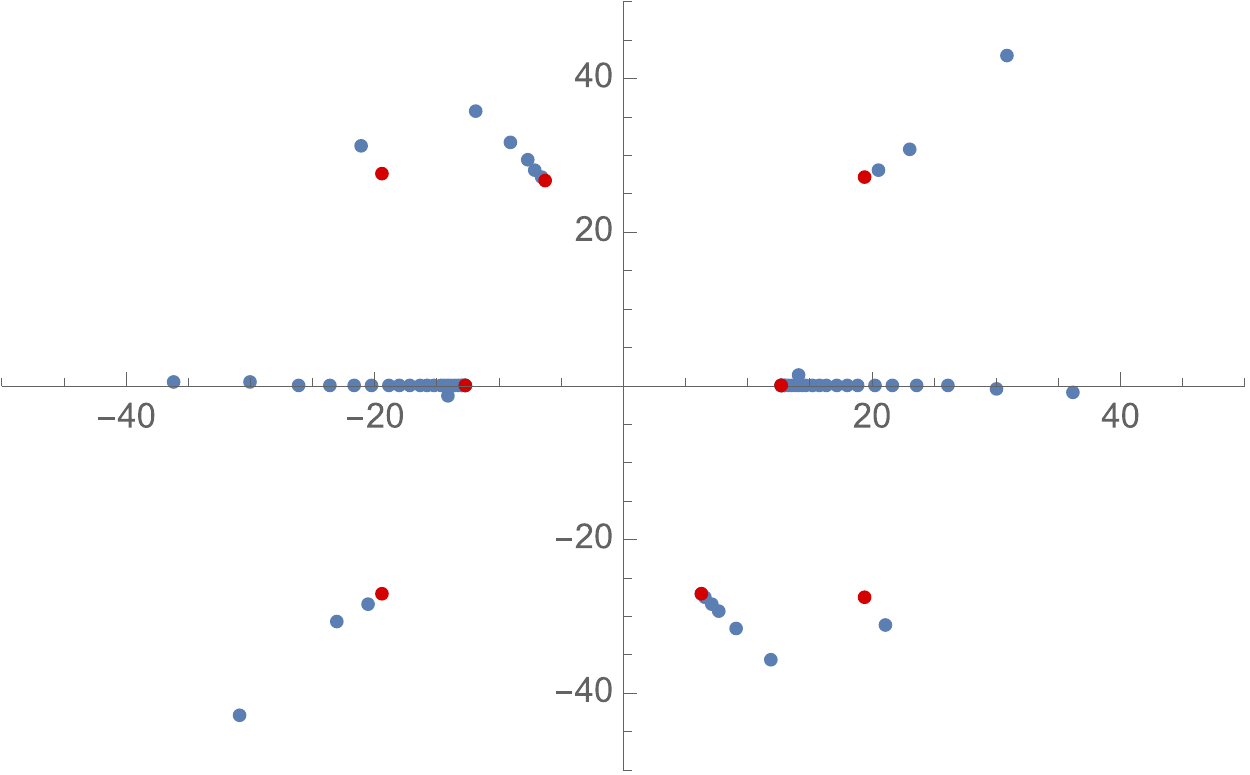}
             \caption{}
  \label{fig:qppole23}
\end{subfigure}
\begin{subfigure}{.3\textwidth}
  \centering
        \includegraphics[width=\textwidth]{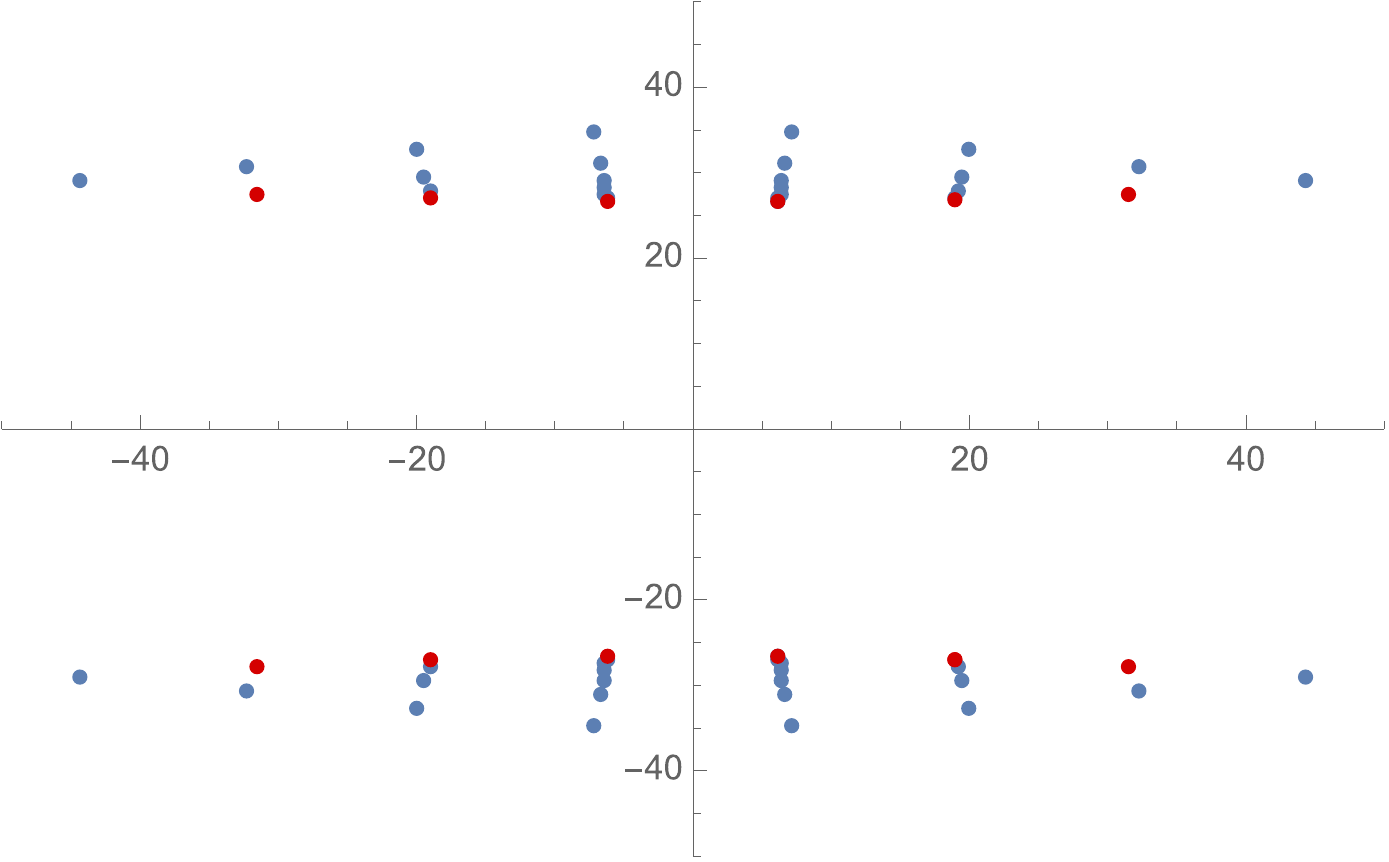}
             \caption{}
  \label{fig:qppole22}
\end{subfigure}
\begin{subfigure}{.3\textwidth}
  \centering
    \includegraphics[width=\textwidth]{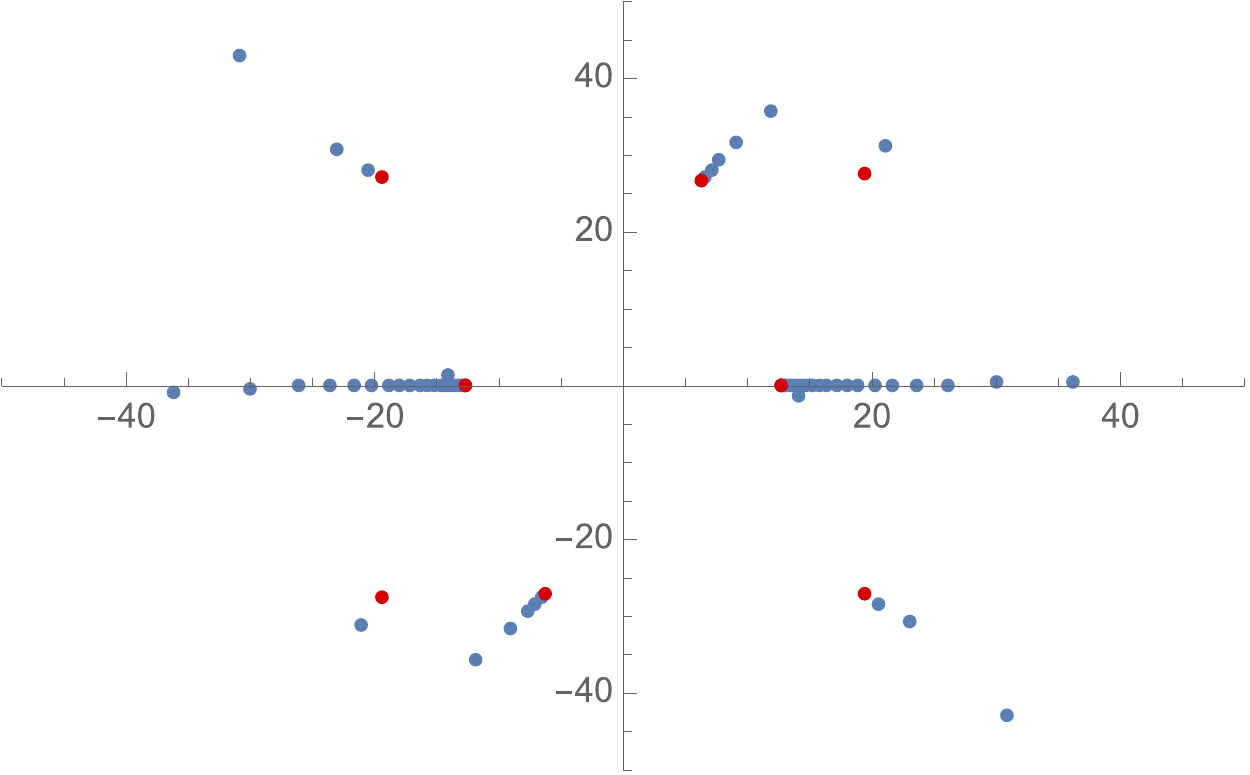}
     \caption{}
  \label{fig:qppole21}
\end{subfigure}
\\
\begin{subfigure}{.3\textwidth}
  \centering
        \includegraphics[width=\textwidth]{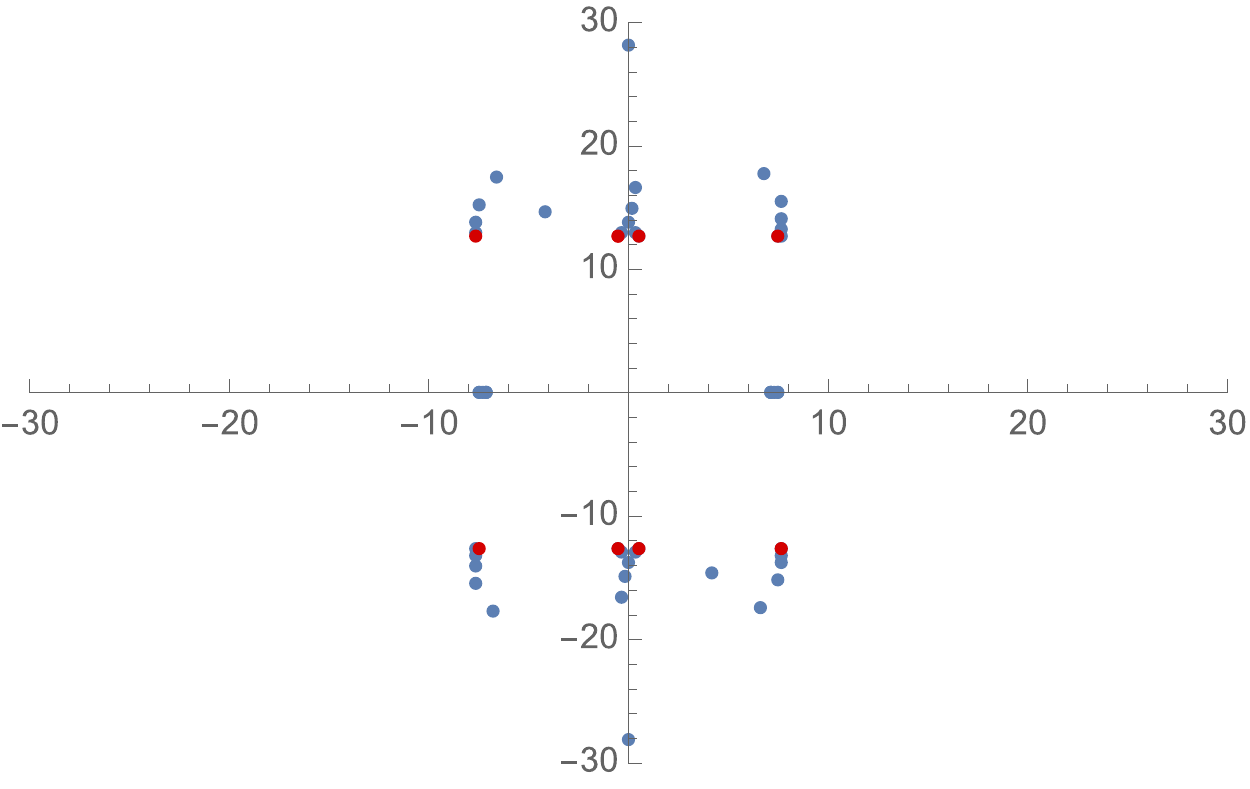}
              \caption{}
  \label{fig:qppole32}
\end{subfigure}
\begin{subfigure}{.3\textwidth}
  \centering
           \includegraphics[width=\textwidth]{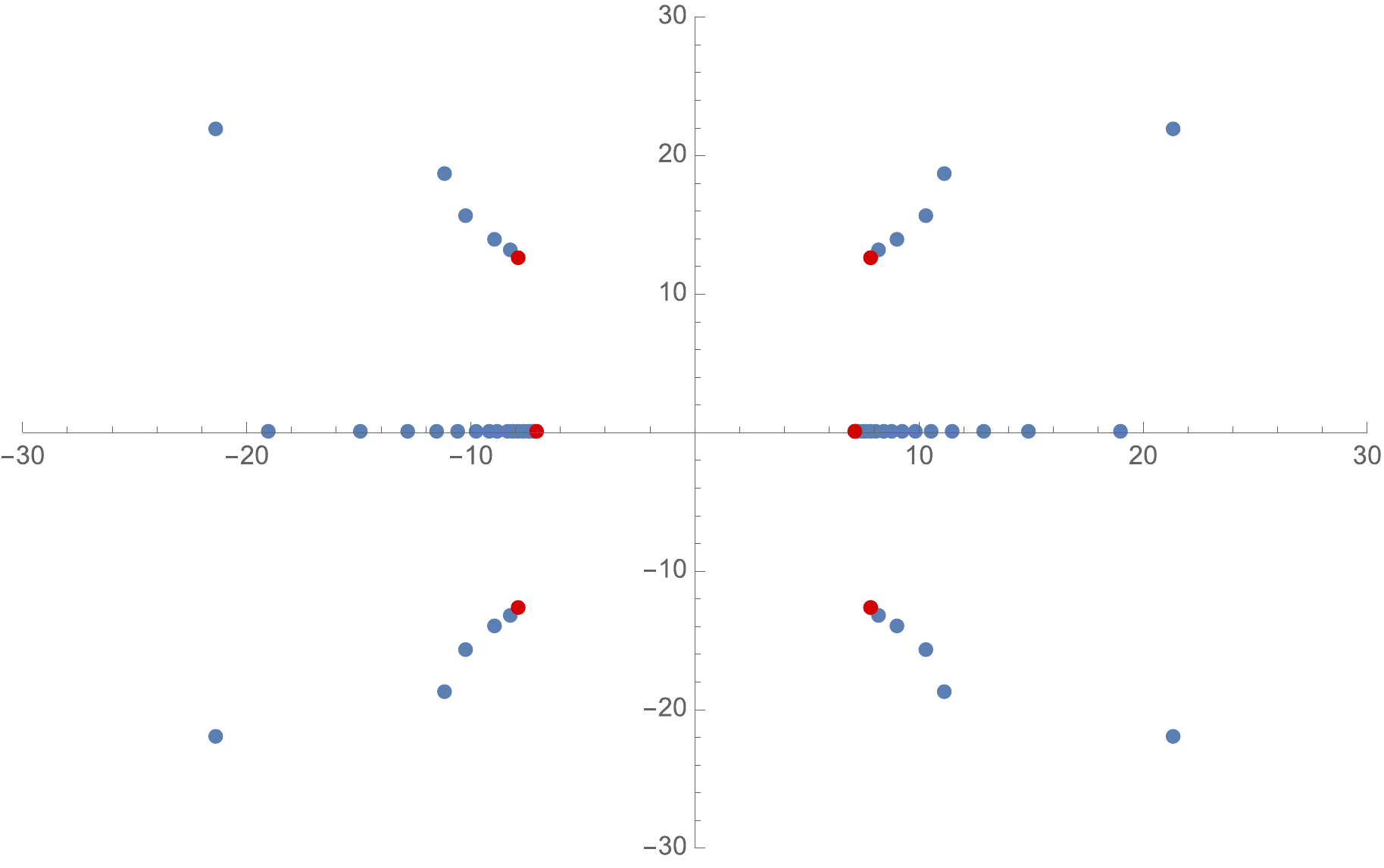}
              \caption{}
  \label{fig:qppole31}
\end{subfigure}
\begin{subfigure}{.3\textwidth}
  \centering
         \includegraphics[width=\textwidth]{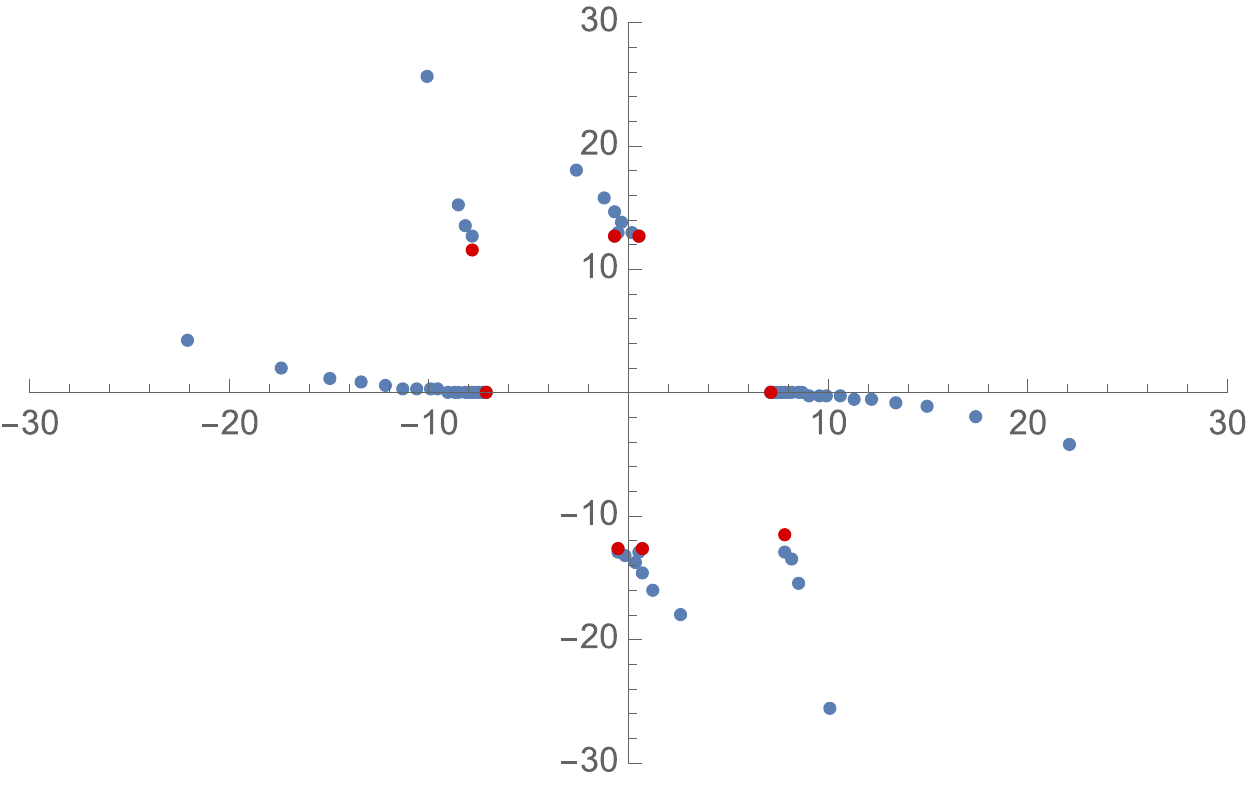}
               \caption{}
  \label{fig:qppole33}
\end{subfigure}
 \caption[Caption for LOF]{Poles of the Pad\'e-Borel transform for various quantum periods. The first row shows the poles for $\Pi_{\gamma_{[1,0,0]}}$ (left), $\Pi_{\gamma_{[0,1,1]}}$(center) and $\Pi_{\gamma_{[-1-1,1]}}$ (right) in strong coupling region at $u=0$, $m=-1/10$ and $\Lambda=1$. The second row shows the poles for $\Pi_{\gamma_{[1,0,0]}}$(left), $\Pi_{\gamma_{[0,2,0]}}$(center) and $\Pi_{\gamma_{[-1,-1,1]}}$ (right)   in weak coupling region for $u=2$, $m=0$ and $\Lambda=1/5$. The third row shows the poles for  $\Pi_{\gamma_{[1,0,0]}}$(left), $\Pi_{\gamma_{[0,2,0]}}$(center) and $\Pi_{\gamma_{[-1,-1,1]}}$ (right)   in weak coupling region for $u=-2$, $m=1/10$ and $\Lambda=1$. The first poles in each accumulated series are denoted by red dots. The position of such poles correspond approximately to the central charges of BPS states with non-zero intersection numbers w.r.t.~the charge $\gamma_{[-,-,-]}$ under consideration. }
 \label{poleQP}
\end{figure}

Let us comment briefly on the singularities of the Borel transform $\hat\Pi_\gamma$.
As predicted in \cite{Grassi:2019coc}, $\hat\Pi_\gamma$ is expected to have singularities at $\zeta = Z_{\gamma'}$ when $\gamma'$ is the charge 
of a BPS particle existing in the theory at  the particular point in moduli space that we are studying,
and $\langle\gamma,\gamma'\rangle\neq 0$. 
In \autoref{poleQP} we show numerical checks of this phenomenon
at strong and weak coupling; we indeed find the poles at the expected places.
These singularities are responsible for the fact that $\Pi_\gamma^{\mathrm{PB}}$
is only piecewise analytic, as we reviewed in \autoref{analyticstructures}.

\section{Computation by small sections}
\label{sec:ssection}

\subsection{Rules for writing down Wronskians}\label{rulesw}

In this chapter we explain how to use Wronskians of local solutions of \eqref{eq:our-eq-intro} 
 to compute quantum periods, following a general formalism 
 laid out in \cite{Gaiotto:2009hg,Hollands:2019wbr}.

 The basic ingredients needed are: 
\begin{enumerate}
\item Three solutions $\psi_1$, $\psi_2$ and $\psi_3$ of \eqref{eq:our-eq-intro}, which decay exponentially along three paths approaching the irregular singularities of
the equation at $z = 0$, $z = \infty$.
\item The monodromy operator $M$ giving evolution of the solutions of \eqref{eq:our-eq-intro} under $x \to x + 2 \pi \I$, and its eigenfunctions $\psi_4$, $\psi_5$.\footnote{We emphasize that $M$ is the monodromy of \eqref{eq:our-eq-intro}, not the version \eqref{dez} transformed
to the $z$-plane; the latter differs by a minus sign.}
\end{enumerate}
Below we give some brief insight into what $\psi_1$, $\psi_2$, $\psi_3$, $\psi_4$, $\psi_5$ are, and why we use them in our calculation.

\subsubsection{Global structure of the Stokes graph}

We need to discuss a bit about the global structure of the Stokes graph (see e.g. \cite{Gaiotto:2009hg} for more).

Recall that the Stokes curves of type $ij$ are examples of \ti{WKB curves} of type $ij$, i.e. curves along which
$\re^{-\ri\vartheta}(Y^{0,(i)}-Y^{0,(j)}) \de z$ is real and positive.
When any WKB curve of type $ij$ approaches an irregular singularity, it asymptotes to one of the distinguished \ti{Stokes directions} 
around the singularity with the same implicit label $ij$. In a local coordinate
$\tilde{z}$, each Stokes direction is a 
ray going into $\tilde{z} = 0$, characterized by the property
that at $\tilde{z} \to 0$ along the ray
\be
\label{stokedr}
\arg \left(\int^{\tilde{z}}_{\tilde{z}_0}(Y^{0,(i)}-Y^{0,(j)}) \de \tilde{z}' \right) \to \vartheta \text{ as } \tilde{z}\rightarrow 0.
\ee
To determine these directions explicitly it is sufficient to consider the
leading-order behavior of $(Y^{0,(i)}-Y^{0,(j)})$. Moreover, because the integral
diverges as $\tilde{z} \to 0$ we see that the choice of
$\tilde{z}_0$ is unimportant. We will show examples shortly, and see
that in the $SU(2)$ $N_f = 1$ theory there is one Stokes direction going into
$z = 0$ and two going into $z = \infty$.

We can also consider generic WKB curves;
each such curve is determined given a point $z\in C\backslash \mathcal{W}$ it goes through. Generic WKB curves never intersect each other, nor do they intersect the 
Stokes curves; 
the Stokes curves and generic WKB curves together make up a foliation of the Riemann surface $C$.

For generic $\vartheta$, following any generic WKB curve in either direction
leads asymptotically to one of the Stokes directions around one of the irregular
singularities, and likewise for following a Stokes curve in the forward direction.
For special $\vartheta$, $\vartheta=\arg(-Z_{\gamma_{\rm BPS}})$ 
with $\gamma_{\rm BPS}$ the charge
of a BPS particle, 
there are two other possibilities.
First, we can have a saddle connection, a Stokes curve connecting two turning points;
this corresponds to a BPS hypermultiplet.
Explicit examples are shown in  \autoref{fig:SNmlwk} and \autoref{fig:SNmlwk02}.
 Second, we can have a ring domain,
a 1-parameter family of closed WKB curves; this corresponds to a BPS vectormultiplet.
A ring domain has two boundaries; each boundary is generically a single Stokes curve emanating from a turning point and returning to the same turning point, but in special cases (such as
the $m=0$ case in  \autoref{fig:SNmlwk}) it can be broken into multiple saddle connections between
multiple turning points. 

\subsubsection{Exponentially decaying solutions}

Following the discussion above, in each 
Stokes direction going into a singularity, we get for free a
corresponding exponentially decaying solution (up to a constant). Recall that the ansatz we use in the WKB method is 
\be
\psi(z)=\re^{\hbar^{-1}\int_{z_0}^{z} Y(z) \de z}.
\ee
Now suppose $z$ approaches an irregular singularity along a Stokes 
direction with label $ij$. Then
we have
\be
\frac{\psi^{(j)}}{\psi^{(i)}} \sim \re^{\hbar^{-1}\left(\int_{z_0}^{z} (Y^{0,(j)}(z)-Y^{0,(i)}(z)) \de z \right )}
 \to \re^{\frac{-\infty\re^{\ri\vartheta}}{\hbar}}.
\ee
Since we require $\rm Re(\re^{-\ri\vartheta}\hbar)>0$, $\psi^{(j)}$ is exponentially small compared to $\psi^{(i)}$ when $z$ approaches the singularity.
Moreover, since $Y^{(i)} = -Y^{(j)}$ we can also write this ratio as
\be
\re^{\hbar^{-1}\int_{z_0}^{z} (Y^{0,(j)}(z)-Y^{0,(i)}(z))\de z}=\re^{\hbar^{-1}\int_{z_0}^{z} 2Y^{0,(j)}(z)\de z} \sim {\psi^{(j)}}^2.
\ee
Thus $\psi^{(j)}$ is exponentially decaying as $z$ goes into the singularity
along a Stokes direction with label $ij$. 

We look at the two singularities separately:
\begin{enumerate}
\item  \label{exSD}When $z\rightarrow0$, $\int Y^{0,(i)}-Y^{0,(j)}\de z\sim \pm\frac{\Lambda}{2z^{1/2}}$. Given a phase $\vartheta$, we have 1 Stokes direction. \autoref{QPFGf} is an example at $\vartheta=0$; then the Stokes direction at $z=0$ is $0^+$ with label $12$. The exponentially decaying solution along Stokes direction is \be\psi_1\equiv\label{exSD1}\psi^{(2)}=\re^{\hbar^{-1}\int_{z_0}^zY^{(2)}\de z}. \ee 
\item When $z\rightarrow\infty$, we use $\tilde{z}=\frac{1}{z}$, and then $\int Y^{0,(i)}-Y^{0,(j)} \de z\sim \pm\frac{\ri\Lambda}{\tilde{z}}$. Given a phase $\vartheta$, there are 2 Stokes directions with opposite labels. In \autoref{QPFGf}, which is at $\vartheta=0$, the Stokes directions at $z=\infty$ are $\infty\re^{\pi\ri/2}$ with label $12$ and $\infty\re^{-\pi\ri/2}$ with label $21$. The exponentially decaying solution along Stokes direction $21$ is\footnote{This looks similar to the exponentially decaying solution given in \eqref{exSD1}, but note: 1) $z_0$ in \eqref{exSD1} can be not lying on WKB paths with one end at $z_\infty$ at all; 2) $z$ are different in two cases.} 
\be\label{exSD2}\psi_2\equiv\psi^{(1)}=\re^{\hbar^{-1}\int_{z_0}^zY^{(1)}\de z}, \ee and the exponentially decaying solution along Stokes direction $12$ is \be\label{exSD3}\psi_3\equiv\psi^{(2)}=\re^{\hbar^{-1}\int_{z_0}^zY^{(2)}\de z}.\ee
\end{enumerate}

Now suppose we look at a generic WKB curve. In one direction it goes into 
a Stokes direction with label $ij$, and in the other direction it goes
into a Stokes direction with label $ji$. Thus the two WKB solutions
$\psi^{(i)}$, $\psi^{(j)}$ along this curve can also be described
as the exponentially decaying solutions in these two Stokes directions.

The Stokes graph $\mathcal{W}^\vartheta$ separates $C\backslash \mathcal{W}^\vartheta$ into domains. In each domain, all generic WKB curves run between
the same pair of asymptotic Stokes directions. Thus the basis of exponentially
decaying solutions which we obtain in a given domain 
is the same no matter which generic
WKB curve we consider. When we move across a Stokes
curve to a different domain, one or both of the asymptotic
Stokes directions along generic WKB curves changes, 
and thus the basis of exponentially decaying
solutions jumps. This is consistent with the 
fact  we reviewed in \autoref{padeborelsol}, that the basis of solutions 
$\psi^{(1)}$, $\psi^{(2)}$ is well defined up to constant multiple
in each domain, and jumps
(due to the presence of singularities in the Borel plane)
when we cross a Stokes curve from one domain to another.

In some cases a generic WKB curve goes to the 
\ti{same} asymptotic Stokes direction at both ends. This might
seem like a contradiction: there is only one exponentially decaying
solution at that Stokes direction, so how will we get a basis
of solutions from exponentially decaying solutions along this
WKB curve? The resolution is that we need to be careful about the global
monodromy: when we speak of the two-dimensional 
space of global solutions of the equation,
we really mean the space of solutions on the complement of some branch
cut (``monodromy cut'') in the $z$-plane. When we transport a solution
across this cut we transform it by the action of the monodromy $M$ or
$M^{-1}$. We will see examples below.

Inside ring domains, the Stokes curves do not go to any irregular singularity. Thus we cannot find a
basis of solutions there using our exponential-decay prescription. Instead, the appropriate basis consists of the eigenfunctions of the monodromy around the ring, as described e.g. in \cite{Hollands:2013qza}.

As we have just reviewed, in each domain of  $C \setminus \cW$ we have a distinguished
pair of solutions determined up to scalar multiple.
We will call the exponentially decaying solutions approaching singularities $\psi_1$, $\psi_2$ and $\psi_3$ as indicated in \eqref{exSD1}, \eqref{exSD2}, \eqref{exSD3}.
Then, see \autoref{QPFGf}, \autoref{resolst}, \autoref{QPFN}, \autoref{quantizationcd} 
for some examples of
explicit local bases.

Some convenient rules for writing down the local bases of solutions in all domains are \cite{Hollands:2019wbr}:

\begin{enumerate}
\item We use the notation $(\psi,\psi')$ to represent a local basis of solutions 
$(\psi^{(1)}, \psi^{(2)})$ labeled by the two sheets of the covering $\Sigma \to C$.
$\psi$ and $\psi'$ are swapped when crossing a branch cut:
\begin{equation}
(\psi,\psi^\prime)\rightarrow(\psi^\prime,\psi)
\end{equation}

\item We choose a monodromy cut for the differential equation as discussed above, 
and when we cross this cut, the basis gains a factor of the monodromy matrix $M$. 
\begin{equation}
(\psi,\psi^\prime)\rightarrow(M\psi,M\psi^\prime)
\end{equation}

\item When we cross a Stokes curve of type $ij$, the solution $\psi^{(j)}$ is unchanged,
while $\psi^{(i)}$ generally changes.

(This can be understood as: shifting the exponentially growing solution by a multiple of the exponentially decaying solution
gives another exponentially growing solution. 
But we will not use this point of view explicitly, and will simply treat $\psi^{(i)}$ on the two sides of the wall 
as different solutions.)

\item  When we cross a double wall, both $\psi^{(1)}$ and $\psi^{(2)}$ generally change.

\item In a ring domain, the local basis of solutions is not constructed directly
from the exponentially decaying solutions; rather they are eigenfunctions of the monodromy matrix $M$.
(Thus, strictly speaking, they are not single-valued solutions on the ring domain, but rather
on the complement of a monodromy cut.)
\end{enumerate}
In terms of these local bases, the quantum period associated to a cycle $\gamma$ 
is obtained as a product of factors as follows. Let $[\psi,\psi']$ denote the Wronskian of the 
two solutions $\psi$ and $\psi'$.
\begin{enumerate}
\item When $\gamma$ crosses a single Stokes curve of type $ij$ on sheet $i$, from side $L$ to $R$, add a factor $\frac{[\psi_i^L,\psi_j^L]}{[\psi_i^R,\psi_j^L]}$ to the product. 

\item When $\gamma$ crosses a double Stokes curve on sheet $i$, add a factor $\sqrt{\frac{[\psi_i^L,\psi_j^L]}{[\psi_i^R,\psi_j^R]}\frac{[\psi_i^L,\psi_j^R]}{[\psi_i^R,\psi_j^L]}}$ to the product.
(This factor is ambiguous as written, because we have not specified a branch of the square
root. This sign can in principle be determined following a rule given in \cite{Hollands:2019wbr}, but 
this rule is cumbersome in practice; in this paper we will not try to implement it,
and just live with this sign ambiguity in cases involving double Stokes curves. All 
formulas in this paper which involve a square root of a product of Wronskians have to be understood as
having this sign ambiguity; in numerical comparisons we just fix the sign by hand
when necessary.)

\item When $\gamma$ crosses a monodromy cut in a ring domain, add a factor $\mu$ or $\frac{1}{\mu}$ to the product, where $\mu$, $\frac{1}{\mu}$ denote the eigenvalues of the monodromy operator $M$.

\item We include an overall sign factor $\exp(-\frac14 \oint_{\gamma} \frac{\de P}{P})$,
where $P(x)$ is the potential in the cylinder coordinate, given in \eqref{P-cylinder}.
(This factor was not included in the rules in \cite{Hollands:2019wbr}; 
we saw it 
in \autoref{sec:one-loop-sign}, and
we discuss it more in \autoref{ssabeli}.)

\end{enumerate}

Combining all factors as described above when going along some loop $\gamma$, we obtain the Wronskian expression $\cX_\gamma^{\rm SS,\vartheta}$. The result only depends on the homology 
class $\gamma$, although the individual factors may well depend on the precise choice
of representative.

\subsection{Quantum periods in strong coupling region}
\label{QPFG}
\begin{figure}[htb]
  \begin{centering}
  \includegraphics[width=0.5\linewidth]{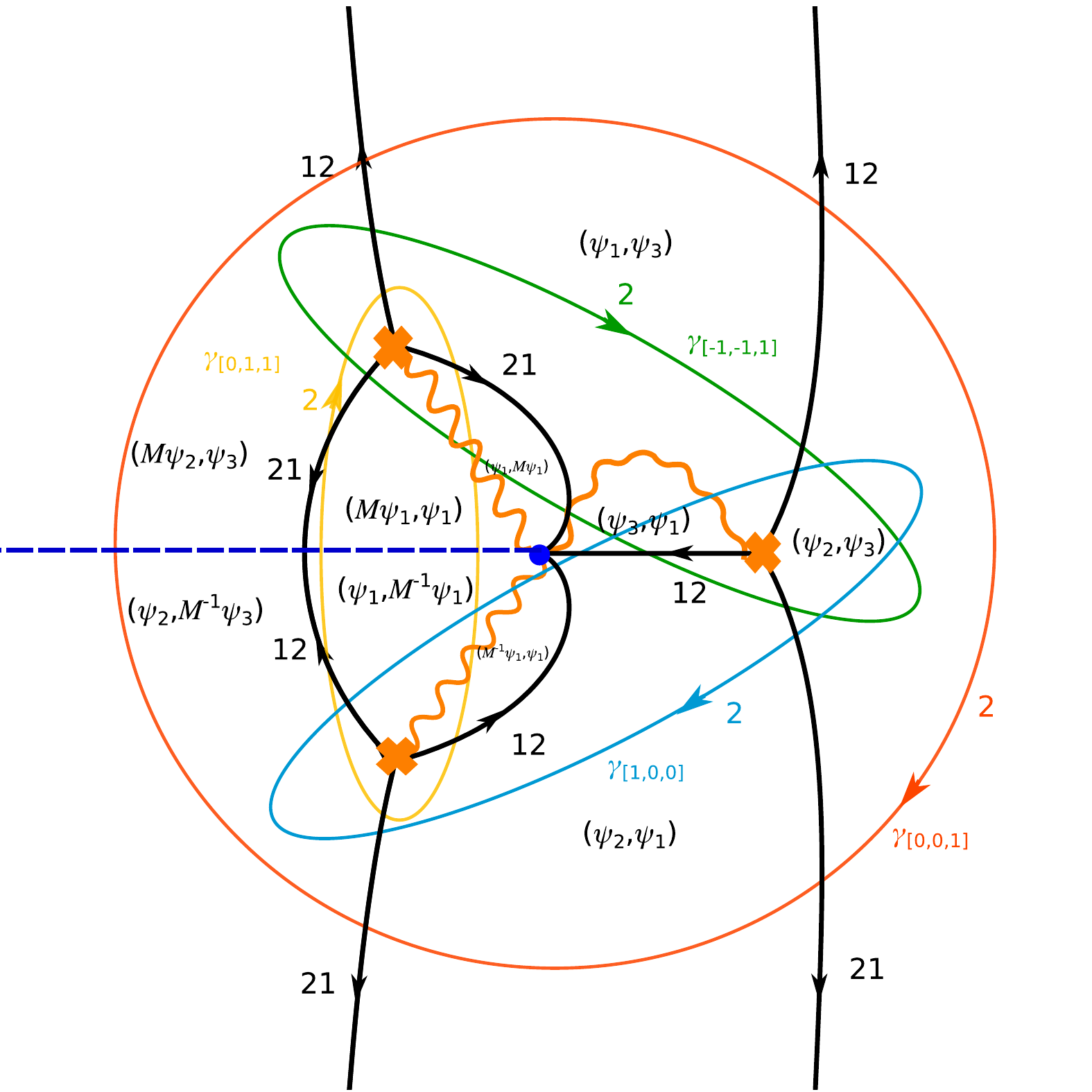}
  \caption{$\mathcal{W}^{\vartheta=0}$ for $u=0,\ m=-1/10,\ \Lambda=1$ in the strong coupling region. Blue dotted line is the monodromy cut. The local basis of solutions for each domain is listed as $(\psi^{(1)},\psi^{(2)})$.}
   \label{QPFGf}
 \end{centering}
\end{figure}

As an example we can choose the parameters to be $u=0$, $m=-1/10$ and $\Lambda=1$. The corresponding spectral network $\mathcal{W}^{\vartheta=0}$ is shown in \autoref{QPFGf}. 
By applying the rules given in \autoref{rulesw} we obtain the following Wronskian expressions for the quantum periods:
\begin{equation}\label{s0}
\cX^{\rm SS,\vartheta=0}_{\gamma_{[0,1,1]}}=-\frac{[\psi_3,\psi_1]}{[M\psi_1,\psi_1]}\frac{[\psi_1,\psi_2]}{[M^{-1}\psi_3,\psi_2]} \, ,
\end{equation}
\begin{equation}\label{s1}
\mathcal{X}^{\rm SS,\vartheta=0}_{\gamma_{[-1,-1,1]}}=\frac{[\psi_2,\psi_3]}{[\psi_2,\psi_1]}\sqrt{\frac{[M\psi_1,\psi_1]}{[\psi_2,M^{-1}\psi_3]}\frac{[M\psi_2,\psi_1]}{[M\psi_1,\psi_3]}}\, ,
\end{equation}
\begin{equation}\label{s2}
\mathcal{X}^{\rm SS,\vartheta=0}_{\gamma_{[1,0,0]}}=\frac{[\psi_3,\psi_2]}{[\psi_3,\psi_1]}\sqrt{\frac{[\psi_1,M\psi_1]}{[M^{-1}\psi_3,\psi_2]}\frac{[\psi_3,M\psi_1]}{[\psi_1,M\psi_2]}}\, .
\end{equation}
For completeness we also show the Wronskian expression for the  period corresponding to flavor mass:
\be
\cX^{{\rm SS},\vartheta=0}_{\gamma_{[0,0,1]}}=-\frac{[\psi_3,\psi_2]}{[M^{-1}\psi_3,\psi_2]}=\re^{\frac{-2\pi m}{\hbar}}\, .
\ee
 \begin{figure}[htbp]
\begin{center}
\includegraphics[width=0.5\linewidth]{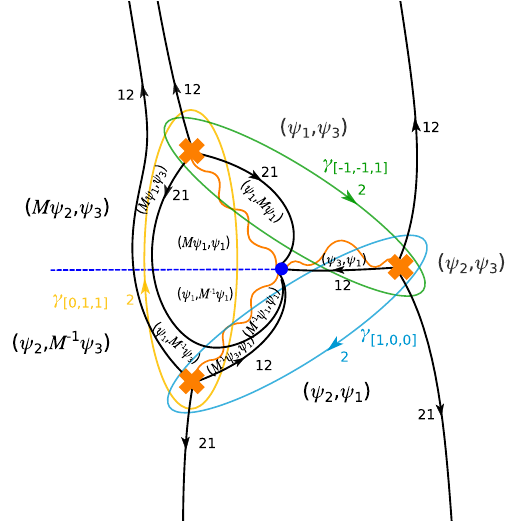}
\caption{Spectral network and local bases of solutions corresponding to $\vartheta=\arg(1+\frac{1}{10}\ri)$, with
other parameters as in \autoref{QPFGf}.}
\label{resolst}
\end{center}
\end{figure}

Alternatively we can consider
 $\mathcal{W}^{\vartheta=\arg(1+\frac{\ri}{10})}$ as shown in \autoref{resolst}. The resulting expressions are
\be\cX^{{\rm SS},\vartheta=\arg(1+\frac{\ri}{10})}_{\gamma_{[0,1,1]}}=-\frac{[\psi_3,\psi_1]}{[M\psi_1,\psi_1]}\frac{[\psi_1,\psi_2]}{[M^{-1}\psi_3,\psi_2]}\, ,\ee
\be\cX^{{\rm SS},\vartheta=\arg(1+\frac{\ri}{10})}_{\gamma_{[-1,-1,1]}}=-\frac{[\psi_1,M\psi_1]}{[\psi_3,M\psi_1]}\frac{[\psi_3,\psi_2]}{[\psi_1,\psi_2]}\, ,\ee
\be\cX^{{\rm SS},\vartheta=\arg(1+\frac{\ri}{10})}_{\gamma_{[1,0,0]}}=-\frac{[\psi_3,\psi_2]}{[M^{-1}\psi_3,\psi_2]}\frac{[M^{-1}\psi_3,\psi_1]}{[\psi_3,\psi_1]}\, .\ee
These periods are related by\footnote{We emphasize again that we have not fixed
the branch of the square root.}
\be\label{exKSss}\ba 
\frac{\cX^{{\rm SS},\vartheta=\arg(1+\frac{\ri}{10})}_{\gamma_{[-1,-1,1]}}}{\mathcal{X}^{\rm SS,\vartheta=0}_{\gamma_{[-1,-1,1]}}}={}&\sqrt{\frac{[M\psi_1,\psi_1]}{[M\psi_1,\psi_3]}\frac{[M\psi_2,\psi_3]}{[M\psi_2,\psi_1]}}\\
={}&\frac{1}{\sqrt{1-\frac{[\psi_1,\psi_3][\psi_2,\psi_1]}{[\psi_1,M\psi_1][\psi_2,M^{-1}\psi_3]}}}\\
={}&(1+\cX^{{\rm SS},\vartheta=\arg(1+\frac{\ri}{10})}_{\gamma_{[0,1,1]}})^{-\frac{1}{2}}=(1+\cX^{{\rm SS},\vartheta=\arg(0)}_{\gamma_{[0,1,1]}})^{-\frac{1}{2}}.
\ea \ee
This relation is precisely the KS transformation \eqref{eq:pi-transform}-\eqref{eq:pi-transform3}
corresponding to the hypermultiplet of charge $\gamma_{[0,1,1]}$ in the spectrum.

We evaluated these Wronskians numerically; some results are shown in \autoref{tab:introresultsnew}.

\subsection{Quantum periods in weak coupling region}
\label{QPWWK}

\begin{figure}
     \centering
     \begin{subfigure}[b]{0.45\textwidth}
         \centering
         \includegraphics[width=\textwidth]{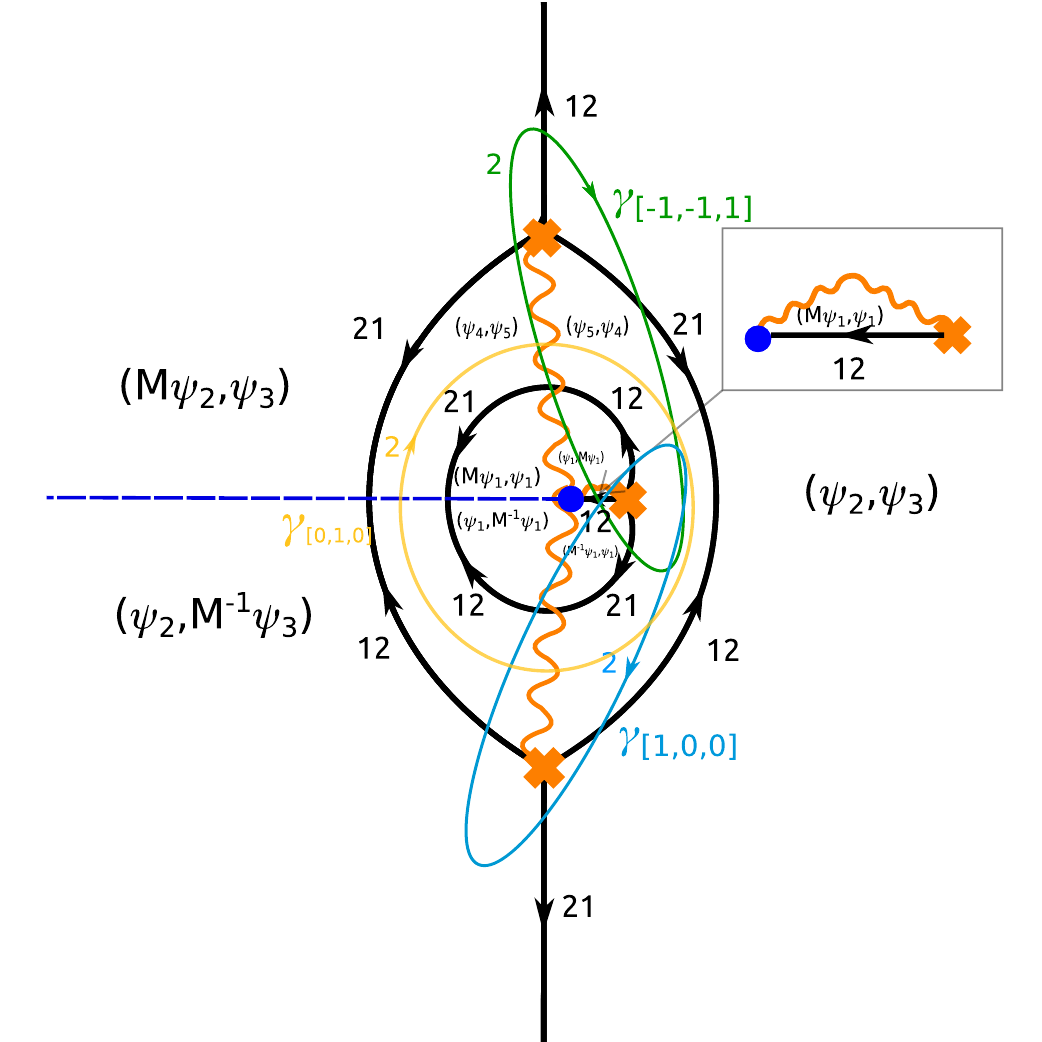}
         \caption{}
         \label{QPFNf}
     \end{subfigure}
          \begin{subfigure}[b]{0.45\textwidth}
         \centering
         \includegraphics[width=\textwidth]{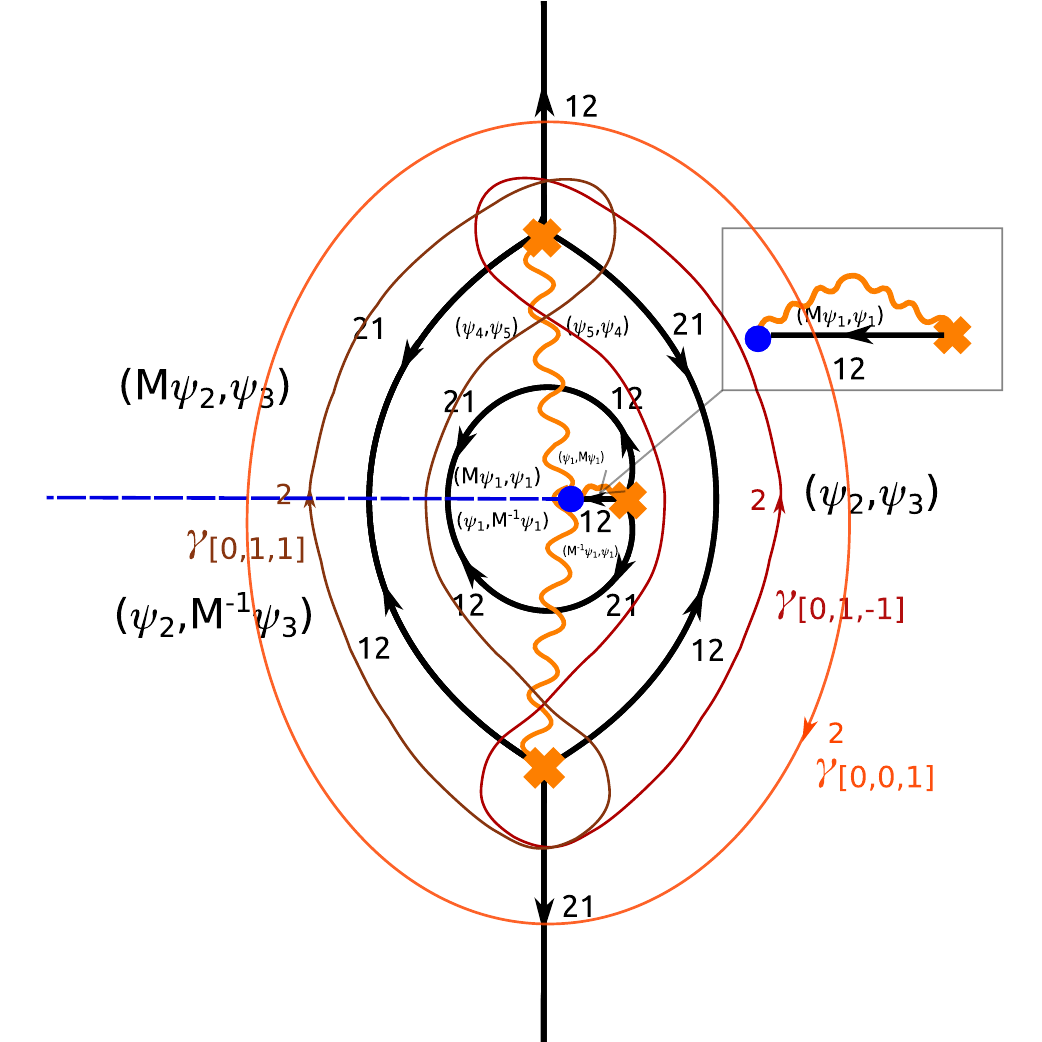}
         \caption{}
         \label{QPFN2hyper}
     \end{subfigure}
            \caption{The spectral network $\mathcal{W}^{\vartheta=0}$ for $u=2$, $m=0$, $\Lambda=1/5$ in the weak coupling region. The blue dotted line is the monodromy cut.  The local basis of solutions in each domain is 
            listed as $(\psi^{(1)},\psi^{(2)})$. The two figures are identical
            except that we show different bases for the charge lattice.}
        \label{QPFN}
\end{figure}

Now let us consider the point $u=2,\ m=0,\ \Lambda=1/5$ in the weak coupling region.
We use the spectral network  $\mathcal{W}^{\vartheta=0}$ shown in \autoref{QPFN}. Again using the 
rules of \autoref{rulesw}, we obtain Wronskian expressions for the quantum periods:
\begin{equation}\label{w1}
\cX^{\rm SS,\vartheta=0}_{\gamma_{[-1,-1,1]}}=-\sqrt{-\frac{1}{\mu}}\frac{[\psi_1,\psi_4]}{[\psi_1,\psi_5]}\sqrt{\frac{[\psi_3,\psi_2][\psi_3,\psi_5][\psi_5,\psi_2]}{[\psi_4,\psi_2][\psi_3,M\psi_2][\psi_3,\psi_4]}}\, ,
\end{equation}
\begin{equation}\label{w2}
\cX^{\rm SS,\vartheta=0}_{\gamma_{[1,0,0]}}=\sqrt{-\frac{1}{\mu}}\frac{[\psi_1,\psi_5]}{[\psi_1,\psi_4]}\sqrt{\frac{[\psi_3,\psi_2][\psi_4,\psi_2][\psi_3,\psi_4]}{[\psi_3,M\psi_2][\psi_3,\psi_5][\psi_5,\psi_2]}}\, ,
\end{equation}
\begin{equation}\label{w0}
\cX^{\rm SS,\vartheta=0}_{\gamma_{[0,1,0]}}=\mu \, ,
\end{equation}
where $\mu$ is the eigenvalue of the monodromy matrix $M$ chosen by comparing with the same quantum period gotten in other method. $\psi_4$ and $\psi_5$ are eigenfunctions of $M$ such that $M\psi_4=\frac{1}{\mu}\psi_4$ and $M\psi_5=\mu\psi_5$. 
For completeness, we also show
\be
\cX^{{\rm SS},\vartheta=0}_{\gamma_{[0,0,1]}}=-\frac{[\psi_3,\psi_2]}{[M^{-1}\psi_3,\psi_2]}\, .\ee
For later convenience we also show the $\cX_\gamma$ for a few other charges:
\be
\cX^{{\rm SS},\vartheta=0}_{\gamma_{[0,1,-1]}}=-\mu\frac{[M^{-1}\psi_3,\psi_2]}{[\psi_3,\psi_2]}=\mu\frac{1}{\cX^{{\rm SS},\vartheta=0}_{\gamma_{[0,0,1]}}}=\mu\re^{\frac{2\pi m}{\hbar}}\, ,
\ee
\be
\cX^{{\rm SS},\vartheta=0}_{\gamma_{[0,1,1]}}=-\mu\frac{[\psi_3,\psi_2]}{[M^{-1}\psi_3,\psi_2]}=\mu\cX^{{\rm SS},\vartheta=0}_{\gamma_{[0,0,1]}}=\mu\re^{-\frac{2\pi m}{\hbar}}\, .
\ee
Notice that if $m=0$, 
\be\cX^{{\rm SS},\vartheta=0}_{\gamma_{[0,1,-1]}}=\cX^{{\rm SS},\vartheta=0}_{\gamma_{[0,1,1]}}=\cX^{{\rm SS},\vartheta=0}_{\gamma_{[0,1,0]}}=\mu.\ee 

\subsection{Quantization condition}
\label{section:qcd}
In this section we study the quantization condition for
 the operator \eqref{eq:our-eq-intro}, in the particular situation where the potential is real, convex and confining along some path in the $z$-plane.  In this case the bound states correspond to solutions decaying exponentially when approaching both ends of the path. Such a solution exists when the solutions decaying on Stokes directions at the two ends of the path are linearly dependent. 
We will see below that the
resulting quantization condition for the energy spectrum takes the simple form
\begin{equation}
\mathcal{X}_{\gamma_{[1,0,0]}}=-1.	
\end{equation}
 
\begin{figure}
\begin{centering}
\includegraphics[width=0.5\linewidth]{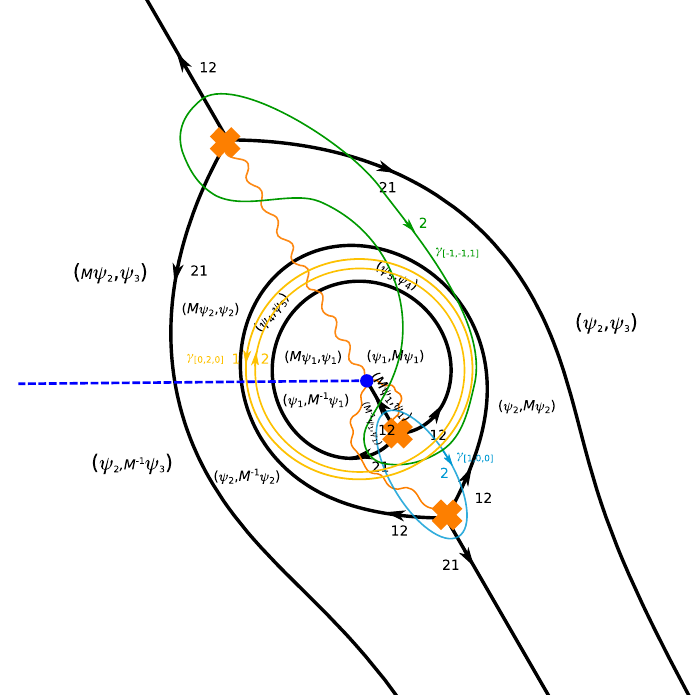}
\caption{$\mathcal{W}^{\vartheta=0}$ at $u=2,\ m=\ri/10,\ \Lambda=\re^{-\ri\pi/6}$. Decaying solutions are along $\mathbb{R}\re^{-\pi \ri/3}$ approaching $\re^{-\pi\ri/3}\infty$, $\re^{-\pi\ri/3} 0$ and $\re^{2\pi\ri/3}\infty$.}
\label{quantizationcd}
\end{centering}
\end{figure}

 As an example we take 
 \be\label{exq} \hbar > 0, \quad m=\frac{\ri}{10}, \quad   \Lambda=\re^{-\frac{\pi \ri}{6}}\,. \ee  
 The corresponding spectral network and local solutions for $\mathcal{W}^{\vartheta=0}$ at $u=2$ are shown in \autoref{quantizationcd}. More generally, keeping $\Lambda$ and $m$ fixed, we allow $u$ to vary in a range on $\mathbb{R}$ containing $u=2$ so long as the spectral network maintains the same topology. 
The locus we are considering is in the weak coupling region and at the special phase $\vartheta=0$, which is the 
phase of the central charge of the BPS vectormultiplet; thus we see a ring domain in \autoref{quantizationcd}.
 
 With the parameters \eqref{exq}, the Schr\"odinger equation \eqref{eq:our-eq-intro} becomes
\begin{equation} \label{eq:schrodinger-specialized}
\left(-\hbar^2\partial_x^2+(\frac{\re^{-\frac{\pi \ri}{3}}\re^{-x}}{2}+\frac{\re^{\frac{\pi \ri}{3}}\re^x}{5}-\re^{-\frac{\pi \ri}{3}}\re^{2x})-2u\right)\psi(x)=0.
\end{equation}
If we choose a path parametrized  by
\be
x=t-\frac{\pi \ri}{3}\quad t\in \mathbb{R}\, ,
\ee
 then \eqref{eq:schrodinger-specialized} becomes
\begin{equation}\label{conf}
\left(-\hbar^2\partial_t^2+(\frac{\re^{-t}}{2}+\frac{\re^t}{5}+\re^{2t})-2u\right)\psi(t)=0,
\end{equation}
with real convex confining potential along $t\in\mathbb{R}$. Bound states are solutions $\psi(t)$ decaying as $t\rightarrow\pm\infty$. 
The corresponding path in the $z$ plane is 
\be
z=\re^x=\re^{-\frac{\pi \ri}{3}}\re^{t}.
\ee
Hence a bound state corresponds to a solution in the $z$-plane which decays exponentially as we approach $z=\re^{-\frac{\pi \ri}{3}}\re^{\infty}=\re^{-\frac{\pi \ri}{3}}\infty$ and $z=\re^{\frac{-\pi \ri}{3}}\re^{-\infty}=\re^{\frac{-\pi \ri}{3}}0$. 
In the labeling shown in \autoref{quantizationcd}, the decaying solutions in these two directions are named $\psi_2$ and $\psi_1$ respectively; thus the bound state condition is that $\psi_1$ and $\psi_2$ are linearly dependent.

The Wronskian expressions for the quantum periods in this parameter chamber,
again obtained by the rules of \autoref{rulesw}, are
\be\label{X010}
\cX^{\rm SS,\vartheta=0}_{\gamma_{[0,1,0]}}( E,\Lambda=\re^{-\ri \pi/6}, m={\frac{\ri}{10}}, \hbar)=\mu\, ,\ee
\be\label{X100}
\cX^{\rm SS,\vartheta=0}_{\gamma_{[1,0,0]}}( E,\Lambda=\re^{-\ri \pi/6}, m={\frac{\ri}{10}}, \hbar)=\frac{[\psi_1,\psi_5][\psi_4,\psi_2]}{[\psi_4,\psi_1][\psi_5,\psi_2]}\, .
\ee
Bound states exist when $\psi_1=\lambda\psi_2$. Substituting this condition in \eqref{X100} we see that it implies
\be
\label{qceqn}
\cX^{\rm SS,\vartheta=0}_{\gamma_{[1,0,0]}}( E,\Lambda=\re^{-\ri \pi/6}, m={\frac{\ri}{10}}, \hbar) = -1\, .
\ee
This is the exact quantization condition.
It has a discrete set of solutions $\{E_n\}_{n\geq 0}$, which give the bound state energies for \eqref{conf}.

For completeness, we also show another quantum period at this point (although this one does not participate in
the quantization condition):
\be
\ba
\cX^{\rm SS,\vartheta=0}_{\gamma_{[-1,-1,1]}}( E,\Lambda=\re^{-\ri \pi/6}, m={\frac{\ri}{10}}, \hbar)&=-\frac{[\psi_4,\psi_1]}{[\psi_1,\psi_5]}\frac{[\psi_5,M\psi_2]}{[\psi_4,\psi_2]}\frac{[\psi_3,\psi_2]}{[\psi_3,M\psi_2]}\, ,\\
&=-\frac{1}{\mu}\frac{[\psi_4,\psi_1]}{[\psi_1,\psi_5]}\frac{[\psi_5,\psi_2]}{[\psi_4,\psi_2]}\frac{[\psi_3,\psi_2]}{[\psi_3,M\psi_2]}\, .
\ea
\ee

\section{TBAs in the strong coupling region}

\subsection{The GMN TBA}
 \label{sec:GMNTBA}

In this section we briefly review the TBA of \cite{Gaiotto:2014bza} (conformal limit of 
the GMN TBA \cite{gmn,Gaiotto:2009hg}), 
focusing on the example of the $SU(2)$ $N_f=1$ theory.  As we will see, this system
of integral equations provides us with another way to compute quantum periods.

We consider spectral coordinates
$\mathcal{X}_\gamma^{\rm TBA}$ obeying
\begin{equation}\label{gmn}
\mathcal{X}_\gamma^{\rm TBA}(\hbar)=\exp\left(\frac{Z_\gamma}{\hbar}+\frac{1}{4\pi \ri}\sum_{\gamma'\in \Gamma }\Omega (\gamma', u)\langle\gamma,\gamma' \rangle \mathcal{I}_{\gamma'}(\hbar)\right),
\end{equation}
where $\Omega (\gamma', u)$ denotes the BPS index of BPS states with charge $\gamma'$ and
\be \label{idef} \mathcal{I}_{\gamma}(\hbar)=\int_{r_\gamma}\frac{d\hbar^\prime}{\hbar^\prime}
\frac{\hbar^\prime+\hbar}{\hbar^\prime-\hbar}\log\left(1-\sigma_{\rm can}(\gamma)\cX_{\gamma}^{\rm TBA}(\hbar')\right)\, , \ee
where \be r_\gamma=\left\{\hbar': {\frac{Z_\gamma}{\hbar'}}\in \IR_-\right\}.\ee

The reason for considering the equation \eqref{gmn} is that it guarantees that the $\cX_\gamma^{\rm TBA}$ will have the expected 
 $\hbar \to 0$ asymptotics $\cX_\gamma \sim \exp(Z_\gamma / \hbar)$, 
 and also the expected jumps as the phase of $\hbar$ is varied.

Since we always have $\Omega(\gamma) = \Omega(-\gamma)$, we can use \cite{Gaiotto:2014bza}
\be 
\label{chargesymmetry}
\cX_\gamma^{\rm TBA}(\hbar)=\cX^{\rm TBA}_{-\gamma}(-\hbar)\, ,
\ee
to write \eqref{gmn} as\footnote{By summing over $\gamma^\prime>0$, we mean that we take either $+\gamma$ or $-\gamma$ but not both.}
\begin{equation}\label{gmn2}
\mathcal{X}_\gamma^{\rm TBA}(\hbar)=\exp\left(\frac{Z_\gamma}{\hbar}+\frac{1}{4\pi \ri}\sum_{\gamma' >0}\Omega (\gamma', u)\langle\gamma,\gamma' \rangle \mathcal{C}_{\gamma'}(\hbar)\right),
\end{equation}
where
\be \label{cdef}\mathcal{C}_{\gamma}(\hbar)=\mathcal{I}_{\gamma}(\hbar)-\mathcal{I}_{-\gamma}(\hbar)= 4 \hbar\int_{r_{\gamma}}
\frac{d\hbar'}{(\hbar^\prime)^2-\hbar^2}\log\left(1-\sigma_{\text{can}}(\gamma)\cX_{\gamma}^{\rm TBA}(\hbar')\right)\, . \ee
Let us denote \be Z_\gamma=|Z_\gamma|\E^{\ri\phi_\gamma}.\ee
We perform a change of variables in \eqref{cdef},
\be \hbar=-\re^{\ri \phi-\theta}, \quad \hbar'=-\re^{\ri \phi_{\gamma}-\theta'}, \ee
and write
\be  \mathcal{C}_{\gamma}(\hbar)= \tilde{\mathcal{C}}_{\gamma}(\theta)=2\int_{\IR}\rd \theta'
 {\log\left(1-\sigma(\gamma)\cX_{\gamma}^{\rm TBA}(-\re^{\ri \phi_{\gamma}-\theta'})\right) \over \sinh(\theta-\theta'+i\phi_{\gamma}-i\phi)}\, .\ee
In the weak coupling region, because of the infinite tower of BPS states, \eqref{gmn} leads to an infinite tower of coupled TBA equations which are hard to solve.
Therefore, in this section we will simply focus on the solution to \eqref{gmn} in the strong coupling region,
where the BPS spectrum consists of hypermultiplets with charges
\be\label{strongBPS} \pm\gamma_{[-1,-1,1]}, \quad \pm\gamma_{[1,0,0]}, \quad\pm\gamma_{[0,1,1]}. \ee
All of these charges have $\sigma_{\rm can}(\gamma)=-1$.
We define
$\tilde{\epsilon}_\gamma$ as \begin{equation}
\mathcal{X}_\gamma^{\rm TBA}(-\re^{\ri\phi_\gamma-\theta})=\exp(-\tilde{\epsilon}_\gamma(\theta)).
\end{equation}
Using this definition we rewrite \eqref{gmn} as 
\begin{equation} 
\tilde {\epsilon}_\gamma(\theta)=|Z_\gamma|\re^{\theta}{- } \frac{1}{2\pi \ri}\sum_{\gamma' >0}\langle \gamma,\gamma'\rangle\Omega(\gamma',u)\int_{\mathbb{R}}\frac{\log(1+\re^{-\tilde{\epsilon}_{\gamma'}(\theta')})}{\sinh(\theta-\theta'+\ri \phi_{\gamma'}-\ri \phi_\gamma)} \, \de \theta'.
\end{equation}
We have
\be \Pi_\gamma^{\rm TBA}(\hbar)={-\tilde{\epsilon}_\gamma(\theta)}, \quad \hbar=-\re^{\ri \phi_\gamma- \theta}.\ee
If the TBA has singularities along a given $\hbar$ direction, then we simply denote
\be  \Pi_\gamma^{\rm TBA}(\hbar)={1\over 2}\left(\Pi_\gamma^{\rm TBA}(\hbar^+) +\Pi_\gamma^{\rm TBA}(\hbar^-)\right)\, , \ee   
where $\hbar^+$ stands for $\hbar\re^{\ri 0^+}$ and $\hbar^-$ stands for $\hbar\re^{\ri 0^-}$.
Following \cite{Gaiotto:2014bza,Hollands:2016kgm,Grassi:2019coc}, we expect that
 \be \label{tbaborel}\Pi_\gamma(\hbar)=\Pi_\gamma^{\rm TBA}(\hbar)\,, \ee
 where we are using the definition \eqref{wkbqm}.

We solved the TBA equations numerically to test this hypothesis; the results are in \autoref{tab:introresultsnew}. Notice that the convergence of the TBA is a bit slow, it may be possible to obtain a few more digits  by implementing explicitly some boundary condition at ${\hbar\over \Lambda} \to \infty$ similar to \cite{Grassi:2019coc}.

\subsection{A special point}

\begin{figure}
\begin{centering}
\includegraphics[width=0.3\linewidth]{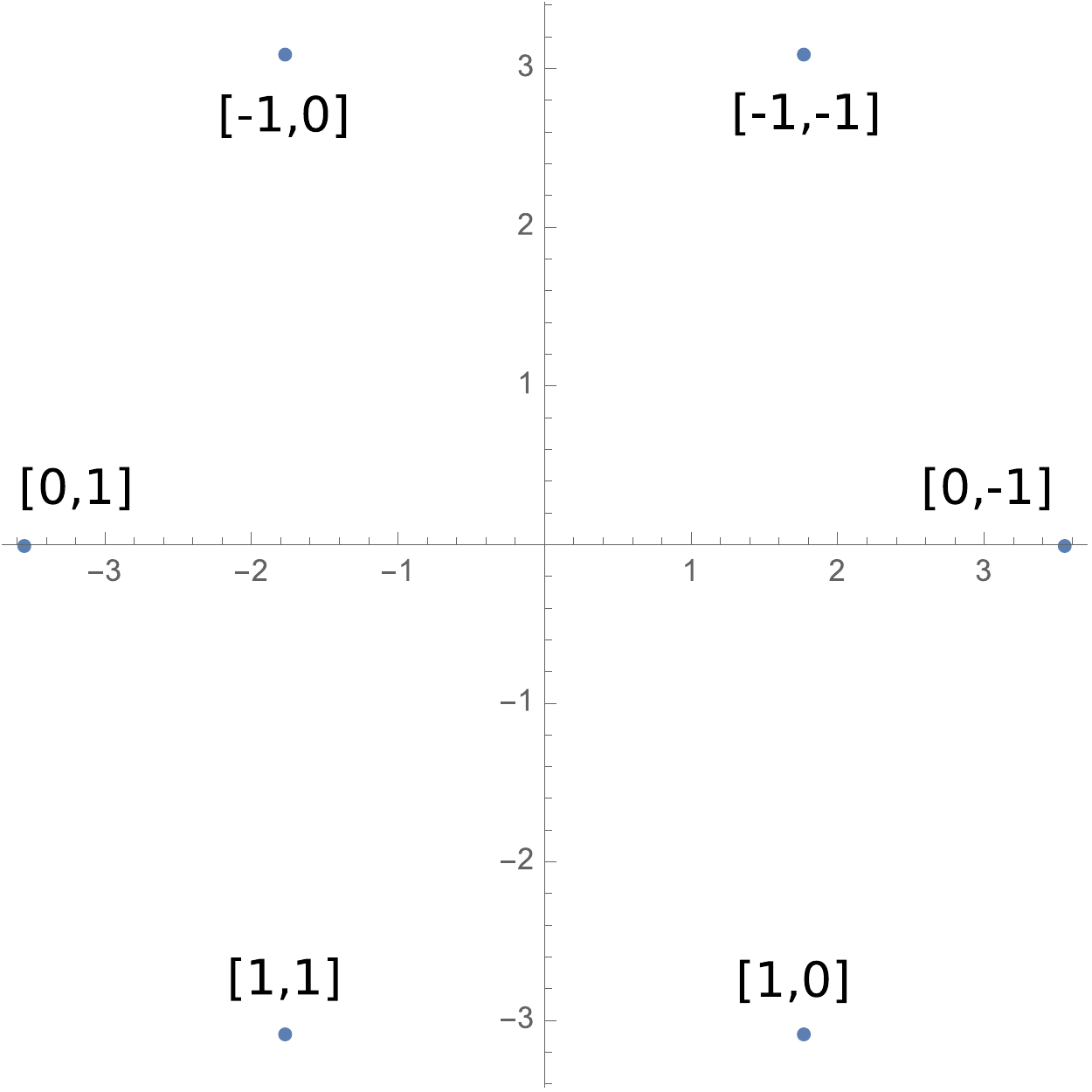}
\caption{Central charges of BPS particles in the theory at $m=0$, $u=0$, $\Lambda=1$.
}
\label{ZoTBA}
\end{centering}
\end{figure}

We now move to another interesting point.
It has been noted that the GMN TBA equations often simplify at particularly symmetric points
of the Coulomb branch. In particular, in the pure $SU(2)$ theory at $u=0$, 
the two GMN TBA equations collapse to a single equation \cite{Gaiotto:2014bza,Grassi:2019coc}. 
Moreover, it was observed in \cite{Grassi:2019coc} that this single equation coincides 
with the one used by Zamolodchikov in \cite{post-zamo} to compute the Fredholm determinant of the 
(modified) Mathieu operator; see also \cite{Fioravanti:2019vxi} for related work.
We will see that a similar phenomenon happens in the $SU(2)$ $N_f=1$ theory at the point
$m = 0$, $u = 0$. 
This point is special because the distribution of central charges of BPS states 
has $\mathbb{Z}_6$ symmetry: see \autoref{ZoTBA}.

Since the flavor charge does not play a role in the TBA in the massless case, 
in the following we will omit this third component and use the notation $\gamma_{[a,b]}$ for the charges.
The BPS spectrum then consists of 3 hypermultiplets with central charges\footnote{For simplicity, we consider $\Lambda=1$.} 
\begin{equation}\ba
Z_{\gamma_{[1,0]}}=&1.77829 - 3.08009 \ri \, ,
\\
Z_{\gamma_{[1,1]}}=&-1.77829 - 3.08009 \ri \, ,
\\
Z_{\gamma_{[0,1]}}=&-3.55658 \, .
\ea
\end{equation}
All the central charges above have the same absolute value
\be \label{Zdef}|Z|= \frac{2^{2/3} \pi ^{3/2}}{\Gamma \left(\frac{1}{3}\right) \Gamma \left(\frac{7}{6}\right)}\ee 
while their arguments are 
\be \phi_{\gamma_{[1,0]}}=-{\pi\over 3}, \quad \phi_{\gamma_{[1,1]}}=-{2\pi\over 3},   \quad \phi_{\gamma{[0,1]}}=\pi. \ee
Therefore the 3 TBAs  at $u=0=m$ read \begin{equation}\ba
\tilde{\epsilon}_{\gamma_{[1,0]}}(\theta)=&|Z|\re^{\theta}{- }\frac{1}{2\pi \ri}\left(\int_{\mathbb{R}}\frac{\log(1+\re^{-\tilde{\epsilon}_{\gamma_{[0,1]}}(\theta')})}{\sinh(\theta-\theta'-\ri\frac{2\pi}{3})}\rd\theta'+\int_{\mathbb{R}}\frac{\log(1+\re^{-\tilde{\epsilon}_{\gamma_{[1,1]}}(\theta')})}{\sinh(\theta-\theta'-\ri\frac{\pi}{3})}\rd\theta'\right),
\\
\\
\tilde{\epsilon}_{\gamma_{[0,1]}}(\theta)=&|Z|\re^{\theta}{-}\frac{1}{2\pi \ri}\left(\int_{\mathbb{R}}\frac{\log(1+\re^{-\tilde{\epsilon}_{\gamma_{[1,0]}}(\theta')})}{\sinh(\theta-\theta'-\ri\frac{\pi}{3})}\rd\theta'+\int_{\mathbb{R}}\frac{\log(1+\re^{-\tilde{\epsilon}_{\gamma_{[1,1]}}(\theta')})}{\sinh(\theta-\theta'-\ri\frac{2\pi}{3})}\rd\theta'\right),\\
\\
\tilde{\epsilon}_{\gamma_{[1,1]}}(\theta)=&
 |Z|\re^{\theta}{- }\frac{1}{2\pi \ri}\left(\int_{\mathbb{R}}\frac{\log(1+\re^{-\tilde{\epsilon}_{\gamma_{[1,0]}}(\theta')})}{\sinh(\theta-\theta'-\ri\frac{2\pi}{3})}\rd\theta'+\int_{\mathbb{R}}\frac{\log(1+\re^{-\tilde{\epsilon}_{\gamma_{[0,1]}}(\theta')})}{\sinh(\theta-\theta'-\ri\frac{\pi}{3})}\rd\theta'\right).
\ea
\end{equation}
If we make the ansatz  \be \tilde{\epsilon}_{\gamma_{[1,0]}}(\theta)=\tilde{\epsilon}_{\gamma_{[0,1]}}(\theta)=\tilde{\epsilon}_{\gamma_{[1,1]}}(\theta)\equiv\epsilon(\theta)\ee
these 3 equations collapse to a single one,
\begin{equation}
\epsilon(\theta)= |Z| \re^{\theta}-\frac{1}{2\pi}\left(\int_{\mathbb{R}}\frac{\log(1+\re^{-\epsilon(\theta')})}{\cosh(\theta-\theta'-\ri\frac{\pi}{6})}\rd\theta'+\int_{\mathbb{R}}\frac{\log(1+\re^{-\epsilon(\theta')})}{\cosh(\theta-\theta'+\ri\frac{\pi}{6})}\rd\theta'\right).
\label{ZTBAeqnc}
\end{equation}
This equation has appeared before in the work of Zamolodchikov; see \cite[eq.~(4.1)]{post-zamo} for $b=1/\sqrt{2}$,
where it is used to compute Fredholm determinants. We discuss this in more detail in \autoref{Zamolodchikov} below.

\section{Computation by instanton counting}
\label{sec:QPI}
Another interesting way of  computing the quantum periods is by using  the  NS limit of the instanton counting partition function \cite{Braverman:2004cr,ns,nrs,mirmor, Zenkevich:2011zx}. As compared to the other methods presented above, the instanton counting has the advantage of providing analytic expressions in $(E,\hbar,m, \Lambda)$.
Nevertheless, the set of spectral problems which we can currently treat within this method is more limited. For example,
we do not know how to approach spectral problems associated to non-Lagrangian theories such as the ones in \cite{ddt,Ito:2018eon,Gaiotto:2014bza}.\footnote{The all-order WKB expansion can be computed within the gauge theory/topological string framework: see, for instance, \cite{cm-ha}. However, at present, we do not know how to perform the analytic resummation of this WKB expansion into an instanton-counting-like expression. Some progress in this direction was also made recently in \cite{Lisovyy:2021bkm}. }

In this section we use the instanton counting approach 
to analyze the spectral problem \eqref{dex2}.

\subsection{Definitions}

The main quantity that appears in this approach is the  Nekrasov-Shatashvili free energy of the $SU(2)$ $N_f=1$ theory:
\be\label{NSde} F_{\rm NS}(a, \hbar,m,\Lambda)=\sum_{n\geq 1} c_n(a,m,\hbar)\Lambda^{3n}, \ee
where the coefficients $c_n$ have a closed form definition in terms of combinatorics of Young diagrams; we refer to \cite[Appendix.~A]{Aminov:2020yma} for the definition and a more exhaustive list of references. For example, the first few terms read
\be \ba c_1 &= \frac{2 m}{4 a^2+\hbar ^2} \, ,\\
c_2&=  \frac{ \left(-48 a^4+80 a^2 m^2-24 a^2 \hbar ^2-28 m^2 \hbar ^2-3 \hbar ^4\right)}{16 \left(a^2+\hbar ^2\right) \left(4 a^2+\hbar ^2\right)^3}. \ea\ee
It is important to note that \eqref{NSde} is exact in $(\hbar, a , m)$ and it is a well-defined convergent sum in the parameter $\Lambda$. Strictly speaking this convergence property has not been proven mathematically (the proofs of \cite{felder,bsu,ilt} do not apply to the $\epsilon_2=0$ background); nevertheless, there is considerable numerical evidence for it. In addition one should note that the convergence is for $ 2 \ri a/\hbar \notin {\mathbb{Z}}$, since at these values the NS free energy diverges; see for example \cite{Gorsky:2017ndg} for a discussion of the structure of these poles. Note also that in four dimensions these poles have a physical meaning, and they should not be confused with the unphysical poles appearing in the five-dimensional uplift of the Nekrasov partition function which were first discussed in \cite{Hatsuda:2012dt}. 

We define $a(E, \hbar,m, \Lambda)$ implicitly via the quantum Matone relation  \cite{matone,francisco},
\be\label{mat} E=\left(a^2+{1\over 3} \Lambda \partial_\Lambda F_{\rm NS} (a, \hbar,m,\Lambda)\right).\ee
This relation defines $a$ only up to an overall sign. We work in the convention where we pick the sign such that ${\rm Re}\left({a\over \hbar}\right)\geq0$.  If ${\rm Re}\left({a\over \hbar}\right)=0$, then we take ${\rm Im}\left({a\over \hbar}\right)>0$. 
We also define (see for instance  \cite{Zenkevich:2011zx})
\be\label{ad} \ba a_D(a,\hbar,m, \Lambda )= {} &-4 \gamma (2 a,\hbar )+\gamma \left(a-m-\frac{\ri \hbar }{2},\hbar \right)+\gamma \left(a+m-\frac{\ri \hbar }{2},\hbar \right)+\partial_a F_{\rm NS} (a, \hbar,m,\Lambda)\\
&+ \ri \pi  a-\ri \pi  m+\frac{1}{2} \ri \hbar  \log \left(\E^{\frac{2 \pi  (m-a)}{\hbar }}+1\right)-\frac{1}{2} \ri \hbar  \log \left(\E^{-\frac{2 \pi  (a+m)}{\hbar }}+1\right)\\
&+ 6 a \log (\Lambda )+\frac{1}{2} \ri \hbar  \log \left(a-m-\frac{\ri \hbar }{2}\right)+\frac{1}{2} \ri \hbar  \log \left(a+m-\frac{\ri \hbar }{2}\right), \ea
\ee
where 
\be  \gamma (a,\hbar )=\left(-\frac{1}{2} a \log \left(\frac{1}{\hbar ^2}\right)-\frac{\pi  \hbar }{4}\right)-\frac{1}{2} \ri \hbar  \left(\log\Gamma\left(\frac{\ri a}{\hbar }+1\right)-\log\Gamma\left(1-\frac{\ri a}{\hbar }\right)\right).\ee
{Note that\footnote{Up to possible $\hbar \pi$ factors.} \be a_D(a,\hbar,m, \Lambda )=a_D(-a,-\hbar,m, \Lambda )+\ri \hbar  \left(\log \left(\re^{\frac{2 \pi  a}{\hbar }}+\re^{\frac{2 \pi  m}{\hbar }}\right)-\log \left(\re^{\frac{2 \pi  (a+m)}{\hbar }}+1\right)\right) . \ee}
  For later purposes it is useful to define
  \be \ba a_{m} (a, \hbar,{m},\Lambda)={} &\gamma \left(a+{m}-\frac{\ri \hbar }{2},\hbar \right)-\gamma \left(a-{m}-\frac{\ri \hbar }{2},\hbar \right)+\partial_{m} F_{\rm NS} (a, \hbar,{m},\Lambda)\\
  & -\frac{1}{2} \ri \hbar  \log \left(a-{m}-\frac{\ri \hbar }{2}\right)+\frac{1}{2} \ri \hbar  \log \left(a+{m}-\frac{\ri \hbar }{2}\right)-\frac{1}{2} \ri \hbar  \log \left(\E^{\frac{2 \pi  ({m}-a)}{\hbar }}+1\right)\\
  &-\frac{1}{2} \ri \hbar  \log \left(\E^{-\frac{2 \pi  (a+{m})}{\hbar }}+1\right)+2 \ri \hbar  \log \left(1-\E^{-\frac{4 \pi  a}{\hbar }}\right), \ea\ee
and the normalized quantities \be \label{instP}\ba{\tt a}(E,\hbar,m,\Lambda):=&-{4 \pi a\over \hbar}(E, \hbar,m, \Lambda),
\\
{\tt a}_D(E,\hbar,m,\Lambda):=&-  { \ri  \over \hbar}a_D(a, \hbar,m, \Lambda)\Big|_{a=a(E, \hbar,m, \Lambda)},\\
{\tt a}_{m}(E,\hbar,m,\Lambda):=&  { \ri  \over \hbar}a_{m}(a, \hbar,m, \Lambda)\Big|_{a=a(E, \hbar,m, \Lambda)} .
\ea\ee
Then we have 
\be{\tt a}(E,\hbar,m,\Lambda)\quad \xrightarrow {\hbar\to 0} \quad \sum_{n\geq 0}\hbar^{2n-1} {\tt a}^{(n)}(E,m,\Lambda),\ee
\be{\tt a}_D(E,\hbar,m,\Lambda)\quad \xrightarrow {\hbar\to 0} \quad \sum_{n\geq 0}\hbar^{2n-1}  {\tt a}_D^{(n)}(E,m,\Lambda).\ee
Let us note that, even though ${\tt a}$ and ${\tt a}_D$ are well defined both in the strong and weak coupling region, the series expansion coefficients  ${\tt a}^{(n)}$ and $ {\tt a}_D^{(n)}$ converge only in the weak coupling region\footnote{To get  a convergent expression for ${\tt a}^{(n)}$ and $ {\tt a}_D^{(n)}$ in the strong coupling region one should use the holomorphic anomaly equation as in \cite{coms}.}. 

Another important observation, due to   \cite{mirmor, Zenkevich:2011zx}, is that the series expansion coefficients  ${\tt a}^{(n)}$ and $ {\tt a}_D^{(n)}$ coincide with the WKB coefficients $\Pi_\gamma^{(n)}$ of the quantum periods in equation \eqref{wkbqm} if the cycle $\gamma$ is suitably chosen. 
Hence instanton counting provides an exact, analytic resummation of the WKB series.

\subsection{Quantum periods}

Now we want to describe the relation between ${\tt a}$, ${\tt a}_D$ and the quantum periods $\Pi_\gamma$ as we have
defined them.

As we reviewed in \autoref{analyticstructures}, the $\Pi_\gamma$ are piecewise analytic; to make an honest analytic function
from them we have to fix a basepoint $(\vartheta, u_0, m_0, \Lambda_0)$. Thus, to compare them to ${\tt a}$, ${\tt a}_D$
we need to pick such a basepoint.
Experimentally, we find that the right basepoint to choose (within the weak coupling region) is one leading to a Schr\"odinger equation \eqref{dex2} with a real convex confining potential along some $x$-direction. 
This condition is fulfilled for
\be\ba
2\arg(\Lambda) -{\rm Im} (x)-2\arg(\hbar)={}&2\pi \ell_1\, ,
\\
\arg(m)+\arg(\Lambda)+{\rm Im} (x)-2\arg(\hbar)={}&2\pi \ell_2\, ,
\\
2\arg(\Lambda) +2{\rm Im} (x)-2\arg(\hbar)+\pi={}&2\pi \ell_3\, ,
\\
 \arg(u)-2\arg(\hbar)={}&2\pi \ell_4\, ,
\ea \ee
where $\ell_i \in \IZ$. This leaves us with a variety of possibilities, but we expect the final result to be independent of this choice.  
 Hence we can take (hopefully without loss of generality)\footnote{In this  case $x \in \re^{-\ri \pi/3}\IR_+$} 
\be \label{topii}m_0^{\rm inst}={\ri \over 10}, \quad u_0^{\rm inst}=2, \quad \Lambda_0^{\rm inst}=\re^{-\ri \pi/6} \, . \ee
Moreover, inspired by   \cite{nrs,Hollands:2019wbr,Grassi:2019coc, Hollands:2013qza,Hollands:2017ahy}, we expect that  instanton counting  corresponds to Fenchel-Nielsen coordinates, namely to spectral coordinates at 
 \be \label{topii2} \vartheta^{\rm inst}_0 ={\arg}(-Z_{\gamma_a}(u_0^{\rm inst},m_0^{\rm inst},\Lambda_0^{\rm inst}))=0\ee where $Z_{\gamma_a}$ is the central charge of the vectormultiplet in the BPS spectrum at weak coupling.   
Hence we propose the identification
\be  \label{xia}\Pi_{\gamma_{[0,2,0]}}^{{\rm inst}, \vartheta_0^{\rm inst}, m_0^{\rm inst}, u_0^{\rm inst}, \Lambda_0^{\rm inst}}(E,\hbar,m,\Lambda)= {\tt a} (E,\hbar,m,\Lambda), \ee
as well as 
\be  \label{xib}\Pi_{\gamma_{[1,0,0]}}^{{\rm inst}, \vartheta_0^{\rm inst},  m_0^{\rm inst}, u_0^{\rm inst}, \Lambda_0^{\rm inst}}(E,\hbar,m,\Lambda)= {\tt a}_D (E,\hbar,m,\Lambda), \ee 
where the charges $\gamma_{[1,0,0]}$ and $\gamma_{[0,2,0]}$ are taken relative to the basis in
\autoref{quantizationcd}.\footnote{Both sides of these equations are multivalued functions
of $(E, \hbar, m, \Lambda)$, because of the logarithms appearing in the definition of $a$, $a_D$, and the choice of sign in solving 
the Matone relation. We take the principal branch
in the logarithms, and fix the sign of $a$ by picking 
${\mathrm{Re}}(\frac{a}{\hbar}) \ge 0$,
when $(E,\hbar,m,\Lambda) \in C^{\rm inst}$. At
other values of $(E,\hbar,m,\Lambda)$ the functions are
defined by analytic continuation.}

We also have as usual the exact formula
\begin{equation}
	\Pi^{{\rm inst},\vartheta_0^{\rm inst},  m_0^{\rm inst}, u_0^{\rm inst}, \Lambda_0^{\rm inst}}_{\gamma_{[0,0,1]}} = -2 \pi m / \hbar.
\end{equation}

We will denote by ${C}^{\rm inst}$ the locus specified by \eqref{topii}, \eqref{topii2}.

\section{Comparisons} \label{sec:comparisons}

In the last few sections we have described four different ways of understanding, and numerically computing,
the quantum periods in the $SU(2)$ $N_f =1$ theory.
Our expectation is that all four methods are approximating the same functions $\Pi_\gamma$, 
so that our four numerical computations should agree, up to the inherent 
numerical error in the various methods.
The results of various such comparisons are reported in \autoref{tab:introresultsnew}.
In this section we discuss in more detail some examples, and
some subtleties that arise in making the comparison.

\subsection{Comparisons in the weak coupling region}
\subsubsection{Instanton counting versus small section method}

We first compare instanton counting with the SS method of \autoref{sec:ssection}, i.e.
we compare $\Pi^{\rm inst}$ with $\Pi^{\rm SS}$.

We recall that when we use the SS method we need to specify two set of parameters:
\begin{itemize}
\item One set of parameters \be
\label{SNp} \vartheta, u_0, m_0, \Lambda_0 \ee specifying  the spectral network $\mathcal W$ that we use in the computation.
\item One set of parameters \be\label{eva} \hbar, u, m, \Lambda\ee specifying the values at which we evaluate the spectral coordinates $\cX^{\rm SS}_\gamma$. \end{itemize}
In \autoref{sec:ssection}
 we always used $(u_0, m_0, \Lambda_0) = (u, m, \Lambda)$ and hence we only needed to specify the parameter $\vartheta$. This is the reason why we used the notation $ \cW^{\vartheta}$ instead of $ \cW^{\vartheta, u_0, m_0, \Lambda_0}$.
For comparing with instanton counting, it is more convenient to choose $(\vartheta, u_0, m_0, \Lambda_0)$
lying in the instanton locus $C^{\rm inst}$. The spectral network which appears then is the one we used in
\autoref{section:qcd}, and the concrete Wronskian formulas we use are the ones in that section as well.

As an example, we take
\be \label{eg2}E= 3+{\ri/5}, \quad m={\ri\over 2}+{3\over 10}, \quad \Lambda={1\over 5}+{\frac{\ri}{10}}, \quad \hbar=1+{\ri\over 2}. \ee
We evaluate $\Pi_{\gamma_{[1,0,0]}}$ by using instanton counting and compare with the result from the
SS method (concretely, \eqref{X100}).
We find good agreement:
\be \ba \Pi_{\gamma_{[1,0,0]}}^{\rm inst, \vartheta=0, m_0^{\rm inst}, u_0^{\rm inst}, \Lambda_0^{\rm inst}} = {\tt a}_D=&~ 14.5926368914\ldots+ 11.9576187144~\ri \ldots\\
\Pi_{\gamma_{[1,0,0]}}^{\rm SS, \vartheta=0, m_0^{\rm inst}, u_0^{\rm inst}, \Lambda_0^{\rm inst}}=&~ 14.5926368914  \ldots  + 11.9576187144~\ri \ldots \ea \ee

More comparisons are shown in \autoref{tab:introresultsnew}. In that table, though, we report the canonical $\Pi_\gamma$,
or said otherwise, we take $(\vartheta, u_0, m_0, \Lambda_0) = (\arg \hbar, u, m, \Lambda)$. Thus,
in particular, the quantities
$\Pi_\gamma^{\rm inst}$ appearing in that table are in general not equal to ${\tt a}$, ${\tt a}_D$; rather they
are related by the appropriate KS transformations to move to the correct chamber.
We discuss this procedure in more detail in the next section.

\subsubsection{Instanton counting versus Borel summation}\label{try}

Let us now look at Borel summation. As we have recalled in \autoref{analyticstructures},
Borel summation always produces the canonical $\Pi_\gamma$, i.e. the ones at 
$(\vartheta,u_0, m_0, \Lambda_0) = (\arg \hbar,u, m, \Lambda)$.
 Therefore, if we pick
 $ (\arg \hbar, u, m, \Lambda) \in C^{\rm inst}$, then instanton counting and (median) Borel summation should 
agree. This is the analogue of what was done in \cite{Grassi:2019coc}. 
However, for generic values of the parameters we need to implement an appropriate KS transformation.

 As an example we can take
\be \label{chBorel} E=4, \quad m=0,\quad \Lambda= {1\over 5}, \quad \hbar=1+{\frac{\ri}{10}}\, .\ee 
We consider the period $\Pi_\gamma$ for $\gamma=\gamma_{a_D}=\gamma_{[1,0,0]}$. 
The point \eqref{chBorel} is not on any wall; in particular, the WKB series $\Pi^{\mathrm {WKB}}_\gamma$ 
at this point is Borel summable,
without requiring median summation, and gives $\Pi_\gamma$ at the point \eqref{chBorel}.
However, since the point \eqref{chBorel} $\notin C^{\rm inst} $,
to compare this Borel sum with instanton counting we have to work out the transformation between 
$\Pi^{\eqref{chBorel}} $ and $\Pi^{C^{\rm inst}} $. For that purpose 
we choose a particular path connecting these two points, 
and apply the transformations \eqref{eq:pi-transform}-\eqref{eq:pi-transform3} for all the BPS states which we encounter along this path.
 
 As a first step, we vary $\hbar$ along a path from \eqref{chBorel} to an intermediate chamber
 \be \label{chBorelp} E=4, \quad m=0,\quad \Lambda= {1\over 5}, \quad \hbar=0\, .\ee
On this path we encounter the BPS states whose central charges satisfy
    \be0\leq -Z_{\gamma_{\rm BPS}}\leq \arg(1 + \frac{\ri}{10}); \ee 
these central charges are \footnote{Since $\vartheta_i=0$ passes through these singularities, each of them only contributes $\frac{1}{2}\IP{\mu,{\gamma}} \Omega({\gamma}) \log(1 - \sigma_{\mathrm{can}}({\gamma}) \cX_{\gamma}^{\vartheta})$.}
    \be -Z_{\gamma_{[0,2,0]}}, -Z_{\gamma_{[0,1,1]}},  -Z_{\gamma_{[0,1,-1]}}, \quad  \ee
    as well as
    \be \ba
  -  Z_{\gamma_n}={}&-Z_{-\gamma_{a_D}}-n Z_{\gamma_a}, \quad n\geq 11,\\
     -  Z_{\overline{\gamma_m}}={}&-\overline{Z_{\gamma_{a_D}}}-m Z_{\gamma_a}, \quad m\geq 11.
     \ea\ee
     Some of them can be seen explicitly in  \autoref{fig:qppole23}. {Note that there will be infinite BPS states in \autoref{fig:qppole23} if we take infinite terms $\Pi^n$ in the series expansion\footnote{As we will discuss later, we know from other methods that the red dot on  the positive real axis represent the central charges for 2 distinct hypermultiplets: $\gamma_{[0,1,-1]}$ and $\gamma_{[0,1,1]}$.  There is also an invisible vectormultiplet at twice the length with charge $\gamma_{[0,2,0]}$. They all contribute to the KS transform.}.}
     We get    %
\be \label{ex3}\ba
\Pi_{\gamma_{a_D}}^{\eqref{chBorelp}}&=\Pi_{\gamma_{a_D}}^{\eqref{chBorel}}
-\frac{1}{2}\log(1+\cX^{\vartheta=0}_{\gamma_{[0,1,1]}})-\frac{1}{2}\log(1+\cX^{\vartheta=0}_{\gamma_{[0,1,-1]}})
-\frac{1}{2}\log(1-\cX^{ \vartheta=0}_{\gamma_{[0,2,0]}})^{-4}\\&- \sum_{n\geq 11} 2n \log\left(\left(1+ \cX^{\vartheta={\arg}(-Z_{\gamma_n})}_{\gamma_n}\right)\left(1+\cX^{ \vartheta={\arg}(-{Z_{\overline{\gamma_n}}})}_{\overline{\gamma}_n}\right)\right). \ea  \ee
At the practical level, we can neglect the last sum since the 
  largest contribution in this infinite series  can be approximated by \be\label{neglect} \log(\sqrt{(1+
  \exp\left[-  Z_{\overline{\gamma_{11}}}(1+\ri/10)^{-1}\right])})\approx 10^{-120} \ri \ee which is much smaller than the accuracy we can reach with Borel summation of 80 even terms.
  
\begin{figure}[htbp]
\begin{center}
\includegraphics[width=0.35\linewidth]{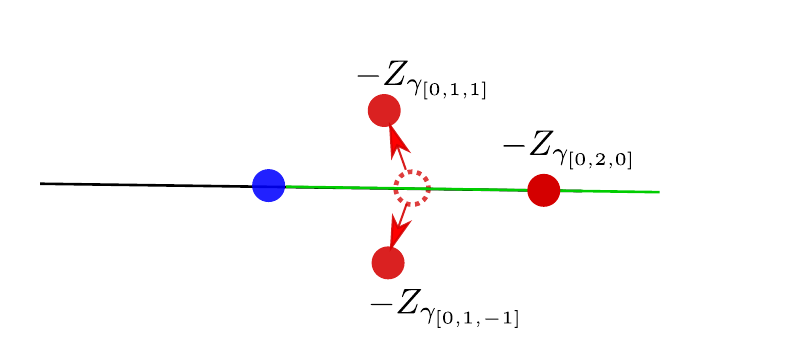}
\caption{Zoom in around the real axis.
The dashed circle represents the singularities from the 2 hypermultiplets with $-Z_{\gamma_{[0,1,1]}}=-Z_{\gamma_{[0,1,-1]}}$ in massless case. 
 Changing from \eqref{chBorelp} to ${C^{\rm inst}}$ corresponds to split the dashed circle into two red dots representing the singularities at $-Z_{\gamma_{[0,1,-1]}}$ and the $-Z_{\gamma_{[0,1,1]}}$. The green ray represents $\vartheta=0$.} 
\label{instc}
\end{center}
\end{figure}

As a second step on our path we go from \eqref{chBorelp} to $C^{\rm inst}$, by turning on a small $m \in \I \R_+$. In this procedure only $-Z_{\gamma_{[0,1,-1]}}$ and $-Z_{\gamma_{[0,1,1]}}$ contribute to \eqref{eq:pi-transform}-\eqref{eq:pi-transform3}. We sketch this in \autoref{instc}. We have
 \be \label{ex3p}
\Pi^{\rm C^{\rm inst}}_{\gamma_{a_D}}=\Pi_{\gamma_{a_D}}^{\rm \eqref{chBorelp}}
-\frac{1}{2}\log(1+\cX^{\rm \vartheta=0}_{\gamma_{[0,1,1]}})+\frac{1}{2}\log(1+\cX^{\rm\vartheta=0}_{\gamma_{[0,1,-1]}}).
 \ee
Now combining \eqref{ex3} and \eqref{ex3p}, we have
\be \label{ex3t}
\Pi^{C^{\rm inst}}_{\gamma_{a_D}}\approx\Pi_{\gamma_{a_D}}^{\eqref{chBorel}}
-\log(1+\cX^{\vartheta=0}_{\gamma_{[0,1,1]}})
-\frac{1}{2}\log(1-\cX^{\vartheta=0}_{\gamma_{[0,2,0]}})^{-4}\, ,\ee
where $\approx$ means that we are neglecting second line of \eqref{ex3}.
 The final contribution from each KS transformation are listed in \autoref{table:transinst}.

\begin{table}[htp] 
\begin{center}
\begin{tabular}{|c|c|}
\hline
$\Pi_{\gamma_{a_D}}^{\rm PB}(E,\hbar, m,\Lambda)$ & $\underline{8.87942}7505929998 + \underline{25.70889}962041928 \ri$ \\
\hline
$\Pi_{\gamma_{a_D}}^{\rm PB}(E,\hbar, m,\Lambda)$ &\\
$-\log(1+\cX^{\rm PB,\vartheta=0}_{\gamma_{[0,1,1]}})$ & $\underline{8.8794262388}60653
   + \underline{25.708895879}81361 \ri$ \\
\hline
$\Pi_{\gamma_{a_D}}^{\rm PB}(E,\hbar, m,\Lambda)$ &\\
$-\log(1+\cX^{\rm PB,\vartheta=0}_{\gamma_{[0,1,1]}})$ &\\
$+2\log({1-\cX^{\rm PB,\vartheta=0}_{\gamma_{[0,2,0]}}})$ & \underline{8.879426238885426} + \underline{25.70889587979465} \ri \\
\hline

\hline
${\tt a}_D(E,\hbar, m,\Lambda) $& 8.879426238885426 + 25.70889587979465 \ri \\
\hline
\end{tabular}
\end{center}
 \caption{Transformation to compare Pad\'e-Borel summation $\Pi_{\gamma_{a_D}}^{\rm PB}$ with ${\tt a}_D$ according to \eqref{ex3t}. The parameters used are $E=4,   m=0, \Lambda= {1\over 5},  \hbar=1+{\frac{\ri}{10}}$. }
 \label{table:transinst}\end{table}%

 Another example we use in this paper is\footnote{The Schr\"{o}dinger operator is invariant under $u\rightarrow -u,\ m\rightarrow \ri m,\ \Lambda\rightarrow \Lambda\re^{-\pi\ri/6},\ \hbar\rightarrow\ri\hbar,\ z\rightarrow\re^{\frac{2\pi\ri}{3}}z$. For numerical calculation, it was more convenient to take $\Lambda$ real; thus we did Borel summation at $m=1/10,u=-2,\Lambda=1,\hbar=-\ri/2$ and used the cycles related to $\gamma_{[0,2,0]}$ and $\gamma_{[1,0,0]}$ under $z\rightarrow\re^{\frac{2\pi\ri}{3}}z$.} 
\be \label{chBorel2} E=4, \quad m=\ri/10,\quad \Lambda=\re^{-\pi\ri/6}, \quad \hbar=1/2\, .\ee 
The results are shown in \autoref{table:transinst2}.  

Here we mention some technical problems encountered in the Pad\'e-Borel method and show how we can nicely overcome  such obstacles by taking advantage of  the KS transformation and the knowledge about the underlying gauge theory.  

Let us consider for example \autoref{fig:qppole21} and \autoref{fig:qppole23}. 
The red dot on the positive real axis represents the two hypermultiplets with central charges $Z_{\gamma_{[0,1,-1]}}=Z_{\gamma_{[0,1,1]}}$, indeed in this example $Z_{\gamma_{[0,0,1]}}=-2\pi m=0$. However, there should also be a singularity at the central charge of the vectormultiplet $\gamma_{[0,2,0]}$. Nevertheless, we do not see it  explicitly in the Borel plane because the singularity at $Z_{\gamma_{[0,1,-1]}}=Z_{\gamma_{[0,1,1]}}$  spans a full series of poles which  mix with the poles corresponding to $Z_{\gamma_{[0,2,0]}}$. 
As a consequece the discontinuities at $Z_{\gamma_{[0,1,1]}}=Z_{\gamma_{[0,1,-1]}}$ and $Z_{\gamma_{[0,2,0]}}$ will  also mix. 
 However in our computations, in particular when implementing the KS transform as in Table \ref{table:transinst}, we need to take this singularity into account even if we do not see it explicitly in the Borel plane. 
  It could be that by combining  Pad\'{e}-Borel with some more advanced techniques we might be able to see these three singularities separately. 
A more general phenomenon also appears when $m\neq 0$. For example, when we study the Pad\'{e}-Borel transform for $\gamma_{[1,0,0]}$ at \eqref{chBorel2}. Such transform has two series of poles close to the real axis, see \autoref{fig:qppole32}\footnote{ Such Figure is rotated by $\ri \pi /2$, so the imaginary axis on this Figure corresponds to the real axis in the current discussion}, which correspond to the 2 hypermultiplets $\gamma_{[0,1,1]}$ and $\gamma_{[0,1,-1]}$. These series of poles point towards the real axis where they collide and then carry on as  a unique  series close to the real axis. Because of their existence, we can not see the series of poles representing the vector multiplet $\gamma_{[0,2,0]}$. In addition in \autoref{fig:qppole33}, when the series of poles for finite number of terms from the two hypermultiplets seems rotated to slightly different direction from real axis, the interference still makes the vectormultiplet hard to be seen.  The mixing of these poles makes it impossible  to find a  path of integration for the lateral  Pad\'{e}-Borel summation around $\arg(\hbar)=0$
which is not affected by the poles of 2 hypermultiplets and take half effect of the vectormultiplet.
We overcome this technical obstacle by taking advantage of the KS transformation. We take the direction of integration  to be $\IR_+ \re^{\ri/10 }$ so that we are in the chamber $\vartheta=1/10$.  Pad\'{e}-Borel summation work in this chamber and  then we use KS transformation to transform it back to chamber  \eqref{chBorel2}.  The leading contribution to such transformation comes from the hypermultiplet with charge $\gamma_{[0,1,1]}$.  
The results are shown in \autoref{table:transinst2}.  

\begin{table}[htp] 
\begin{center}
\begin{tabular}{|c|c|}
\hline
$\Pi^{\rm PB, \vartheta=\frac{1}{10}}_{\gamma_{a_D}}(E,\hbar, m,\Lambda)$ & $3.9*10^{-12}+14.0834766258633 \ri$ \\
\hline
$\Pi^{\rm PB, \vartheta=\frac{1}{10}}_{\gamma_{a_D}}(E,\hbar, m,\Lambda)$ &\\
$-\log(1+\cX^{\rm PB,\vartheta=\frac{1}{10}}_{\gamma_{[0,1,1]}})$ & $14.0834766258755\ri$ \\
\hline
${\tt a}_D(E,\hbar, m,\Lambda') $& $14.0834766258755\ri$  \\
\hline
\end{tabular}
\end{center}
 \caption{Transformation to compare Pad\'e-Borel summation $\Pi_{\gamma_{a_D}}^{\rm PB}$ with ${\tt a}_D$. The parameters used for $E,\hbar, m,\Lambda$ are those in \eqref{chBorel2}.  }
 \label{table:transinst2}\end{table}%
 
 We conclude this part with a comment on the poles of the four dimensional NS free energy. As anticipated earlier the NS free energy  (and in particular the  ${\tt a}_D=\Pi_{\gamma_{[1,0,0]}}^{C^{\rm inst}}$ period) has poles if $2 a \ri/\hbar \in \IZ$, i.e.~if $(E, \Lambda, m, \hbar)$ are such that \be \label{pole} \Pi_{\gamma_{[0,2,0]}}^{C^{\rm inst}}(E, \Lambda, m, \hbar)= 2 \pi \ri \IZ .\ee  
 On the other hand we know that at such values of  $(E, \Lambda, m,\hbar)$ Borel summation gives a well defined number for $\Pi_{\gamma_{[1,0,0]}}(E, \Lambda, m,\hbar)$. The reason for such different behaviour is  due to the fact that when \eqref{pole} holds, the KS transform for the vector multiplet is singular. Hence the transformation between instanton and Borel chamber is singular.

\subsection{Comparison in strong coupling region}
\label{instst}

In the strong coupling region the relation between instanton counting and small section method is much more complicated. The reason is that the topology of the spectral network at strong coupling is very different from the one making contact with instanton counting, namely  \eqref{topii}. Indeed, if we follow a path from the instanton locus to strong coupling
which avoids singularities in the Coulomb branch, we will have to cross an infinite number of walls, 
and thus encounter an infinite tower of transformations 
\eqref{eq:pi-transform}-\eqref{eq:pi-transform3}. To avoid such a complicated calculation, we can 
use the following procedure.

At $\vartheta_0^{\rm inst},u_0^{\rm inst},{m}_0^{\rm inst},\Lambda_0^{\rm inst}$,
we choose a basis in which the monodromy matrix and its eigenvectors are
\be \ba 
M=\begin{pmatrix}\frac{1}{\mu} & 0\\ 0 &\mu\end{pmatrix}
,\quad
    \psi_4 = \left(\begin{matrix}
           1 \\
           0 \\
         \end{matrix}\right), \quad \psi_5= \left(\begin{matrix}
           0 \\
           1 \\
         \end{matrix}\right)\ea\ee
We start from the weak coupling region at $E_0^{\rm inst}=4, ~ \Lambda^{\rm inst}_0=\exp(-\ri\pi/6), ~ m_0^{\rm inst}=\frac{\ri}{10}$, with $\cW^{\vartheta=\vartheta_0^{\rm inst}=0}$. 
By matching the small section result  \eqref{X010}, \eqref{X100} $$\cX^{\rm SS,\vartheta=\vartheta_0^{\rm inst},u=u_0^{\rm inst},{m}={m}_0^{\rm inst},\Lambda=\Lambda_0^{\rm inst}}(\hbar,u,m,\Lambda)$$ with the instanton counting result  \eqref{xia}, \eqref{xib}  $$\cX^{\rm inst,\vartheta_0^{\rm inst},u_0^{\rm inst},{m}_0^{\rm inst},\Lambda_0^{\rm inst}}(\hbar,u,m,\Lambda)$$ we deduce that  

\be \label{psiPi}\ba
 \psi_1=& \left(\begin{matrix}
           \psi_{1,1} \\
           1 \\
         \end{matrix}\right),\\
\psi_2=& \left(\begin{matrix}
             -\psi_{1,1}\xadi \\
           1 \\
         \end{matrix}\right),\\
          \psi_3=& \left(\begin{matrix}
            \psi_{1,1} \left(-\re^{\adi+\frac{\ai}{2}}\right) \cosh \left(\frac{\ai}{4}-\frac{\pi  m}{h}\right)
   \text{sech}\left(\frac{\ai}{4}+\frac{\pi  m}{h}\right)\\
           1 \\
         \end{matrix}\right),
        \ea \ee
         for some $\psi_{1,1}$ which we do not need to determine for our purpose.
             Even though these expressions  have been derived at weak coupling, by analytic continuation they hold at strong coupling as well.
  Therefore the spectral coordinates at strong coupling, given in \eqref{s0}, \eqref{s1}, \eqref{s2}, become\footnote{Again, since there is a square root in the expression, we have to choose a branch; we do not give a rule for fixing the branch here, instead just living with the sign ambiguity.}
\be \label{xxs-1} \cX_{\gamma_{[1,0,0]}}^{\rm inst, \vartheta=0}(\hbar,u,m,\Lambda)=\frac{\re^{\adi} \left(\re^{\ai}-1\right) \sqrt{\frac{\re^{-\adi} \left(\left(\re^{\adi}+1\right) \re^{\frac{\ai}{2}-\frac{2 \pi 
   m}{h}}+\re^{\adi}+\re^{\ai}\right)}{e^{\adi+\ai}+1}}}{\left(\re^{\adi}+1\right) \re^{\frac{\ai}{2}+\frac{2 \pi 
   m}{h}}+\re^{\adi+\ai}+1},
\ee
 \be
\label{xxs-2}\cX_{\gamma_{[0,1,1]}}^{\rm inst, \vartheta=0}(\hbar,u,m,\Lambda)= \frac{\re^{-\adi+\frac{\ai}{2}-\frac{2 \pi  m}{h}} \left(\left(\re^{\adi}+1\right)^2 \re^{\frac{\ai}{2}+\frac{2 \pi 
   m}{h}}+\re^{\adi} \left(\re^{\adi+\ai}+\re^{\ai}+1\right)+1\right)}{\left(\re^{\ai}-1\right)^2}\, ,\\
\ee
 \be \label{xxs-3} \cX_{\gamma_{[-1,-1,1]}}^{\rm inst, \vartheta=0}(\hbar,u,m,\Lambda)=-\frac{\re^{\adi} \left(\re^{\ai}-1\right) \sqrt{\frac{\left(\re^{\adi+\ai}+1\right) }{\left(\re^{\adi}+1\right) \re^{\adi+\frac{3 \ai}{2}+\frac{2 \pi  m}{h}}+\left(\re^{\adi}+\re^{\ai}\right)
   \re^{\adi+\ai+\frac{4 \pi  m}{h}}}}}{\left(\re^{\adi}+1\right)}\, ,\ee
   where on the r.h.s.~we used the shortcut notation
   \be {\tt a}={\tt a}(\hbar,u,m,\Lambda),\quad {\tt a}_D={\tt a}_D(\hbar,u,m,\Lambda)\, .\ee

We also show expressions for $m=0$:
 \be \label{xxs} \cX_{\gamma_{[1,0,0]}}^{\rm inst, \vartheta=0}(\hbar,u,m=0,\Lambda)=\frac{\sinh\left(\frac{\ai}{4}\right) \sqrt{\left(\cosh \left(\frac{\adi-\ai}{2}\right)+\cosh
   \left(\frac{\adi}{2}\right)\right) \text{sech}\left(\frac{\adi+\ai}{2}\right)}}{\sinh \left(\frac{1}{4} (\ai+2 \adi)\right)},\ee
 \be \label{xxs2} \cX_{\gamma_{[0,1,1]}}^{\rm inst, \vartheta=0}(\hbar,u,m=0,\Lambda)=\frac{1}{2} \cosh \left(\frac{\adi}{2}\right) \text{csch}^2\left(\frac{\ai}{4}\right) \text{sech}\left(\frac{\ai}{4}\right) \cosh
   \left(\frac{1}{4} (2 \adi+\ai)\right),\ee
 \be \label{xxs3} \cX_{\gamma_{[-1,-1,1]}}^{\rm inst, \vartheta=0}(\hbar,u,m=0,\Lambda)=- \frac{\sinh \left(\frac{\ai}{4}\right)  \sqrt{2\cosh \left(\frac{\ai}{4}\right) \cosh \left(\frac{\ai+\adi}{2}\right) \text{sech}\left(\frac{1}{4} (\ai-2 \adi)\right)}}{\text{cosh}\left(\frac{\adi}{2}\right)}.\ee
 The equations above are the transformations between Fock-Goncharov coordinates and Fenchel-Nielsen coordinates (instanton counting).

 The results are in \autoref{tab:introresultsnew}, where the spectral coordinates reported  are at $\vartheta=\arg(\hbar)$. For our examples with
$u=0$, $m\in \{0,-1/10\}$, $ \Lambda\in\{1,{1\over 5}\}$
 and $\arg(\hbar)=\arg(1+{\frac{\ri}{10}})$, we have
\be\label{transf}\ba  \Pi_{\gamma[-1,-1,1]}^{}={}&  \Pi_{\gamma[-1,-1,1]}^{ \vartheta=0}-{1\over 2}\log \left(1+\cX^{ \vartheta=0}_{[0,1,1]}\right),\\
 \Pi_{\gamma[1,0,0]}={}&  \Pi_{\gamma[1,0,0]}^{ \vartheta=0}+{1\over 2}\log \left(1+\cX^{ \vartheta=0}_{[0,1,1]}\right),\\
  \Pi_{\gamma[0,1,1]}={}&  \Pi_{\gamma[0,1,1]}^{ \vartheta=0}. \ea\ee
Notice that the r.h.s.~of \eqref{xxs-1}-\eqref{xxs-3} provides an analytic solution to the GMN TBA at strong coupling (after we implement \eqref{transf}). 

We conclude this section with a brief comment on Painlev\'e equations.
It was suggested in \cite{Coman:2020qgf}, based on \cite{ilt}, that the transformations between Fock-Goncharov and Fenchel-Nielsen coordinates  are useful in the study of the connection problems in Painlev\'e equations. From this point of view  the  transformations \eqref{xxs}-\eqref{xxs3} could play and interesting role  in the study of the connection problem for Painlev\'e $\rm III_2$.

\section{Fredholm determinant}
\label{sec:fredholm}

\subsection{From Topological String}
\label{sec:TS}
A convenient way to encode the spectral properties  
of a given trace class operator $\rm O^{-1}$ is by using the Fredholm determinant \be \label{sd}\det\left(1-{E\over {\rm O}}\right).\ee
 Such object is also interesting from a physical point of view since it is an entire function in $E$ which is usually identified with the Coulomb branch parameter. 
The computation of these determinant is a challenging question. In some situations we can use (numerical) TBA techniques  or WKB analysis \cite{post-zamo, voros-zeta, voros, voros-zq,voros-quartic}. However if the operator $\rm O$ has an interpretation in terms of quantum mirror curve to toric CY manifolds, then its  Fredholm determinant  can be computed explicitly and exactly \cite{ghm,cgm,cgm8}. Such construction was originally formulated only in some particular slice of the moduli space where the operators have a positive discrete spectrum with bound states. This was later generalised in \cite{cgum,gm17,gm3} to include operators with complex eigenvalues and in particular resonance states. 
The  operator  we are studying in this paper \eqref{dex2} does not correspond to a quantum mirror curve to toric manifolds. Nevertheless it can be obtained from such construction after implementing the geometric engineering limit \cite{kkv,selfdual}, similar to what was done in \cite{Grassi:2019coc} for the (modified) Mathieu example.
 
We are interested in the toric CY geometry that engineers the $SU(2)$ 4d $N_f=1$ theory, namely local $\IF_2$ blown up at 1 point  as in \cite{kkv}. We denote  this geometry as local ${\CB}\IF_2$. 
The  the quantum mirror curve for this CY is 
\be\label{eq:5d} a_1+ a_2 \re^{\hat y}+  \re^{\hat x}+ a_4 \re^{-\hat x-\hat y}+ a_5 \re^{-2\hat x- \hat y}+  \re^{- \hat x}=0, \quad [\hat x, \hat p]=\ri \hbar
\ee
where  $a_i$ are expressed by using the Batyrev coordinates $z_i$ as
\be\label{bat}
\ba
z_1&={ a_5 \over a_1 a_4}, \\
z_2&={a_4  \over a_1 a_5}, \\
z_3&={a_2 a_5 }.
\ea
\ee
Topological string partition function on toric CY manifolds corresponds to Nekrasov function for five dimensional gauge theory on $S_1\times {\IR}^4_{\epsilon_1, \epsilon_2} $, see  for instance \cite{nek5}. Hence it is useful to  use the parametrisation
\be\ba
z_1&={\re^{m R}\over H }, \\
z_2&={ \re^{-m R}\over H }\\
z_3&=\Lambda^3 R^3, 
\ea
\ee
where  $\Lambda$ is the instanton counting parameter of the five dimensional gauge theory, $H$ is a coulomb branch parameter, $m$ plays the role of a mass parameter and $R$ is the radius of $S_1$. 
This parametrisation will be useful later.
The operator we study in the setup of \cite{ghm, cgm}  is \be\label{thodef} {\rm O_{{\CB}\IF_2}}= a_2\re^{\hat p}+  \re^{\hat x} +a_4  \re^{-\hat x-\hat p}+ a_5 \re^{-2 \hat x-\hat p}+  \re^{-\hat x}, \quad [\hat x, \hat p]=\ri \hbar .\ee
We think of $ \rm O_{{\CB}\IF_2}$ as an operator on  functions in $L^2(\IR)$ which admit an analytic continuation on the strip $\pm \ri \hbar$. Then, according to \cite{ghm, cgm},  we have
\be \label{sd5d}\det(1-\re^{\mu}\rho)=\sum_{n\in \IZ}\re^{{\rm J}(\mu+2\pi \ri n +\ri \pi, \mu_1+\ri \pi, \mu_2 ,\mu_3, \hbar)},\ee
where $\rho={\rm O}_{{\CB}\IF_2}^{-1}$ and we use the following dictionary 
\be a_1=\re^{ \mu+\ri \pi}, \quad a_4=\re^{ \mu_1+\ri \pi}, \quad a_5=\re^{\mu_2},\quad a_2=\re^{\mu_3}.\ee
The quantity ${\rm J}$ in \eqref{sd5d} is the topological string grand potential associated to the local ${\CB}\IF_2$ geometry.  A self-contained definition is given for instance in \cite[eq.~(93)]{mmrev} or \cite[Sec.~3.1]{cgm}.

We now implement the geometric engineering limit on \eqref{sd5d}.
Let us first look at the operator $\rm O_{{\CB}\IF_2}$.  In this limit we rescale
\be\hat x\to R \hat x , \quad  \hbar \to R \hbar\ee
and  take $R\to 0$.
After removing the overall $R^2$ factor, we are left with the following operator \be \left( \Lambda^{3}\re^{\hat p} +(\hat x+m+{\ri \hbar \over 2})\re^{-\hat p} \right)+\hat x^2, \quad [\hat x,\hat p]=\ri \hbar .\ee 
The numerical study of this operator is a bit involved. Hence it is  convenient shift the momentum according to \footnote{Note: if the parameters are such that this shift is complex, then the spectral properties of the operator can change.} \be \hat p \to \hat p -{3\over 2}\log \Lambda +{1\over 2}\log\left(m+{\ri \hbar \over 2}\right) \ee
to write it as
\be \label{oo1} {\rm O}_{\rm 4d}= \Lambda^{3/2} \sqrt{m+{\ri \hbar \over 2}}\left( \re^{\hat p} +\re^{-\hat p} \right)+{ \Lambda^{3/2}\over \sqrt{m+{\ri \hbar \over 2}}}\hat x \re^{-\hat p}+\hat x^2, \quad [\hat x,\hat p]=\ri \hbar.\ee
Implementing the $R\to 0$ limit on the r.h.s.~of \eqref{sd5d} is long and cumbersome. Therefore we just report the results (the details of the computation are available upon request). It would be actually great to find a way to compute such determinants directly within the four dimensional theory. However so far this is not possible so we have to start from the topological string setup and then implement the geometric engineering limit. 
After some computations the final result is
\be \label{sd1}\det\left(1-{E \over {\rm O}_{\rm 4d}}\right)= A(\hbar,m, \Lambda)\re^{ \frac{3 }{8 }  {\tt a}}\re^{{\tt a}_{m}\over 2} 2\cosh \left({ {\tt a}_D \over 2 } \right),\ee
where we use \be \ba{\tt a}={}&{\tt a}(E,\hbar,m,\Lambda),\\ {\tt a}_{m}={}&{\tt a}_m(E,\hbar,m,\Lambda), \\ {\tt a}_D={}&{\tt a}_D(E,\hbar,m,\Lambda),\ea\ee
as defined in \eqref{instP}.
We tested \eqref{sd1} numerically, see for instance \autoref{tab:st}.
The term $A(\hbar,m,\Lambda)$ is fixed by the normalisation 
\be \det\left(1-{E \over {\rm O}_{\rm 4d}}\right) \Big|_{E=0}=1 \, ,\ee
which means 
\be  A(\hbar,m,\Lambda)={\re^{- \frac{3   {\tt a}}{8 }}\re^{- {\tt a}_{m} \over 2 } \over 2 \cosh \left({{\tt a}_D \over 2 }\right)}\Big |_{{\tt a}={\tt a}_0},\ee
where  \be {\tt a}_0= {\tt a} (E,\hbar,{m}, \Lambda) \mid_{E=0}.\ee
The quantization condition for the spectrum of the operator \eqref{oo1} is then given by 
\be \det\left(1-{E \over {\rm O}_{\rm 4d}}\right)=0 \, ,\ee
leading to
\be \label{qc}\cosh \left({ {\tt a}_D (E,\hbar,{m}, \Lambda)  \over 2 } \right)=0.\ee
One can easily test that this quantization condition produce the correct numerical spectrum of \eqref{oo1}, see for instance  \autoref{tb2}.  See also \cite{Zenkevich:2011zx} for a WKB analysis. Note that if
\be \label{PT}\Lambda=\re^{-\ri \pi/6}|\Lambda|, \quad {m}=\ri |{m}|, \quad \hbar \in \IR, \ee
then \eqref{oo1} is PT symmetric. By dong a simple change of variable it is easy that the spectral problem \eqref{oo1} with \eqref{PT}  is equivalent to \be \label{Opt}\widehat{{\rm O}}_{\rm 4d}=|\Lambda|^2  \re^{-2 \hat x}+\frac{|\Lambda|^2 \re^ {\hat x}}{2}+2 |\Lambda| |{m}| \re^{- \hat x}+\widehat p^2 \, .\ee
This is the same operator as in \autoref{section:qcd}, equation \eqref{conf}. In particular the quantization condition for \eqref{Opt} reads
\be\label{q1}  \cosh \left({ {\tt a}_D (E,\hbar,\ri |{m}|, \re^{-\ri \pi/6}|\Lambda|)  \over 2 } \right)=0 \, ,\ee
in perfect agreement with \eqref{qceqn}.
 For completeness let us note that one can go from \eqref{oo1} to the operator \eqref{dex2} by using
 \be \ba p &\to x+\frac{1}{2} \log \left(\frac{\Lambda}{2 (2 {m}+i \hbar )}\right), \\
 x &\to -p. \ea\ee
 Under such change (+redefinition of eigenfunctions) we obtain
 \be \hat p^2 +\frac{1}{2} \Lambda^2 \re^{-\hat x}-\Lambda^2 \re^{2 \hat x}+2 \Lambda {m} \re^{\hat x}.\ee

\subsection{From Zamolodchikov's approach}\label{Zamolodchikov}
In \cite{post-zamo} Zamolodchikov proposed a parametric family of TBAs which can be used to compute Fredholm determinant of a class of operators.
Such class include the (modified) Mathieu operator as well as the massless $N_f=1$ operator (see below)\footnote{For the massive case one should use \cite{Fateev:2005kx}, which generalise \cite{post-zamo}. We thank  Daniele Gregori for pointing out this reference.}.
Let us first summarise Zamolodchikov results by following his conventions in \cite{post-zamo}.
We look at the following operator
\be \label{Oz}\left(-{\rm O_{\rm Zamo}}-P^2\right)\phi(x)=\left(\partial_x^2 -\left(\mu_- \exp (-x b)+\mu_+ \exp ( x/b)\right) + P^2\right)\phi(x)=0\ee
where $$P, b, \mu_\pm >0.$$ It is easy to see that if
we set
\be{\label{bvalue} b={1\over \sqrt{2}}}\, ,\ee
as well as
\be 
 {|\Lambda|^2 \over \hbar^2}=2 \mu_+= 4\mu_-\, ,\ee
then we have
\be  \label{rel}  \widehat{{\rm O}}_{\rm 4d}\Big|_{m=0}=2{\rm O_{\rm Zamo}},\ee
where $\widehat{{\rm O}}_{\rm 4d}$ is defined in \eqref{PT}.
Following  \cite{post-zamo} we define\be\ba 
Q&=b+1/b\, ,\\
\mu^2&=\mu_-\mu_+^{b^2}\, ,\\
\mu^2&=\left({\re^{\theta}\over 8 \sqrt{\pi}Q}\Gamma\left({b\over 2 Q}\right)\Gamma\left({1\over 2 b Q}\right)\right)^{ 2 b Q}\, .\ea
\ee\newline
Then we consider the TBA
 \be\label{tbau0} \epsilon(\theta)= {\pi} \re^{\theta} -\int_{\IR}\rd \theta' K(\theta'-\theta)\log\left(1+\re^{- \epsilon(\theta)}\right)\, ,\ee
with \be K(x)= \frac{1}{2 \pi } \left({\frac{1}{\cosh \left(x-\frac{i \pi  {\bf \alpha}}{2}\right)}+\frac{1}{\cosh \left(x+\frac{i \pi {\bf \alpha}}{2}\right)}}\right), \quad { \alpha}= {1-b^2\over 1+b^2}={1 \over 3}\, .\ee
As pointed out in  \cite{post-zamo}, if we want to use this TBA to compute Fredholm determinants we have to supply \eqref{tbau0}  with appropriate boundary conditions as $\theta \to - \infty$. More precisely we ask that
\be \label{BC}\epsilon(\theta)\approx  Q P \theta-C(P)/2, \quad \theta\to -\infty \, ,  \ee
where \be\ba C(P)=&\log\left(16^{b P+\frac{P}{b}} b^{-4 b P} \pi ^{b P+\frac{P}{b}-1} \Gamma\left(\frac{2 P}{b}\right) \Gamma (2 b P) P\right)
\\
&-\frac{2 \left(b^2+1\right) P}{b}\log \left(\Gamma \left(\frac{b^2}{2 b^2+2}\right) \Gamma \left(1+\frac{1}{2 b^2+2}\right)\right)\, .\ea\ee
To implement  such boundary conditions on the TBA \eqref{tbau0} 
we define (recall that we are working with $b=1/\sqrt{2}$)
\be \ba 
f_0(\theta, P)&=3 \sqrt{2} P \log \left(e^{-\theta }+1\right)\\
f_1(\theta, P)&=\frac{C(P)}{e^{\theta }+1}\\
L_0(\theta, P)&=6 \sqrt{2} P \log \left(\frac{e^{\theta }+e^{2 \theta }+1}{e^{\theta }+e^{2 \theta }}\right)\\
L_1(\theta, P)&=\frac{2 \left(2 e^{\theta }+1\right) C(P)}{\left(e^{\theta }+1\right) \left(e^{\theta }+e^{2 \theta }+1\right)}\\
\ea \ee
where $f_i$ are fixed by \eqref{BC} while $L_i$ are obtained by solving 
\be 2 f_i=K\star L_i. \ee
Then the relevant TBA is 
\be\label{tbabc} \ba \epsilon(\theta,P)=&{\pi}\re^{\theta}-2f_0(\theta, P)-2f_1(\theta, P)\\
&-\int_{\IR}\rd \theta' K(\theta'-\theta)\left(\log\left(1+\re^{-\epsilon(\theta')}\right) -L_0(\theta', P)-L_1(\theta', P)\right).\ea\ee
The claim of \cite{post-zamo} is that
\be\label{xdef} \det\left(1+{P^2\over {\rm O_{\rm Zamo}}}\right)=X(\theta,P)/X(\theta,0) \, ,\ee
with
 \be \label{xzamo}X(\theta, P)=\exp\left[-{\pi \re^{\theta}\over \sqrt{3}}+\int {\log\left(1+\re^{-\epsilon(\theta',P)}\right)\over \cosh(\theta-\theta')}{\rd \theta' \over 2 \pi}\right],\ee
 where $\epsilon(\theta,P)$ is the solution to \eqref{tbabc}.
It follows that
\be \label{epX}{X(\theta+{\ri \pi \over 6},P)X(\theta-{\ri \pi \over 6},P)= \re^{- \epsilon(\theta) }.} \ee
Since the operator studied by Zamolodchikov is a particular case of the operator \eqref{PT},  the topological string approach allows to obtain an analytic, closed form solution to \eqref{xzamo}, \eqref{tbabc}. More precisely,
 consistency between our analytic expression \eqref{sd1}
 and the Zamolodchikov's TBA requires that  \be\label{zamoSW1} X(\theta, P)=\re^{ \frac{3  {\tt a}}{8  }}\re^{ {\tt a}_m \over 2 }2\cosh \left({ {\tt a}_D \over 2 } \right) \Big|_{m=0, ~E=- 2 P^2, ~ \Lambda \hbar^{-1}=\pi |Z|^{-1}\re^{\theta-\ri \pi/6}}\, \ee
 where we used \be |Z|=\frac{2^{2/3} \pi ^{3/2}}{\Gamma \left(\frac{1}{3}\right) \Gamma \left(\frac{7}{6}\right)},\ee
 as in \eqref{Zdef}.
    After some algebra, we find that \eqref{zamoSW1} can also be written as
      \be\label{zamoSW}   \re^{-\epsilon(\theta,P)}=\frac{1}{2} {\cosh \left(\frac{{\tt a}_{\rm D}}{2}\right) \cosh \left(\frac{1}{4} ({\tt a}+2 {\tt a}_{\rm D})\right)\over \sinh ^2\left(\frac{{\tt a}}{4}\right) \cosh \left(\frac{{\tt a}}{4}\right)}\Big|_{m=0, ~E=- 2 P^2, ~ \Lambda \hbar^{-1}=\pi |Z|^{-1}\re^{\theta}}\, ~ .\ee
To test the identity \eqref{zamoSW}  we compute the l.h.s
   by solving numerically the TBA \eqref{tbabc} and test that this matches the analytic expression coming from the r.h.s.~of \eqref{zamoSW}.
 For example by solving numerically   \eqref{tbabc} we find
\be\ba \epsilon(0,\sqrt{5})=&-11.41907410144\dots 
  \ea \ee
Therefore instanton counting provides analytic solutions to the TBA \eqref{xzamo},\eqref{tbabc}.

As a final remark we notice that  Zamolodchikov's TBA \eqref{tbau0}  is identical to the  conformal limit of the GMN TBA  \eqref{ZTBAeqnc} at \be \label{deg}u=0=m,\ee
provided we identify
\be \theta_{\eqref{tbau0}}=\theta_{\eqref{ZTBAeqnc}}+\log |Z|-\log \pi \, .\ee
This is why the r.h.s~of \eqref{zamoSW} precisely matches the r.h.s.~of \eqref{xxs2}.  

As discussed in \autoref{sec:GMNTBA}, the  point \eqref{deg} is quite special in the sense that the GMN TBA collapse into one equation.  This type of behaviour, and the corresponding link with the Zamolodchikov's TBA for Fredholm determinants,  was also observed in the example of  the Mathieu operator \cite{Grassi:2019coc}.
However, we do not know if it is always possible to obtain a TBA for Fredholm determinants starting from the conformal 
limit of the GMN TBA at the special point where it collapses.

\appendix

\section{The flavor period}\label{quantummass}
Here we show \eqref{pnm}.  Recall that the definition of the WKB flavor periods is
\be \sum_{n\geq 0}\hbar ^{2n}\oint_{\gamma_f} Y^{2n}(x)\rd x.\ee
By using the fact that $Y$ satisfy the Riccati equation \eqref{ricattix}, we find the following structure
\be Y^{2 n}\sim{\sqrt{P(x)}}{P(x)^{-3n}}
\sum_{\{c_0,\dots,c_n\}\in I_n} \#  \prod_{i=0}^n\left( \frac{\rd^{i} }{\rd x^i}P(x)\right)^{c_i},\ee
where $\#$ denotes  numerical factors and
  $I_n=\{c_0,\dots,c_n | \sum_{i=0}^n c_i=2n \text{ and } \sum_{i=0}^n i c_i=2n\}$. We also use
  \be P(x)=\left(\frac{\Lambda^2 \re^{-x}}{2}+2 m \Lambda \re^{x} - \Lambda^2 \re^{2x}\right)-E.\ee
 We deduce that when $x\to \infty$
 \be Y^{2n} (x) \sim  \sqrt{P(x)} \re^{-2 n x}.\ee
  By using $\tilde z=\re^{-x}$
  we have
  \be \label{yap} \oint_{\gamma_f} Y^{2n}(x)\rd x\sim \oint_0  \rd \tilde z ~ \tilde z ^{2n-2}\sqrt{2 m \Lambda \tilde z-\Lambda^2} ,\ee
   it is easy to see that the last integrand has residue equal to zero around $\tilde z=0$,  unless $n=0$. Therefore 
   \be \sum_{n\geq 0}\hbar ^{2n}\oint_{\gamma_f} Y^{2n}(x)\rd x=\oint_{\gamma_f} Y^{0}(x)\rd x.\ee

\section{Abelianization and signs}\label{ssabeli}

In this appendix we briefly relate the definition of quantum periods in the main text to the
notion of abelianization of flat connections as considered in e.g. \cite{Gaiotto:2012rg,Hollands:2013qza,Hollands:2019wbr}, with particular attention to the role of the one-loop sign.

We work in the more general setting of a Riemann surface $C$ with a spin structure,
a complex projective structure, and a holomorphic quadratic differential $\phi_2$,
which we can write locally as $\phi_2 = P(z) \de z^2$.
The Schr\"{o}dinger equation can be recast as the equation for covariantly 
constant sections of a flat $SL(2)$-connection in the 1-jet bundle 
of sections of $K_C^{-1/2}$ over $C$: in local coordinate patches this looks like

\begin{equation}\label{2x2}
\left[\partial_z+ \hbar^{-1} \begin{pmatrix}
 0 & -P(z) \\
 1 & 0
 \end{pmatrix}\right]
 \begin{pmatrix}
 -\hbar\psi'(z)\\
 \psi(z)
\end{pmatrix}=0
\end{equation}
 
The WKB solutions $\psi^{(i),\vartheta}(z)=\exp\left(\hbar^{-1}\int_{z_0}^z\lambda(z)^{(i),\vartheta} \de z\right)$ are solutions
of another differential equation,
 \begin{equation}
 \left(\partial_z-\hbar^{-1}\lambda^{(i),\vartheta}(z)\right)\psi^{(i),\vartheta}(z)=0.
  \end{equation}
 $\partial_z-\hbar^{-1}\lambda^{(i),\vartheta}(z)$ can be interpreted as a $GL(1)$-connection $\nabla^{\text{ab},\vartheta}$ 
 in a line bundle over $\Sigma$, or more precisely over the complement of the Stokes graph $\mathcal{W}^\vartheta$.  $\psi^{(i),\vartheta}(z)$ is a flat section for  $\nabla^{\text{ab},\vartheta}$. 

 $\nabla^{\text{ab},\vartheta}$ can be extended over the lift of Stokes curves to $\Sigma$, by gluing as follows.
The gluing maps for extending $\nabla^{\text{ab},\vartheta}$ across lifted Stokes curves of type $ij$ to sheet $i$ of $\Sigma$ can be computed in terms of Wronskians of solutions on the two sides of the Stokes curve: 
\begin{equation}
\left(
    \begin{array}{c}
      \psi_i^L \\
       \psi_j^L
    \end{array}
  \right)\rightarrow \begin{pmatrix}
 1 & \beta\\
 0 & 1 
 \end{pmatrix}
 \left(
    \begin{array}{c}
      \psi_i^L \\
       \psi_j^L
    \end{array}
  \right)=
  \left(
    \begin{array}{c}
      \frac{[\psi_i^L,\psi_j^L]}{[\psi_i^R,\psi_j^L]}\psi_i^R \\
       \frac{[\psi_j^L,\psi_i^L]}{[\psi_j^R,\psi_i^L]} \psi_j^R
    \end{array}
  \right)
  \end{equation}
When extending over a double wall (where Stokes curves of types $ij$ and $ji$ overlap) the gluing matrix takes the form
\begin{equation}
\left(
    \begin{array}{c}
      \psi_i^L \\
       \psi_j^L
    \end{array}
  \right)\rightarrow \begin{pmatrix}
 \rho & \beta\\
 \alpha & \rho 
 \end{pmatrix}
 \left(
    \begin{array}{c}
      \psi_i^L \\
       \psi_j^L
    \end{array}
  \right)=
  \left(
    \begin{array}{c}
      \sqrt{\frac{[\psi_i^L,\psi_j^L]}{[\psi_i^R,\psi_j^R]}\frac{[\psi_i^L,\psi_j^R]}{[\psi_i^R,\psi_j^L]}}\psi_i^R \\
      \sqrt{\frac{[\psi_j^L,\psi_i^L]}{[\psi_j^R,\psi_i^R]}\frac{[\psi_j^L,\psi_i^R]}{[\psi_j^R,\psi_i^L]}} \psi_j^R
    \end{array}
  \right),
\end{equation}
where $\rho^2-\alpha\beta=1$.

After the gluing, $\nabla^{\text{ab},\vartheta}$ is flat on $\Sigma\backslash\{\rm branch\ points\}$, and has monodromy $-1$ on loops circling only one branch point on $\Sigma$. We call $\nabla^{\text{ab},\vartheta}$ an almost-flat connection over $\Sigma$, 
 since it is flat except for the monodromy $-1$ around branch points.
Thus we have replaced an $SL(2)$-connection on $C$ by an almost-flat 
$GL(1)$-connection on $\Sigma$. This replacement is an example of the
abelianization of flat connections; we call it almost-flat abelianization.
We can think of it as a map
$$ \cM(C) \to \cM^{\rm al}(\Sigma) $$
where $\cM(C)$ denotes a moduli space of flat $SL(2)$-connections over $C$
and $\cM^{\rm al}(\Sigma)$ a moduli space of almost-flat $GL(1)$-connections over $\Sigma$.
This form of abelianization was used in \cite{Hollands:2013qza}.

One awkward feature of almost-flat abelianization is that the holonomy of $\nabla^{\text{ab},\vartheta}$ along homology classes of paths on $\Sigma$ obtains a $-1$ contribution when we move a loop across a branch point. In other words, the small loops going once around a branch point on $\Sigma$ cannot be considered as trivial. We need a $\mathbb{Z}_2$ extension $\hat{\Gamma}$ of the charge lattice to take care of these extra monodromies. Then we could define $\cX_{\hat{\gamma}}^{\rm al, \vartheta}$ for $\hat{\gamma}\in\hat{\Gamma}$ as 
\begin{equation}
\mathcal{X}_{\hat{\gamma}}^{\rm al, \vartheta}=\text{Hol}_{\hat{\gamma}}\nabla^{\rm ab,\vartheta}.
\end{equation}
$\mathcal{X}_{\hat{\gamma}}^{\rm al, \vartheta}$ constructed this way is a function on $\mathcal{M}^{\rm al}(\Sigma)$. 

It is not convenient to work with $\cX_{\hat{\gamma}}^{\rm al, \vartheta}$ directly. 
Rather, we want to work with variables labeled by 
the actual charge lattice $\gamma$. Fortunately, given 
a charge $\hat\gamma \in \hat\Gamma$
which extends $\gamma \in \Gamma$ one can define a sign $\tau(\hat\gamma)$,
in such a way that the combination
\be
\label{altof}
\cX_\gamma=\tau(\hat\gamma) \cX^{\rm al}_{\hat{\gamma}}
\ee
depends only on $\gamma$, and
the resulting $\cX_\gamma$ obey
 $\cX_\gamma \cX_{\gamma'} = \cX_{\gamma + \gamma'}$.
They can be thought of as
functions on 
the moduli space $\cM(\Sigma)$ 
of flat $GL(1)$-connections over $\Sigma$.

The sharpest definition of $\tau(\hat\gamma)$ is
as follows. Given the spin structure and the quadratic differential
$\phi_2$ on $C$, there is a canonical $GL(1)$-connection in the bundle $K_C^{-1/2}$,
which has local flat sections given by choices of $\phi_2^{-1/4}$.
Pulling this connection back to $\Sigma$ gives an almost-flat connection
in the bundle $\pi^* K_C^{-1/2}$. The holonomy of this connection around
a loop $\hat\gamma \in \hat\Gamma$ is the sign $\tau(\hat\gamma)$.
To compute $\tau(\hat\gamma)$ in practice, one can use local coordinate patches
$z$ on $C$, and local choices of $\sqrt{\de z}$, both pulled back to $\Sigma$.
Having done so, the connection form in each patch 
is simply the pullback of $-\frac14 \frac{\de P}{P}$, where $\phi_2 = P(z) \de z^2$,
but there
are potential additional signs from comparing the local choices of $\sqrt{\de z}$
on patch overlaps.

In the particular case we consider in the main body of the paper, we are taking the
standard spin structure on the cylinder: this is the one we obtain by choosing a global
$\sqrt{\de x}$ on the plane, then declaring that the shift $x \to x + 2 \pi \I$ acts
on the spin structure by $\sqrt{\de x} \to \sqrt{\de x}$, and taking the quotient.
It follows that in this paper we can compute 
$\tau(\hat\gamma)$ as $\exp(-\frac14 \oint_{\hat \gamma} \frac{\de P}{P})$, 
without any extra signs,
provided that we work 
in the cylinder coordinate, i.e. we use the representation $\phi_2 = P(x) \de x^2$.
In that form, this sign appeared in \autoref{sec:one-loop-sign} and
\autoref{ssabeli}.

The map
\begin{equation}
\cM(C) \to \cM(\Sigma)	
\end{equation}
given by the functions $\cX_\gamma$ comes from a
version of abelianization which we could 
call modified abelianization. Modified abelianization 
is the version which was used e.g. in \cite{Gaiotto:2009hg}.
The difference between modified abelianization and
almost-flat abelianization is just the sign $\tau(\hat\gamma)$.
(There is yet another version of abelianization, twisted abelianization,
which was used in \cite{Gaiotto:2012rg}; that one plays no role in 
this paper.)

\section{Some miscellany in the pure \texorpdfstring{$SU(2)$}{SU(2)} theory}

\subsection{Projections of the BPS walls}

Here we show two projections of the BPS walls in the pure $SU(2)$ theory, whose SW curve can be represented as
\be\label{SWpure} \left\{\lambda^2-\left(\frac{\Lambda^2}{z^3}-{2 u \over z^2}+{\Lambda^2\over z}\right) \de z^2=0 \right\}\, .\ee
 In the strong coupling region, the BPS spectrum is given by two hypermultiplets and their antiparticles \cite{fb}:
 \be \pm \gamma_{a+a_D}=\pm \gamma_{[1,1]},\quad  \pm \gamma_{a_D}=\pm \gamma_{[0,1]} \, .\ee  In the weak coupling region there is a vector multiplet  \be \pm \gamma_{[1,0]},\ee and an infinite tower of hypermultiplets \be \pm \gamma_{[n,1]}, \quad n\in \IZ.\ee 
  The elliptic integral expressions for $a$ and $a_D$ are given in \cite[eq.~(2.18), (2.19)]{Grassi:2019coc}.
  
\begin{figure}
     \centering
            \includegraphics[width=0.9\textwidth]{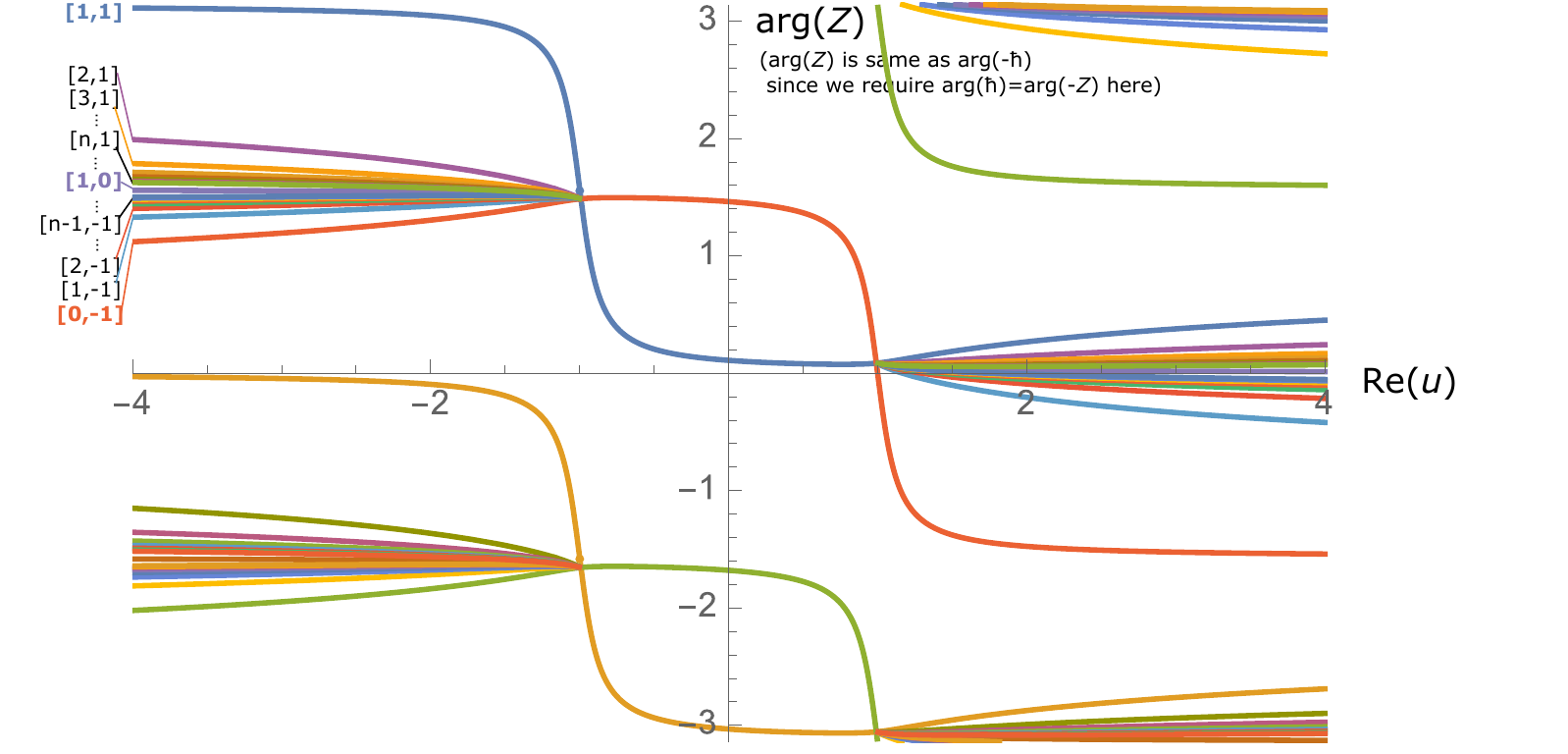}
           
         \caption{ The BPS walls when we fix  $\Lambda=1$ and ${\rm Im}(u)=\frac{\ri}{10}$. The horizontal axis is ${\rm Re}(u)$.
       There are two special points around  
          at $u_1\approx-1+\frac{\ri}{10}$ and $u_2\approx1+\frac{\ri}{10}$ where
          the wall of marginal stability is encountered: 
           $\arg(Z_{\gamma_{[1,1]}})=\arg(Z_{\gamma_{[0,-1]}})$. 
                 For ${\rm Re}(u) \in (u_1,u_2)$ (strong coupling) we have 2 hypermultiplets and their antiparticles, so there are 4 walls in total shown in this region. For ${\rm Re}(u)< {\rm Re} (u_1)$ or ${\rm Re} (u)> {\rm Re}(u_2)$ (weak coupling), we have infinitely many hypermultiplets. The hypermultiplet walls accumulate around the vectormultiplet wall. 
  We list the electromagnetic charges of the particles next to their corresponding walls.}
         \label{su2argz_reu}
\end{figure}

\begin{figure}
\centering
     \includegraphics[width=0.5\textwidth]{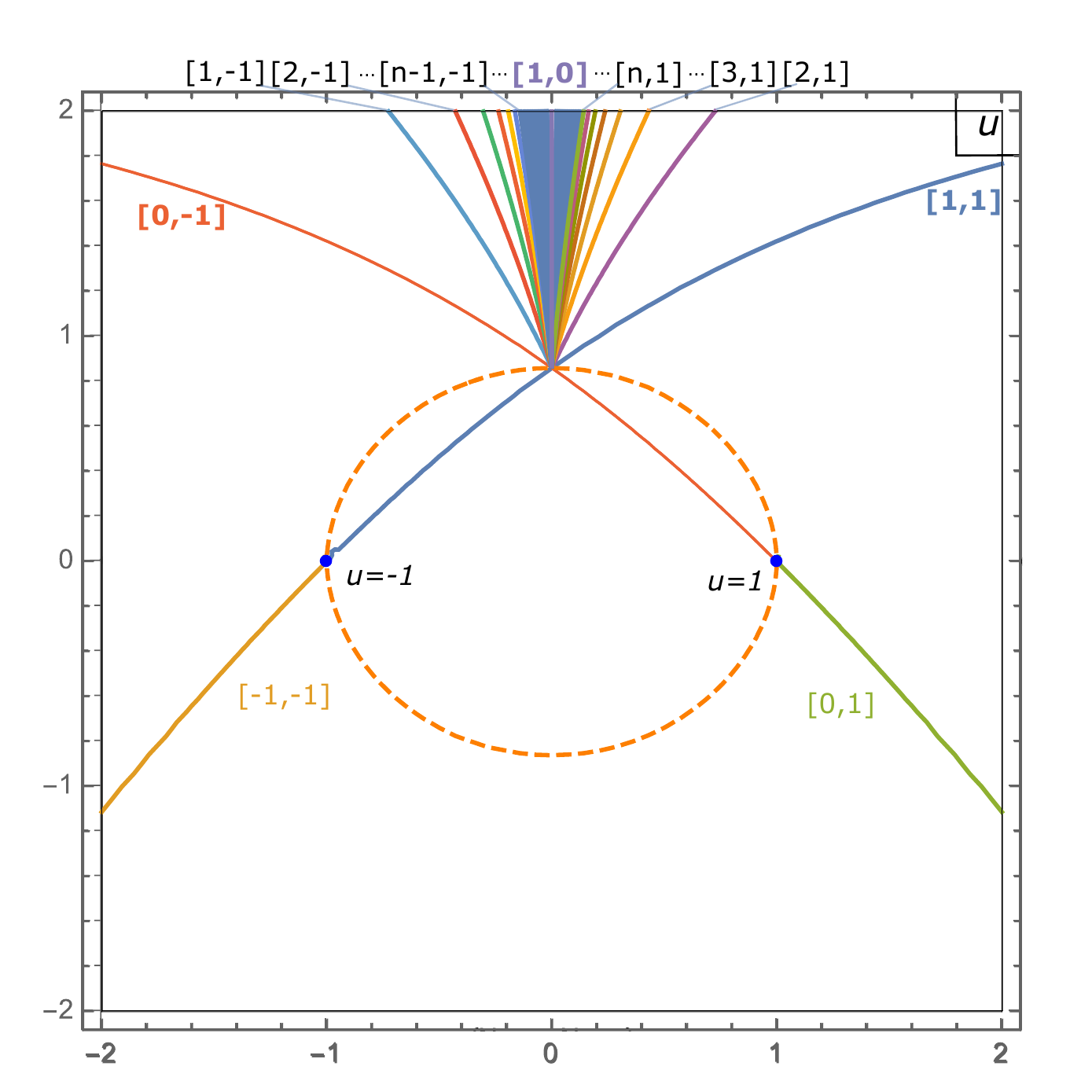}
     \caption{The $u$ plane for the pure $SU(2)$ SYM theory with  $\Lambda=1$.   The wall of marginal stability is shown as an orange dashed curve passing through  the magnetic and dyonic singularities at $u=\pm1$.
      We plot $\arg(Z_\gamma(u,\Lambda=1))=\pi/4$ for different $\gamma$.  The notation $[i , j]$ stands for a path obeying $\arg(Z_{\gamma_{[i,j]}}(u,\Lambda=1))=\pi/4$. 
     In the strong coupling region we have only two paths corresponding to
      $\arg(Z_{\gamma_{[0,-1]}}(u,\Lambda=1))=\pi/4$  and $\arg(Z_{\gamma_{ [1,1]}}(u,\Lambda=1))=\pi/4$.
            Outside the wall of marginal stability, there is a vector multiplet $[1,0] $ and an infinite tower of hypermultiplets: $[n,-1]$ and $[n+1,1]$ with $n\geq 1$. These accumulate  around $[1,0]$ when $n\rightarrow \infty$. At the boundaries of this infinite tower we have $[0,-1]$ and $[1,1]$.
      }
 \label{su2counter}
\end{figure}

 In \autoref{su2argz_reu} we show the projection to a slice in
the $\arg(-\hbar)$-$\mathrm{Re}(u)$ plane, with all the other variables held fixed;
this is the analogue of \autoref{argz_reuwhole} in the $SU(2)$ $N_f = 1$ theory.
In \autoref{su2counter} we show the projection to a slice in
the $u$-plane, with all the other variables held fixed.

\subsection{Borel plane poles of WKB solutions}\label{localsolpure}

We have seen that in \autoref{padeborelsol}, the singularities in the Borel transform
$\hat Y(z,\zeta)$
 correspond to central charges of solitons on the surface defect parametrized by $z$. We want to test this also  for the simpler example of the pure $SU(2)$ theory. 
In this case the relevant %
Schr\"{o}dinger equation is (we take $ \Lambda^2=1/2$)
\be\label{schp}(-\hbar^2\partial_x^2+\cosh x-2 u)\psi(x)=0. \ee 
 
\begin{figure}
\begin{centering}
\includegraphics[width=1\linewidth]{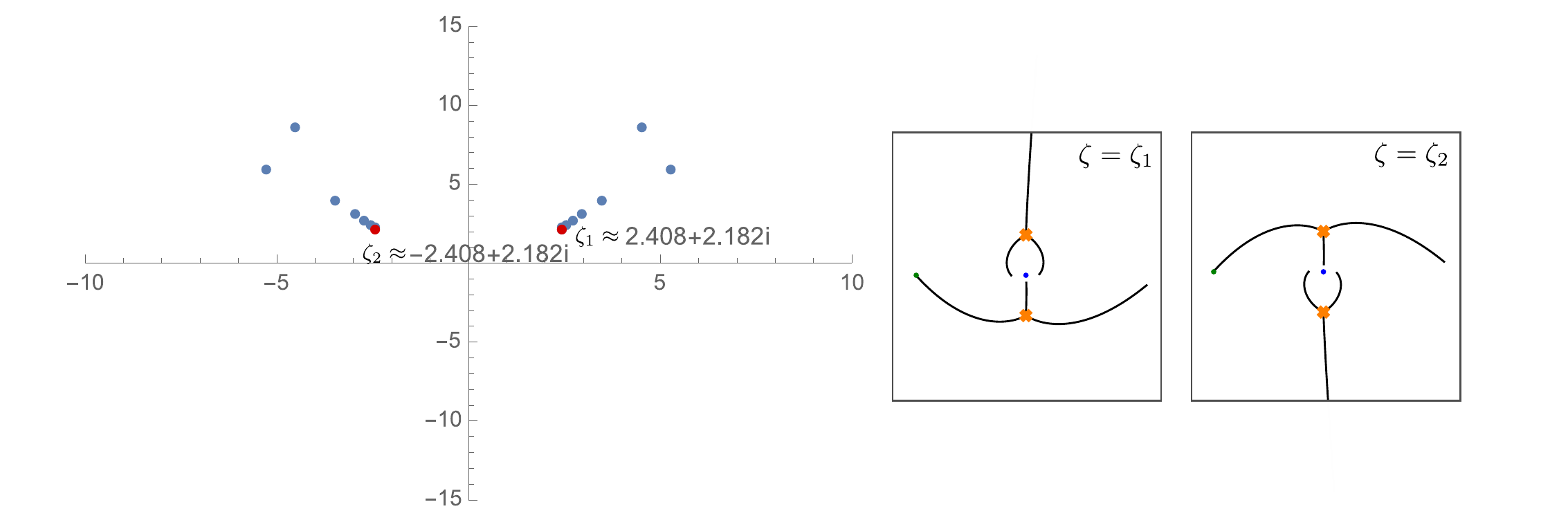}
 \caption[Caption for LOF]{
  Left: singularities of the Pad\'e-Borel transform of the ${Y}^{(2)}$ in \eqref{wkbY}  for the operator \eqref{schp}. We take  $u=0$,  $z=\re^x=-\re$ and use $N=40$ terms in the series.
The leading singularities $\zeta_i$, marked in red, match well with \eqref{polesSol}; thus as expected 
    they are the central charges of BPS solitons on the
    surface defect with parameter $z$. (For $Y^{(1)}$ the singularities would be at the opposite points $\zeta = -\zeta_i$.)
     Right: for fixed $\zeta = \zeta_i$ we plot a cutoff version 
     of the Stokes graphs with $\vartheta = \arg(\zeta)$, 
     plotting the Stokes curves only up
     to $|2\int_{z_0}^{z} \lambda| = \abs{\zeta}$.
    As explained around \eqref{polesSol}, the cutoff Stokes curves $\cW^{\vartheta=\arg(\zeta_i)}$ run exactly up to the point $z=-\re$, which is plotted as a green dot. 
}

\label{polestructure3}
\end{centering}
\end{figure}

\begin{figure}
\begin{centering}
\includegraphics[width=1\linewidth]{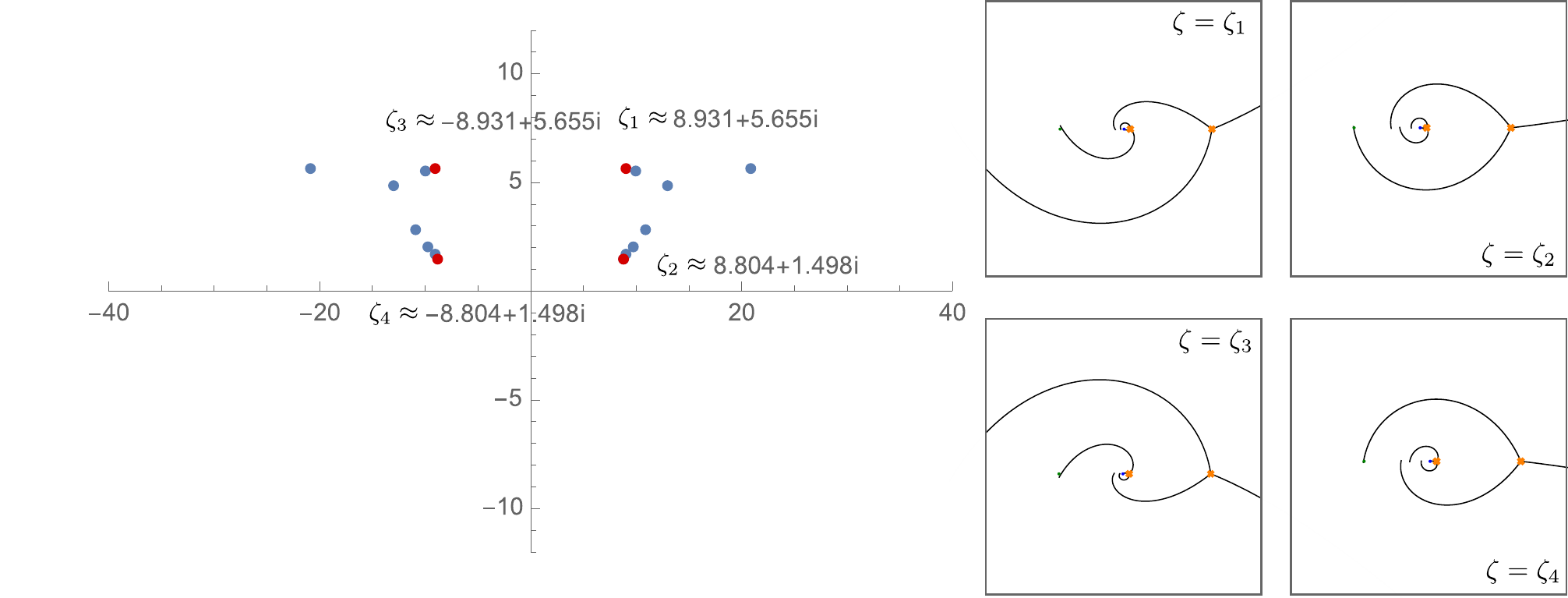}
\caption[Caption for LOF]{
    Left: singularities in the Pad\'e-Borel transform of the ${Y}^{(2)}$ in \eqref{wkbY}  for the operator \eqref{schp}. We take $u=1$,   $z=\re^x=-\re$ and use $N=40$ terms in the series.
      There are infinitely many poles of $\hat{Y}^{(2)}$, corresponding to the central charges of solitons on  the surface defect parameterized by $z$. Only finitely many of them are visible in our
  approximation. (Again, for $Y^{(1)}$ the singularities would be at the opposite points $\zeta = -\zeta_i$.)
  Right: cutoff Stokes graphs at $\zeta = \zeta_1, \zeta_2, \zeta_3,\zeta_4$,
  as in \autoref{polestructure3} above.
}
\label{polestructure2}
\end{centering}
\end{figure}

The results are shown in \autoref{polestructure3}  and \autoref{polestructure2}.  We find perfect agreement with the statement presented in \autoref{padeborelsol}.

\subsection{Relation between Fock-Goncharov coordinates and Fenchel-Nielsen coordinates in the pure \texorpdfstring{$SU(2)$}{SU(2)} theory}
\label{su2fgfn}

In \autoref{instst} we use the analytic continuation of exponentially decaying solutions, as well as the Wronskians expressions for the quantum periods, to relate Fock-Goncharov coordinates to Fenchel-Nielsen coordinates in the $SU(2)$ $N_f=1$ theory. 
This enables us to compare quantum periods gotten by instanton counting (which are Fenchel-Nielsen coordinates) with all the other Fock-Goncharov coordinates.

Here we work out the  relation between Fock-Goncharov coordinates and Fenchel-Nielsen coordinates in the case of the pure $SU(2)$ theory.

\begin{figure}
\begin{subfigure}{.5\textwidth}
  \centering
  \includegraphics[width=1\linewidth]{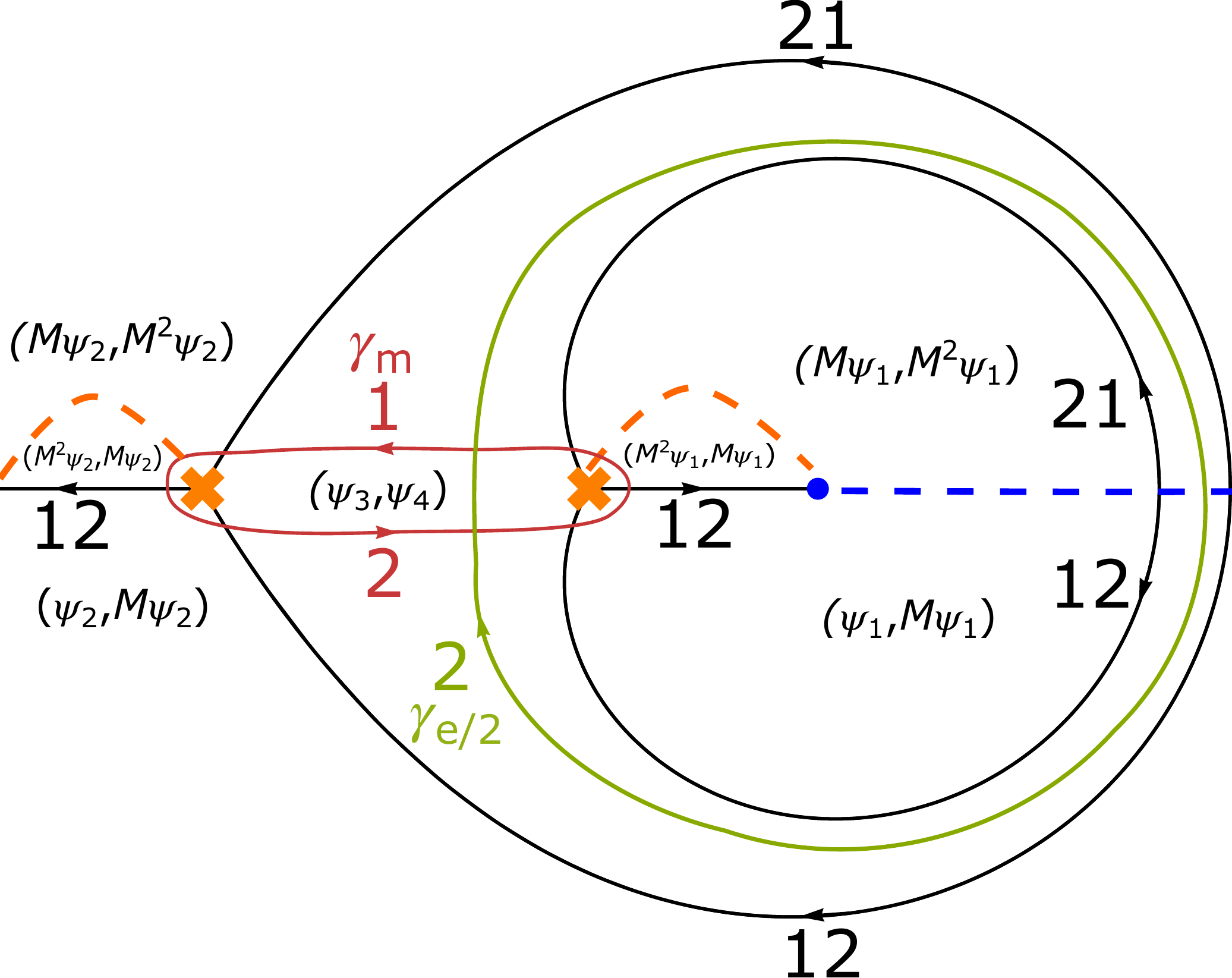}
 \caption{}
  \label{fig:SNsu2vector}
\end{subfigure}%
\begin{subfigure}{.5\textwidth}
  \centering
  \includegraphics[width=0.8\linewidth]{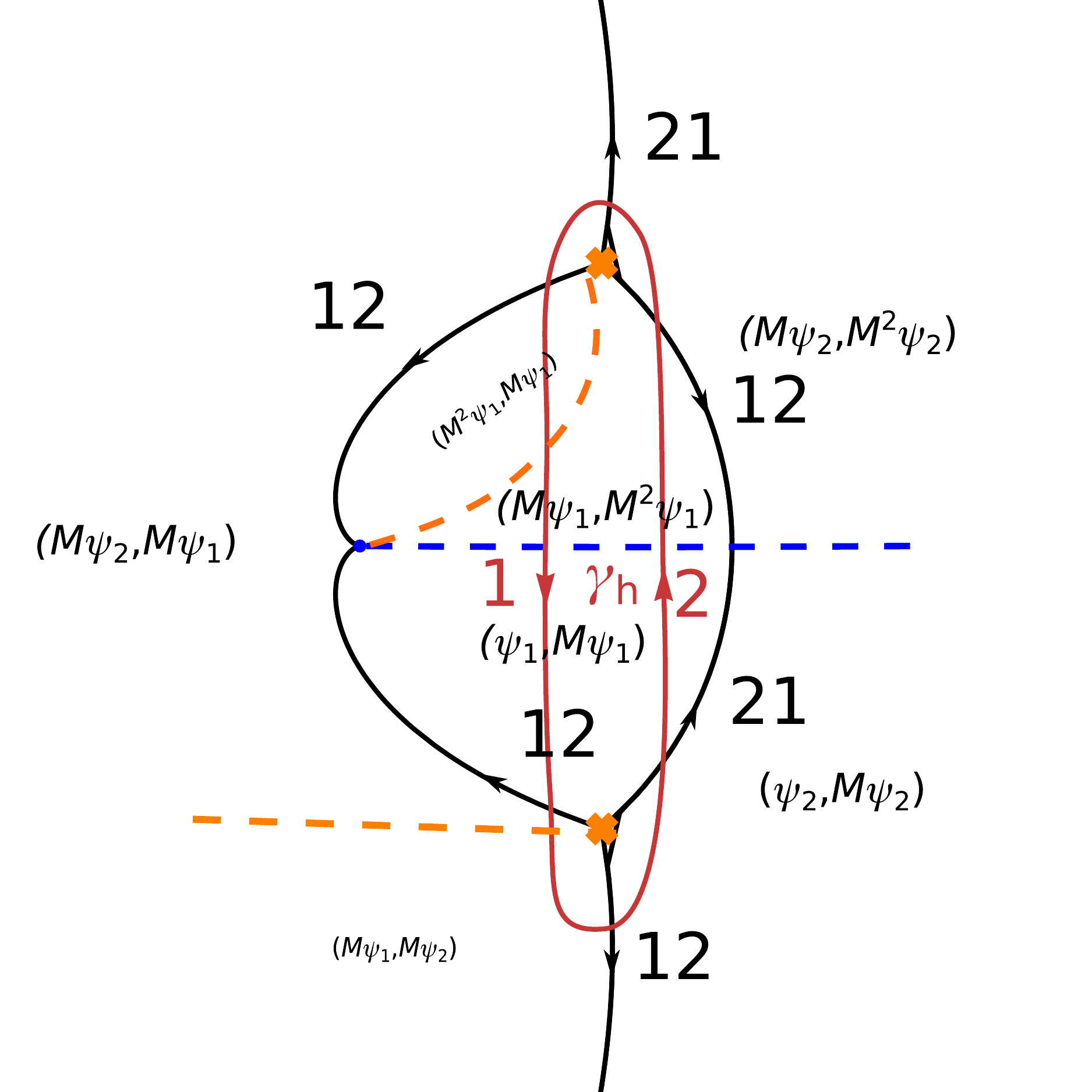}
  \caption{}
  \label{fig:SNsu2hyperp}
\end{subfigure}%
  \caption{The spectral networks used for writing the Wronskians for Fenchel-Nielsen and Fock-Goncharov coordinates. The Seiberg-Witten curve is given in \eqref{SWpure}.
  In \autoref{fig:SNsu2vector}  we choose $\Lambda=1$, $u=-9/8$. 
  In \autoref{fig:SNsu2hyperp} 
  we choose $\Lambda=1$, $u=13/20$.}
  \label{fig:su2fgfn}
\end{figure}

 For Fenchel-Nielsen coordinates we use the same spectral network $\mathcal{W}$ as in \cite{Hollands:2019wbr}, this is shown in \autoref{fig:SNsu2vector}. More precisely,  the  Fenchel-Nielsen coordinates for the ring domain and the hypermultiplet are $\mathcal{X}^{\rm FN}_{\gamma_e/2}$ and $\mathcal{X}_{\gamma_m}^{\rm FN}$ respectively  (see  \autoref{fig:SNsu2vector}).
 We also denote  $\mathcal{X}_{\gamma_h}^{\rm FG}$ the Fock-Goncharov coordinate corresponding to the hypermultiplet shown in \autoref{fig:SNsu2hyperp}. 
 
By using the spectral networks in \autoref{fig:su2fgfn}
we can obtain the Wronskian expressions for these coordinates
\footnote{\autoref{fig:SNsu2vector} and the corresponding Wronskian expressions are taken from \cite{Hollands:2019wbr}}. We get
\begin{equation}
\mathcal{X}^{\rm FN}_{\gamma_e/2}=\mu,
\end{equation}
\begin{equation}
\mathcal{X}_{\gamma_{m}}^{\rm FN}=-\frac{[\psi_1,\psi_4][\psi_2,\psi_3]}{[\psi_1,\psi_3][\psi_2,\psi_4]}\, ,
\end{equation}
\begin{equation}
\mathcal{X}_{\gamma_{h}}^{\rm FG}=-\frac{[\psi_1,M\psi_1][\psi_2,M\psi_2]}{[\psi_2,\psi_1]^2}\, 
\end{equation}
where $M\psi_1$ and $M\psi_2$ are the exponentially decaying solutions approaching $0^-$ and $-\infty$ respectively; we refer to \cite{Hollands:2019wbr} for the details. Following the same techniques described in \autoref{instst}, we get
\be
\mathcal{X}_{\gamma_{h}}^{\rm FG}=\frac{\sinh^2(\Pi^{\rm FN}_{\gamma_e}/2)}{\cosh^2(\Pi^{\rm FN}_{\gamma_m}/2)}.
  \ee
Thus
\be
\re^{\Pi_{\gamma_{h}}^{\rm FG}/2}=\frac{\sinh(\Pi^{\rm FN}_{\gamma_e}/2)}{\cosh(\Pi^{\rm FN}_{\gamma_m}/2)},
\ee
up to a sign\footnote{Once again, we do not choose a branch for the square root or fix the sign.}.
This is precisely the relation obtained in  \cite[eq.~(5.44)]{Grassi:2019coc} by comparing Fredholm determinant expressions with the TBA for quantum periods\footnote{The fact that \cite[eq.~(5.44)]{Grassi:2019coc} could be interpreted as a change of coordinates between Fenchel-Nielsen and 
 Fock-Goncharov was also noted in \cite{Coman:2020qgf}. }.

 \section{Fredholm determinant: numerical tests}
In this part we present some numerical test of formula \eqref{sd1} and its corollary \eqref{qc}. 

We start by  providing some tests of the quantization condition \eqref{qc}.   Some results are shown in \autoref{tb2}.
\begin{table}[h!] 
\centering
   \begin{tabular}{l l l}
  \\
 Nb& $E_0$   \\
\hline
  0 &\underline{1.67}6395409 + \underline{1.52}6799724 \ri \\
  1 &\underline{1.67231}1313 + \underline{1.52}8945186 \ri \\
  2 &\underline{1.675315}205 + \underline{1.5248}99573 \ri \\
  4 &\underline{1.675315}904 + \underline{1.52488}8977 \ri \\
  5 &\underline{1.6753158}35 + \underline{1.524889}216 \ri \\
  8 &1.675315869 + 1.524889085 \ri   \\ 
 \hline
Num &   1.67531587 + 1.52488909 \ri    \\
\end{tabular}\qquad
 \begin{tabular}{l l l}
  \\
 Nb& $E_0$  \\
\hline
  0&\underline{4.3}98570612\\
 1 & \underline{4.30}8165565 \\
 3 & \underline{4.3032}16550\\
 5 &\underline{4.3032}09956\\ 
   6 &        \underline{4.3032101}45     \\ 
    8 &        \underline{4.30321016}5      \\ 
 \hline
Num & 4.303210165     \\
\end{tabular}
\\
\caption{   First energy level of \eqref{oo1} for $\hbar=1$, $m=1/20$, $\Lambda=1$ (left) and $\hbar=1$, $m=2\ri$,   $\Lambda=\re^{-\ri \pi /6}$ (right). Nb denotes the order at which we truncate the instanton counting expression \eqref{NSde}. The numerical value in the last line is obtained by using numerical diagonalization of the operator.
}
 \label{tb2}
 \end{table}
 
We now want to test the expression \eqref{sd1} for the Fredholm determinant.
In principle we can compute the spectrum $E_n$ numerically  and evaluate the full determinant $\prod_{n\geq 0}\left(1-{E\over E_n}\right)$ numerically. However this product converges a bit slow.  
Hence, from a numerical point of view, it is often more convenient to compute the  spectral traces 
\be \label{def:zn}Z_{\ell}= \Tr {\rm O}_{\rm 4d}^{-\ell}=\sum_{n\geq0 } {1\over E_n^{\ell}}.\ee
These appear when we expand the Fredholm determinant at small $E$, namely
\be  \det\left(1-{E \over {\rm O}_{4d}}\right)=\prod_{n\geq 0}\left(1-{E\over E_n}\right)=\sum_{N\geq 0} (-E)^N Z(N)\ee
where the first few terms reads
\be \ba Z(1)=&Z_{1}, \\
 Z(2)=&{1\over 2}\left( Z_1^2- Z_2\right),\\
 Z(3)=&\frac{1}{6} \left(Z_1^3-3 Z_2 Z_1+2 Z_3\right), \\
 Z(4)=&\frac{1}{24} \left(Z_1^4-6 Z_2 Z_1^2+8 Z_3 Z_1+3 Z_2^2-6 Z_4\right),\\
 \cdots
 \ea\ee
 We call $Z(N)$ fermionic spectral traces to distinguish them from  the standard spectral traces $Z_\ell$ defined in \eqref{def:zn} . 
 We can invert the above relation and find
 \be\ba 
 Z_1=&Z({1}), \\
 Z_2=&Z(1)^2 - 2 Z(2), \\
 Z_3=&Z(1)^3 - 3 Z(1) Z(2) + 3 Z(3), \\
 Z_4=&Z(1)^4 - 4 Z(1)^2 Z(2) + 2 Z(2)^2 + 4 Z(1) Z(3) - 4 Z(3), \\
 \ea \ee
From a gauge/string theory perspective  we can compute these traces analytically by using the expression on the r.h.s.~of \eqref{sd1}. More precisely we have 
\be\label{zngauga}Z(N)= A(\hbar,{m},\Lambda){(-1)^N\over N!}{\rd^N\over \rd^N E }\left(
\re^{ \frac{3 }{8 }  {\tt a}}\re^{{\tt a}_{m}\over 2} 2\cosh \left({ {\tt a}_D \over 2 } \right)
\right)\Big|_{E=0}. \ee
In \autoref{tab:st} we test that the specral traces computed numerically match with the analytic expression obtained from \eqref{zngauga}. Note that numerically it is hard to compute $Z_1$ and $Z_2$ because they converge quite slowly. Hence we compute $Z_3$ and $Z_4$.
\begin{table}[h!] 
\centering
   \begin{tabular}{l l  l l}
  \\
 Nb& $Z_3$& $Z_4$  \\
\hline
 1 & \underline{24.0}2225295 & \underline{66.}8251024452\\
 3 & \underline{24.0842}4882 & \underline{66.524}1898596\\
 5 &\underline{24.0842250}1& \underline{66.5243963}962\\  
 7 &                 24.08422501  &                66.5243963159 \\  
 \hline
Num & 24.0842250  &  66.5243963159  \\
\end{tabular}    \quad
\begin{tabular}{l l  l}
  \\
 Nb&  $Z_3$  \\
\hline
 1 & \underline{24.1806}23096276  - \underline{59.72}24031294781 \ri \\
 3 & \underline{24.16864}3262235  - \underline{59.7212305}226401 \ri \\
 5 & \underline{24.168642506}932  - \underline{59.72123051495}33 \ri\\  
 7 &                  24.168642506897 -  59.7212305149523 \ri       \\  
 \hline
Num &              24.16864250 \qquad-  59.721230515 \ri  \\
\end{tabular}   
\\
\caption{ Left: the third and fourth spectral traces as computed from \eqref{zngauga} for $m=0$,  $\hbar=1$, $ \Lambda= \frac{1}{5} \re^{-\frac{1}{6} (i \pi )}$.
Right: the third  spectral trace as computed from \eqref{zngauga} for $m=1/20$,  $\hbar=1$, $ \Lambda= \frac{1}{7} $.
 We denote   by $\rm Nb$ the order at which we truncate the NS partition function \eqref{NSde}.
The numerical result is obtained by  numerical diagonalization of the operator \eqref{oo1}. }
 \label{tab:st}
 \end{table}
\newpage

\bibliography{su2-nf1}

\providecommand{\href}[2]{#2}\begingroup\raggedright\begin{thebibliography}{10}

\bibitem{bpv}
R.~Balian, G.~Parisi, and A.~Voros, ``Quartic oscillator,'' in {\em Feynman
  Path Integrals}, vol.~106, pp.~337--360.
\newblock Springer--Verlag, 1979.

\bibitem{voros-quartic}
A.~Voros, ``The return of the quartic oscillator. {T}he complex {WKB} method,''
  {\em Annales de l'I.H.P. Physique Th\'eorique} {\bf 39} (1983), no.~3,
  211--338.

\bibitem{voros}
A.~Voros, {\em Spectre de l'\'equation de {S}chr\"odinger et m\'ethode {BKW}}.
\newblock Publications Math\'ematiques d'Orsay, 1981.

\bibitem{ddpham}
E.~Delabaere, H.~Dillinger, and F.~Pham, ``Exact semiclassical expansions for
  one-dimensional quantum oscillators,'' {\em J. Math. Phys.} {\bf 38} (1997),
  no.~12, 6126--6184.

\bibitem{reshyper}
H.~Dillinger, E.~Delabaere, and F.~Pham, ``R\'esurgence de {V}oros et
  p\'eriodes des courbes hyperelliptiques,'' {\em Annales de l'Institut
  Fourier} {\bf 43} (1993) 163.

\bibitem{in-exactwkb}
K.~Iwaki and T.~Nakanishi, ``{Exact WKB analysis and cluster algebras},'' {\em
  J. Phys. A} {\bf 47} (2014), no.~47, 474009.

\bibitem{Gaiotto:2014bza}
D.~Gaiotto, ``{Opers and TBA},'' \href{http://www.arXiv.org/abs/1403.6137}{{\tt
  1403.6137}}.

\bibitem{Hollands:2019wbr}
L.~Hollands and A.~Neitzke, ``{Exact WKB and abelianization for the $T_3$
  equation},'' \href{http://www.arXiv.org/abs/1906.04271}{{\tt 1906.04271}}.

\bibitem{ns}
N.~A. Nekrasov and S.~L. Shatashvili, ``{Quantization of integrable systems and
  four dimensional gauge theories},'' in {\em {16th International Congress on
  Mathematical Physics, Prague, August 2009, 265-289, World Scientific 2010}}.
\newblock 2009.
\newblock
\href{http://www.arXiv.org/abs/0908.4052}{{\tt 0908.4052}}.
\newblock

\bibitem{nrs}
N.~Nekrasov, A.~Rosly, and S.~Shatashvili, ``{Darboux coordinates, Yang-Yang
  functional, and gauge theory},'' {\em Theor. Math. Phys.} {\bf 181} (2014),
  no.~1, 1206--1234. [Erratum: Theor.Math.Phys. 182, 368 (2015)].

\bibitem{Nekrasov:2010ka}
N.~Nekrasov and E.~Witten, ``{The Omega Deformation, Branes, Integrability, and
  Liouville Theory},'' {\em JHEP} {\bf 09} (2010) 092,
  \href{http://www.arXiv.org/abs/1002.0888}{{\tt 1002.0888}}.

\bibitem{Drukker:2009id}
N.~Drukker, J.~Gomis, T.~Okuda, and J.~Teschner, ``{Gauge Theory Loop Operators
  and Liouville Theory},'' {\em JHEP} {\bf 1002} (2010) 057,
\href{http://www.arXiv.org/abs/0909.1105}{{\tt 0909.1105}}.

\bibitem{Alday:2009fs}
L.~F. Alday, D.~Gaiotto, S.~Gukov, Y.~Tachikawa, and H.~Verlinde, ``{Loop and
  surface operators in N=2 gauge theory and Liouville modular geometry},'' {\em
  JHEP} {\bf 01} (2010) 113, \href{http://www.arXiv.org/abs/0909.0945}{{\tt
  0909.0945}}.

\bibitem{Maruyoshi:2010iu}
K.~Maruyoshi and M.~Taki, ``{Deformed Prepotential, Quantum Integrable System
  and Liouville Field Theory},'' {\em Nucl. Phys.} {\bf B841} (2010) 388--425,
\href{http://www.arXiv.org/abs/1006.4505}{{\tt 1006.4505}}.

\bibitem{Jeong:2018qpc}
S.~Jeong and N.~Nekrasov, ``{Opers, surface defects, and Yang-Yang
  functional},'' \href{http://www.arXiv.org/abs/1806.08270}{{\tt 1806.08270}}.

\bibitem{mirmor}
A.~Mironov and A.~Morozov, ``{Nekrasov functions and exact Bohr--Sommerfeld
  integrals},'' {\em JHEP} {\bf 1004} (2010) 040,
\href{http://www.arXiv.org/abs/0910.5670}{{\tt 0910.5670}}.

\bibitem{kpt}
A.-K. Kashani-Poor and J.~Troost, ``{Pure $\mathcal{N}=2$ super Yang--Mills and
  exact WKB},'' {\em JHEP} {\bf 08} (2015) 160,
\href{http://www.arXiv.org/abs/1504.08324}{{\tt 1504.08324}}.

\bibitem{Emery:2019znd}
Y.~Emery, M.~Mari\~no, and M.~Ronzani, ``{Resonances and PT symmetry in quantum
  curves},'' {\em JHEP} {\bf 04} (2020) 150,
  \href{http://www.arXiv.org/abs/1902.08606}{{\tt 1902.08606}}.

\bibitem{Dumas:2020zoz}
D.~Dumas and A.~Neitzke, ``{Opers and nonabelian Hodge: numerical studies},''
  \href{http://www.arXiv.org/abs/2007.00503}{{\tt 2007.00503}}.

\bibitem{Ito:2018eon}
K.~Ito, M.~Mari\~no, and H.~Shu, ``{TBA equations and resurgent Quantum
  Mechanics},'' {\em JHEP} {\bf 01} (2019) 228,
  \href{http://www.arXiv.org/abs/1811.04812}{{\tt 1811.04812}}.

\bibitem{Ito:2019twh}
K.~Ito, S.~Koizumi, and T.~Okubo, ``{Quantum Seiberg-Witten curve and
  Universality in Argyres-Douglas theories},'' {\em Phys. Lett. B} {\bf 792}
  (2019) 29--34, \href{http://www.arXiv.org/abs/1903.00168}{{\tt 1903.00168}}.

\bibitem{Fioravanti:2019awr}
D.~Fioravanti, H.~Poghosyan, and R.~Poghossian, ``{$T$, $Q$ and periods in
  $SU(3)$ ${\cal N}=2$ SYM},'' {\em JHEP} {\bf 03} (2020) 049,
  \href{http://www.arXiv.org/abs/1909.11100}{{\tt 1909.11100}}.

\bibitem{Yan:2020kkb}
F.~Yan, ``{Exact WKB and the quantum Seiberg-Witten curve for 4d $N=2$ pure
  $SU(3)$ Yang-Mills, Part I: Abelianization},''
  \href{http://www.arXiv.org/abs/2012.15658}{{\tt 2012.15658}}.

\bibitem{gm3}
A.~Grassi and M.~Mari\~no, ``{A Solvable Deformation of Quantum Mechanics},''
  {\em SIGMA} {\bf 15} (2019) 025,
\href{http://www.arXiv.org/abs/1806.01407}{{\tt 1806.01407}}.

\bibitem{Hatsuda_2018}
Y.~Hatsuda, A.~Sciarappa, and S.~Zakany, ``{Exact quantization conditions for
  the elliptic Ruijsenaars-Schneider model},'' {\em Journal of High Energy
  Physics} {\bf 2018} (Nov, 2018).

\bibitem{Imaizumi:2021cxf}
K.~Imaizumi, ``{Quantum periods and TBA equations for $\mathcal{N}=2\ SU(2)\
  N_f=2$ SQCD with flavor symmetry},''
  \href{http://www.arXiv.org/abs/2103.02248}{{\tt 2103.02248}}.

\bibitem{Emery:2020qqu}
Y.~Emery, ``{TBA Equations and Quantization Conditions},''
  \href{http://www.arXiv.org/abs/2008.13680}{{\tt 2008.13680}}.

\bibitem{Aminov:2020yma}
G.~Aminov, A.~Grassi, and Y.~Hatsuda, ``{Black Hole Quasinormal Modes and
  Seiberg-Witten Theory},'' \href{http://www.arXiv.org/abs/2006.06111}{{\tt
  2006.06111}}.

\bibitem{Ito:2021boh}
K.~Ito, T.~Kondo, K.~Kuroda, and H.~Shu, ``{WKB periods for higher order ODE
  and TBA equations},'' \href{http://www.arXiv.org/abs/2104.13680}{{\tt
  2104.13680}}.

\bibitem{gmn}
D.~Gaiotto, G.~W. Moore, and A.~Neitzke, ``{Four-dimensional wall-crossing via
  three-dimensional field theory},'' {\em Commun. Math. Phys.} {\bf 299} (2010)
  163--224,
\href{http://www.arXiv.org/abs/0807.4723}{{\tt 0807.4723}}.

\bibitem{ddt}
P.~Dorey, C.~Dunning, and R.~Tateo, ``{The ODE/IM Correspondence},'' {\em J.
  Phys.} {\bf A40} (2007) R205,
\href{http://www.arXiv.org/abs/hep-th/0703066}{{\tt hep-th/0703066}}.

\bibitem{nikolaev2020exact}
N.~Nikolaev, ``Exact solutions for the singularly perturbed {R}iccati equation
  and exact {WKB} analysis,'' \href{http://www.arXiv.org/abs/2008.06492}{{\tt
  2008.06492}}.

\bibitem{Gaiotto:2009hg}
D.~Gaiotto, G.~W. Moore, and A.~Neitzke, ``{Wall-crossing, Hitchin Systems, and
  the WKB Approximation},'' \href{http://www.arXiv.org/abs/0907.3987}{{\tt
  0907.3987}}.

\bibitem{Hollands:2013qza}
L.~Hollands and A.~Neitzke, ``{Spectral Networks and
  Fenchel\textendash{}Nielsen Coordinates},'' {\em Lett. Math. Phys.} {\bf 106}
  (2016), no.~6, 811--877, \href{http://www.arXiv.org/abs/1312.2979}{{\tt
  1312.2979}}.

\bibitem{Grassi:2019coc}
A.~Grassi, J.~Gu, and M.~Mari\~no, ``{Non-perturbative approaches to the
  quantum Seiberg-Witten curve},'' {\em JHEP} {\bf 07} (2020) 106,
  \href{http://www.arXiv.org/abs/1908.07065}{{\tt 1908.07065}}.

\bibitem{Dorey:1998pt}
P.~Dorey and R.~Tateo, ``{Anharmonic oscillators, the thermodynamic Bethe
  ansatz, and nonlinear integral equations},'' {\em J. Phys. A} {\bf 32} (1999)
  L419--L425, \href{http://www.arXiv.org/abs/hep-th/9812211}{{\tt
  hep-th/9812211}}.

\bibitem{cgm}
S.~Codesido, A.~Grassi, and M.~Mari\~no, ``{Spectral theory and mirror curves
  of higher genus},'' {\em Annales Henri Poincar\'e} {\bf 18} (2017), no.~2,
  559--622,
\href{http://www.arXiv.org/abs/1507.02096}{{\tt 1507.02096}}.

\bibitem{ghm}
A.~Grassi, Y.~Hatsuda, and M.~Mari\~no, ``{Topological strings from quantum
  mechanics},'' {\em Annales Henri Poincar\'e} {\bf 17} (2016), no.~11,
  3177--3235,
\href{http://www.arXiv.org/abs/1410.3382}{{\tt 1410.3382}}.

\bibitem{mmrev}
M.~Mari\~no, ``{Spectral Theory and Mirror Symmetry},'' {\em Proc. Symp. Pure
  Math.} {\bf 98} (2018) 259,
\href{http://www.arXiv.org/abs/1506.07757}{{\tt 1506.07757}}.

\bibitem{bgt}
G.~Bonelli, A.~Grassi, and A.~Tanzini, ``{Seiberg--Witten theory as a Fermi
  gas},'' {\em Lett. Math. Phys.} {\bf 107} (2017), no.~1, 1--30,
\href{http://www.arXiv.org/abs/1603.01174}{{\tt 1603.01174}}.

\bibitem{Doran:2021hcy}
C.~F. Doran, M.~Kerr, and S.~S. Babu, ``{$K_2$ and quantum curves},''
  \href{http://www.arXiv.org/abs/2110.08482}{{\tt 2110.08482}}.

\bibitem{post-zamo}
A.~B. Zamolodchikov, ``{Generalized Mathieu equations and Liouville TBA},'' in
  {\em Quantum Field Theories in Two Dimensions}, vol.~2.
\newblock World Scientific, 2012.

\bibitem{Hollands:2017ahy}
L.~Hollands and O.~Kidwai, ``{Higher length-twist coordinates, generalized
  Heun\textquoteright{}s opers, and twisted superpotentials},'' {\em Adv.
  Theor. Math. Phys.} {\bf 22} (2018) 1713--1822,
  \href{http://www.arXiv.org/abs/1710.04438}{{\tt 1710.04438}}.

\bibitem{Coman:2020qgf}
I.~Coman, P.~Longhi, and J.~Teschner, ``{From quantum curves to topological
  string partition functions II},''
  \href{http://www.arXiv.org/abs/2004.04585}{{\tt 2004.04585}}.

\bibitem{Gaiotto:2011tf}
D.~Gaiotto, G.~W. Moore, and A.~Neitzke, ``{Wall-crossing in coupled 2d-4d
  systems},'' {\em JHEP} {\bf 12} (2012)
\href{http://www.arXiv.org/abs/1103.2598}{{\tt 1103.2598}}.

\bibitem{acdkv}
M.~Aganagic, M.~C. Cheng, R.~Dijkgraaf, D.~Krefl, and C.~Vafa, ``{Quantum
  geometry of refined topological strings},'' {\em JHEP} {\bf 1211} (2012) 019,
\href{http://www.arXiv.org/abs/1105.0630}{{\tt 1105.0630}}.

\bibitem{eager}
R.~Eager, S.~A. Selmani, and J.~Walcher, ``{Exponential Networks and
  Representations of Quivers},'' {\em JHEP} {\bf 08} (2017) 063,
\href{http://www.arXiv.org/abs/1611.06177}{{\tt 1611.06177}}.

\bibitem{Longhi:2021qvz}
P.~Longhi, ``{On the BPS spectrum of 5d SU(2) super-Yang-Mills},''
  \href{http://www.arXiv.org/abs/2101.01681}{{\tt 2101.01681}}.

\bibitem{longhi}
S.~Banerjee, P.~Longhi, and M.~Romo, ``{Exploring 5d BPS Spectra with
  Exponential Networks},''
\href{http://www.arXiv.org/abs/1811.02875}{{\tt 1811.02875}}.

\bibitem{Banerjee:2020moh}
S.~Banerjee, P.~Longhi, and M.~Romo, ``{Exponential BPS graphs and D-brane
  counting on toric Calabi-Yau threefolds: Part II},''
  \href{http://www.arXiv.org/abs/2012.09769}{{\tt 2012.09769}}.

\bibitem{dfr}
M.~R. Douglas, B.~Fiol, and C.~Romelsberger, ``{The Spectrum of BPS branes on a
  noncompact Calabi-Yau},'' {\em JHEP} {\bf 09} (2005) 057,
\href{http://www.arXiv.org/abs/hep-th/0003263}{{\tt hep-th/0003263}}.

\bibitem{Closset:2019juk}
C.~Closset and M.~Del~Zotto, ``{On 5d SCFTs and their BPS quivers. Part I:
  B-branes and brane tilings},''
  \href{http://www.arXiv.org/abs/1912.13502}{{\tt 1912.13502}}.

\bibitem{Bonelli:2020dcp}
G.~Bonelli, F.~Del~Monte, and A.~Tanzini, ``{BPS quivers of five-dimensional
  SCFTs, Topological Strings and q-Painlev\'e equations},''
  \href{http://www.arXiv.org/abs/2007.11596}{{\tt 2007.11596}}.

\bibitem{wzh}
X.~Wang, G.~Zhang, and M.-x. Huang, ``{New exact quantization condition for
  toric Calabi--Yau geometries},'' {\em Phys. Rev. Lett.} {\bf 115} (2015)
  121601,
\href{http://www.arXiv.org/abs/1505.05360}{{\tt 1505.05360}}.

\bibitem{mz-wv}
M.~Mari\~no and S.~Zakany, ``{Exact eigenfunctions and the open topological
  string},'' {\em J. Phys.} {\bf A50} (2017), no.~32, 325401,
\href{http://www.arXiv.org/abs/1606.05297}{{\tt 1606.05297}}.

\bibitem{Gaiotto:2010be}
D.~Gaiotto, G.~W. Moore, and A.~Neitzke, ``{Framed BPS States},'' {\em Adv.
  Theor. Math. Phys.} {\bf 17} (2013), no.~2, 241--397,
  \href{http://www.arXiv.org/abs/1006.0146}{{\tt 1006.0146}}.

\bibitem{ks}
M.~Kontsevich and Y.~Soibelman, ``{Stability structures, motivic
  Donaldson-Thomas invariants and cluster transformations},''
\href{http://www.arXiv.org/abs/0811.2435}{{\tt 0811.2435}}.

\bibitem{MR2181810}
M.~Kontsevich and Y.~Soibelman, ``Affine structures and non-{A}rchimedean
  analytic spaces,'' in {\em The unity of mathematics}, vol.~244 of {\em Progr.
  Math.}, pp.~321--385.
\newblock Birkh\"{a}user Boston, Boston, MA, 2006.

\bibitem{MR2846484}
M.~Gross and B.~Siebert, ``From real affine geometry to complex geometry,''
  {\em Ann. of Math. (2)} {\bf 174} (2011), no.~3, 1301--1428.

\bibitem{Bridgeland_2018}
T.~Bridgeland, ``Riemann--{H}ilbert problems from {D}onaldson--{T}homas
  theory,'' {\em Inventiones mathematicae} {\bf 216} (Dec, 2018) 69--124.

\bibitem{Gaiotto:2012rg}
D.~Gaiotto, G.~W. Moore, and A.~Neitzke, ``{Spectral networks},'' {\em Annales
  Henri Poincare} {\bf 14} (2013) 1643--1731,
  \href{http://www.arXiv.org/abs/1204.4824}{{\tt 1204.4824}}.

\bibitem{dpham}
E.~Delabaere and F.~Pham, ``Resurgent methods in semi-classical asymptotics,''
  {\em Annales de l'IHP} {\bf 71} (1999) 1--94.

\bibitem{MR3706198}
Y.~Takei, ``W{KB} analysis and {S}tokes geometry of differential equations,''
  in {\em Analytic, algebraic and geometric aspects of differential equations},
  Trends Math., pp.~263--304.
\newblock Birkh\"{a}user/Springer, Cham, 2017.

\bibitem{kawai2005algebraic}
T.~Kawai and Y.~Takei, {\em Algebraic analysis of singular perturbation
  theory}, vol.~227.
\newblock American Mathematical Soc., 2005.

\bibitem{Gaiotto:2009fs}
D.~Gaiotto, ``{Surface Operators in N = 2 4d Gauge Theories},'' {\em JHEP} {\bf
  11} (2012) 090,
\href{http://www.arXiv.org/abs/0911.1316}{{\tt 0911.1316}}.

\bibitem{CecottiVafa}
S.~Cecotti and C.~Vafa, ``{On classification of {$N=2$} supersymmetric
  theories},'' {\em Commun. Math. Phys.} {\bf 158} (1993) 569--644,
  \href{http://www.arXiv.org/abs/hep-th/9211097}{{\tt hep-th/9211097}}.

\bibitem{huangNS}
M.-x. Huang, ``{On gauge theory and topological string in Nekrasov--Shatashvili
  limit},'' {\em JHEP} {\bf 06} (2012) 152,
\href{http://www.arXiv.org/abs/1205.3652}{{\tt 1205.3652}}.

\bibitem{Huang:2014nwa}
M.-x. Huang, A.~Klemm, J.~Reuter, and M.~Schiereck, ``{Quantum geometry of del
  Pezzo surfaces in the Nekrasov-Shatashvili limit},'' {\em JHEP} {\bf 02}
  (2015) 031, \href{http://www.arXiv.org/abs/1401.4723}{{\tt 1401.4723}}.

\bibitem{Hollands:2016kgm}
L.~Hollands and A.~Neitzke, ``{BPS states in the Minahan-Nemeschansky ${E_6}$
  theory},'' {\em Commun. Math. Phys.} {\bf 353} (2017), no.~1, 317--351,
\href{http://www.arXiv.org/abs/1607.01743}{{\tt 1607.01743}}.

\bibitem{Fioravanti:2019vxi}
D.~Fioravanti and D.~Gregori, ``{Integrability and cycles of deformed ${\cal
  N}=2$ gauge theory},'' {\em Phys. Lett. B} {\bf 804} (2020) 135376,
  \href{http://www.arXiv.org/abs/1908.08030}{{\tt 1908.08030}}.

\bibitem{Braverman:2004cr}
A.~Braverman and P.~Etingof, ``{Instanton counting via affine Lie algebras II:
  From Whittaker vectors to the Seiberg-Witten prepotential},''
  \href{http://www.arXiv.org/abs/math/0409441}{{\tt math/0409441}}.

\bibitem{Zenkevich:2011zx}
Y.~Zenkevich, ``{Nekrasov prepotential with fundamental matter from the quantum
  spin chain},'' {\em Phys. Lett.} {\bf B701} (2011) 630--639,
\href{http://www.arXiv.org/abs/1103.4843}{{\tt 1103.4843}}.

\bibitem{cm-ha}
S.~Codesido and M.~Mari\~no, ``{Holomorphic Anomaly and Quantum Mechanics},''
  {\em J. Phys.} {\bf A51} (2018), no.~5, 055402,
\href{http://www.arXiv.org/abs/1612.07687}{{\tt 1612.07687}}.

\bibitem{Lisovyy:2021bkm}
O.~Lisovyy and A.~Naidiuk, ``{Accessory parameters in confluent Heun equations
  and classical irregular conformal blocks},''
  \href{http://www.arXiv.org/abs/2101.05715}{{\tt 2101.05715}}.

\bibitem{felder}
G.~Felder and M.~M\"uller-Lennert, ``{Analyticity of Nekrasov Partition
  Functions},'' {\em Commun. Math. Phys.} {\bf 364} (2018), no.~2, 683--718,
\href{http://www.arXiv.org/abs/1709.05232}{{\tt 1709.05232}}.

\bibitem{bsu}
M.~A. Bershtein and A.~I. Shchechkin, ``{q-deformed Painlev{\'e} $\tau$
  function and q-deformed conformal blocks},'' {\em J. Phys.} {\bf A50} (2017),
  no.~8, 085202,
\href{http://www.arXiv.org/abs/1608.02566}{{\tt 1608.02566}}.

\bibitem{ilt}
A.~Its, O.~Lisovyy, and {\relax Yu}.~Tykhyy, ``{Connection problem for the
  sine-Gordon/Painlev\'e III tau function and irregular conformal blocks},''
  {\em Int. Math. Res. Notices} {\bf 18} (2015) 8903--8924,
\href{http://www.arXiv.org/abs/1403.1235}{{\tt 1403.1235}}.

\bibitem{Gorsky:2017ndg}
A.~Gorsky, A.~Milekhin, and N.~Sopenko, ``{Bands and gaps in Nekrasov partition
  function},'' {\em JHEP} {\bf 01} (2018) 133,
  \href{http://www.arXiv.org/abs/1712.02936}{{\tt 1712.02936}}.

\bibitem{Hatsuda:2012dt}
Y.~Hatsuda, S.~Moriyama, and K.~Okuyama, ``{Instanton Effects in ABJM Theory
  from Fermi Gas Approach},'' {\em JHEP} {\bf 01} (2013) 158,
  \href{http://www.arXiv.org/abs/1211.1251}{{\tt 1211.1251}}.

\bibitem{matone}
M.~Matone, ``{Instantons and recursion relations in $\mathcal{N}=2$ SUSY gauge
  theory},'' {\em Phys. Lett.} {\bf B357} (1995) 342--348,
\href{http://www.arXiv.org/abs/hep-th/9506102}{{\tt hep-th/9506102}}.

\bibitem{francisco}
R.~Flume, F.~Fucito, J.~F. Morales, and R.~Poghossian, ``{Matone's relation in
  the presence of gravitational couplings},'' {\em JHEP} {\bf 04} (2004) 008,
\href{http://www.arXiv.org/abs/hep-th/0403057}{{\tt hep-th/0403057}}.

\bibitem{coms}
S.~Codesido, M.~Mari\~no, and R.~Schiappa, ``{Non-Perturbative Quantum
  Mechanics from Non-Perturbative Strings},'' {\em Annales Henri Poincare} {\bf
  20} (2019), no.~2, 543--603,
\href{http://www.arXiv.org/abs/1712.02603}{{\tt 1712.02603}}.

\bibitem{voros-zeta}
A.~Voros, ``{Zeta-regularisation for exact-WKB resolution of a general 1D
  Schr\"odinger equation},''
\href{http://www.arXiv.org/abs/1202.3100}{{\tt 1202.3100}}.

\bibitem{voros-zq}
A.~Voros, ``{The zeta function of the quartic oscillator},'' {\em Nucl. Phys.}
  {\bf B165} (1980)
209--236.

\bibitem{cgm8}
S.~Codesido, A.~Grassi, and M.~Mari\~no, ``{Exact results in N=8
  Chern-Simons-matter theories and quantum geometry},'' {\em JHEP} {\bf 1507}
  (2015) 011,
\href{http://www.arXiv.org/abs/1409.1799}{{\tt 1409.1799}}.

\bibitem{cgum}
S.~Codesido, J.~Gu, and M.~Mari\~no, ``{Operators and higher genus mirror
  curves},'' {\em JHEP} {\bf 02} (2017) 092,
\href{http://www.arXiv.org/abs/1609.00708}{{\tt 1609.00708}}.

\bibitem{gm17}
A.~Grassi and M.~Mari\~no, ``{The complex side of the TS/ST correspondence},''
  {\em J. Phys.} {\bf A52} (2019), no.~5, 055402,
\href{http://www.arXiv.org/abs/1708.08642}{{\tt 1708.08642}}.

\bibitem{kkv}
S.~H. Katz, A.~Klemm, and C.~Vafa, ``{Geometric engineering of quantum field
  theories},'' {\em Nucl.Phys.} {\bf B497} (1997) 173--195,
\href{http://www.arXiv.org/abs/hep-th/9609239}{{\tt hep-th/9609239}}.

\bibitem{selfdual}
A.~Klemm, W.~Lerche, P.~Mayr, C.~Vafa, and N.~P. Warner, ``{Selfdual strings
  and N=2 supersymmetric field theory},'' {\em Nucl. Phys.} {\bf B477} (1996)
  746--766,
\href{http://www.arXiv.org/abs/hep-th/9604034}{{\tt hep-th/9604034}}.

\bibitem{nek5}
N.~Nekrasov, ``{Five dimensional gauge theories and relativistic integrable
  systems},'' {\em Nucl. Phys.} {\bf B531} (1998) 323--344,
\href{http://www.arXiv.org/abs/hep-th/9609219}{{\tt hep-th/9609219}}.

\bibitem{Fateev:2005kx}
V.~A. Fateev and S.~L. Lukyanov, ``{Boundary RG flow associated with the AKNS
  soliton hierarchy},'' {\em J. Phys. A} {\bf 39} (2006) 12889--12926,
  \href{http://www.arXiv.org/abs/hep-th/0510271}{{\tt hep-th/0510271}}.

\bibitem{fb}
F.~Ferrari and A.~Bilal, ``{The Strong coupling spectrum of the Seiberg-Witten
  theory},'' {\em Nucl. Phys.} {\bf B469} (1996) 387--402,
\href{http://www.arXiv.org/abs/hep-th/9602082}{{\tt hep-th/9602082}}.

\end{thebibliography}\endgroup
\bibliographystyle{utphys}

\end{document}